\documentstyle[11pt,epsfig]{article}

\newcommand{\be}{\begin{equation}}
\newcommand{\ee}{\end{equation}}
\newcommand{\ba}{\begin{eqnarray}}
\newcommand{\ea}{\end{eqnarray}}

\newcommand{\lp}{\left(}
\newcommand{\rp}{\right)}
\newenvironment{technical}{\begin{quotation}\small}{\end{quotation}}
[1999/03/29 PSNFSS v.7.2
Times font as default roman
: S Rahtz]

\begin{document}

\begin{center}
\centering{\LARGE Critical Market Crashes
\footnote{This paper is extracted in part from the book which
develops and documents this theme in depth [Sornette, 2003].
I acknowledge the fruitful and inspiring
discussions and collaborations with J.V. Andersen, S. Gluzman, Y. Huang, K. Ide, 
P. J\"ogi, O. Ledoit, M.W. Lee, Y. Malevergne, V.F. Pisarenko, H. Saleur, D. Stauffer,
W.-X. Zhou and especially A. Johansen.}}

\centering{\Large D. Sornette\\
Institute of Geophysics and Planetary Physics \\
and Department of Earth and Space Science\\
University of California, Los Angeles, California 90095\\
and 
Laboratoire de Physique de la Mati\`ere Condens\'ee\\ CNRS UMR6622 
and Universit\'e des Sciences, B.P. 70, Parc Valrose\\
06108 Nice Cedex 2, France}
\end{center}

\vskip1cm
{\bf Abstract}: This review presents a general theory of financial crashes and of
stock market instabilities that his co-workers and the author have developed 
over the past seven years. We start by discussing
the limitation of standard analyses for characterizing 
how crashes are special. The study of the frequency distribution 
of drawdowns, or runs of successive losses shows that large financial
crashes are ``outliers'': they form a class of their own as can be seen 
from their statistical signatures. If large financial crashes are ``outliers'', 
they are special
and thus require a special explanation, a specific model, a theory of their own.
In addition, their special properties may perhaps be used for their prediction.
The main mechanisms leading to positive feedbacks, i.e., 
self-reinforcement, such as imitative behavior and herding between investors
are reviewed with many references provided to the relevant literature outside
the narrow confine of Physics.
Positive feedbacks provide the fuel for the development of speculative bubbles,
preparing the instability for a major crash. We demonstrate several detailed
mathematical models of speculative bubbles and crashes. A first model
posits that the crash hazard drives the market price. The crash hazard may 
sky-rocket at some times due to the collective behavior of ``noise traders'',
those who act on little information, even if they think they ``know''. A second version
inverses the logic and posits that prices drive the crash hazard. Prices may
skyrocket at some times again due to the speculative or imitative behavior of 
investors. According the rational expectation model, this entails automatically
a corresponding increase of the probability for a crash. We also review
two other models including the competition between imitation and contrarian 
behavior and between value investors and technical analysts.
The most important message is the
discovery of robust and universal signatures of the approach to crashes.
These precursory patterns have
been documented for essentially all crashes on developed as well as emergent
stock markets, on currency markets, on company stocks, and so on.
We review this discovery at length and
demonstrate how to use this insight and the detailed predictions
obtained from these models to forecast crashes. For this, we review 
the major crashes of the past that occurred on the major
stock markets of the planet and describe the
empirical evidence of the universal nature of the critical
log-periodic precursory signature of crashes. 
The concept of an ``anti-bubble'' is also summarized, with the Japanese collapse 
from the beginning of 1991 to present, taken as a prominent example.
A prediction issued and advertised in Jan. 1999 has been until recently
born out with remarkable precision, predicting correctly several changes of trends,
a feat notoriously difficult using standard techniques of 
economic forecasting. We also summarize a very recent analysis 
the behavior of the US S\&P500 index from 1996 to Aug. 2002 and the forecast
for the two following years.
We conclude by presenting our view of the organization
of financial markets.

\pagebreak

\tableofcontents

\pagebreak

\section{Introduction}

The total world market capitalization rose from \$3.38 trillion (thousand
billions) in 1983 to
\$26.5 trillion in 1998 and to \$38.7 trillion in 1999. To put these numbers
in perspective, the 1999 US budget was \$1.7 trillion while its 1983
budget was \$800 billion.
Market capitalization and trading volumes tripled during the 1990s. The volume
of securities issuance was multiplied by six. Privatization has played a
key role in the stock market growth [Megginson, 2000]. Stock market investment
is clearly the big game in town. 

A market crash occurring simultaneously on most of the stock markets of the world
as witnessed in Oct. 1987 would amount to the quasi-instantaneous evaporation of trillions
of dollars. In values of Jan. 2001, 
a stock market crash of $30\%$ indeed would correspond
to an absolute loss of about 13 trillion dollars!
Market crashes can thus swallow years of pension and
savings in an instant. Could they make us suffer even more by being the precursors or 
triggering factors of major recessions as in 1929-33 after the great
crash of Oct. 1929? Or could they lead to 
a general collapse of the financial and banking system as seems to
have being barely avoided several times in the not-so-distant past?

Stock market crashes are also fascinating because they personify the class of
phenomena known as ``extreme events''. Extreme events are characteristic of
many natural and social systems, often
refered to by scientists as ``complex systems''.

Here, we discuss how financial
crashes can be understood by invoking the latest and most sophisticated concepts
in modern science, i.e., the theory of complex systems and of critical phenomena. 
Our aim is to cover a territory bringing us all the way
from the description of how the wonderful organization around us arises, to the
holy grail of crash predictions.

This article is organized in eight parts.  
Section~2 
introduces the fundamental questions:
what are crashes? How do they happen? Why do they occur? When do they occur?
Section 2 outlines the answers we propose, taking as examples some famous,
or we should rather say, infamous historical crashes. 
Section 3 discusses first the limitation of standard analyses for characterizing 
how crashes are special. It presents then the study of the frequency distribution 
of drawdowns, or runs of successive losses, and shows that large financial
crashes are ``outliers'': they form a class of their own as can be seen 
from their statistical signatures. 
If large financial crashes are ``outliers'', they are special
and thus require a special explanation, a specific model, a theory of their own.
In addition, their special properties may perhaps be used for their prediction.
Section 4 reviews the main mechanisms leading to positive feedbacks, i.e., 
self-reinforcement, such as imitative behavior and herding between investors.
Positive feedbacks provide the fuel for the development of speculative bubbles,
preparing the instability for a major crash. 
Section 5 presents two versions 
of a rational model of speculative bubbles and crashes. The first version
posits that the crash hazard drives the market price. The crash hazard may 
sky-rocket at some times due to the collective behavior of ``noise traders'',
those who act on little information, even if they think they ``know''. The second version
inverses the logic and posits that prices drive the crash hazard. Prices may
skyrocket at some times again due to the speculative or imitative behavior of 
investors. According the rational expectation model, this entails automatically
a corresponding increase of the probability for a crash. 
The most important message is the
discovery of robust and universal signatures of the approach to crashes.
These precursory patterns have
been documented for essentially all crashes on developed as well as emergent
stock markets, on currency markets, on company stocks, and so on. Section 5 also 
discusses two simple models of imitation and contrarian behavior of agents, 
leading to a chaotic dynamics of speculative bubbles and crashes and of 
the competition between value investors and technical analysts.
Section 6 takes a step back and presents the general concept
of self-similarity, with complex dimensions and their associated
discrete self-similarity. Section 6 shows how these remarkable geometric
and mathematical objects allow one to codify the information contained in
the precursory patterns before large crashes.
Section 7 analyzes the major crashes of the past that occurred on the major
stock markets of the planet. It describes the
empirical evidence of the universal nature of the critical
log-periodic precursory signature of crashes. It also
presents the concept of an ``anti-bubble'', with the Japanese collapse 
from the beginning of 1991 to present, taken as a prominent example.
A prediction issued and advertised in Jan. 1999 has been until now
born out with remarkable precision, predicting correctly several changes of trends,
a feat notoriously difficult using standard techniques of 
economic forecasting.  We also summarize a very recent analysis 
the behavior of the US S\&P500 index from 1996 to Aug. 2002 and the forecast
for the two following years. Section 8 concludes.

\section{Financial crashes: what, how, why and when?}
\vskip-0.3cm
\subsection{What are crashes and why do we care?}

Stock market crashes are momentous financial events that are fascinating to
academics and practitioners alike. According to the academic world view
that markets are efficient,
only the revelation of a dramatic piece of information can cause a crash, yet
in reality even the most thorough {\em post-mortem} analyses are typically
inconclusive as to what this piece of information might have been. For traders
and investors,
the fear of a crash is a perpetual source of stress, and the onset of the
event itself always ruins the lives of some of them.

Most approaches to explain crashes search for possible mechanisms or effects
that operate at very short time scales (hours, days or weeks at most). 
We propose here a radically different view:
the underlying cause of the crash must be searched months and years before
it, in the progressive increasing build-up of market cooperativity or 
effective interactions between investors, often translated into 
accelerating ascent of the market price (the bubble). According to this 
``critical'' point of view, the
specific manner by which prices collapsed is not the most important problem:
a crash occurs because the market has entered an unstable phase and 
any small disturbance or process
may have triggered the instability. Think of a ruler held up vertically on
your finger: this very unstable position will lead eventually to its
collapse, as a result of a small (or absence of adequate)
motion of your hand or due to any tiny whiff.
The collapse is fundamentally due to the unstable position; 
the instantaneous cause of the collapse is secondary.
In the same vein, the growth of the sensitivity
and the growing instability of the market close to such a critical point might explain
why attempts to unravel the local origin of the crash have been so diverse.
Essentially, anything would work once the system is ripe. We explore here
the concept that a crash has fundamentally an endogenous origin and 
that exogenous shocks only serve as triggering
factors. As a consequence, the origin of crashes is much more subtle than
often thought as it is
constructed progressively by the market as a whole, as a self-organizing process.
In this sense, this could be termed a systemic instability. 

Systemic instabilities are of great concern to goverments, central banks and
regulatory agencies [De Bandt and Hartmann, 2000].
The question that has often arisen in the 1990s is whether the new,
globalized, information technology-driven
economy has advanced to the point of outgrowing the set of rules dating
from the 1950s, in effect creating the need for a
new rule set for the New Economy.  Those who make this call basically point to
the systemic instabilities since 1997 (or even back to Mexico's peso crisis of
1994) as evidence that the old post-world war II rule set is now antiquated, thus
endangering this second great period of globalization to the same fate as the
first.  With the global economy appearing so fragile 
sometimes, how big of a disruption would be needed to throw a wrench
into the world's financial machinery? One of the leading moral authorities,
the Basle
 Committee on Banking Supervision, advises [1997] that,
``in handling systemic issues, it will be necessary to address, on the one hand,
risks to confidence in the financial system and contagion to otherwise sound
institutions, and, on the other hand, the need to minimise the distortion to
market signals and discipline.'' 

The dynamics of confidence and of contagion and decision making
based on imperfect information
are indeed at the core of the present work and will lead us to examine the 
following questions.
What are the mechanisms underlying crashes? Can we forecast crashes?
Could we control them?  Or at least, could we have some influence on them? 
Do crashes point to the existence of
a fundamental instability in the world financial structure?
What could be changed to mollify or suppress these instabilities?

\vskip-0.3cm
\subsection{The crash of October, 1987}

From the opening on October 14, 1987 through the market close on October 19,
major indexes of market valuation in the United States declined by 30 percent
or more. Furthermore, all major world markets declined substantially in the
month, which is itself an exceptional fact that contrasts with the usual modest
correlations of returns across countries and the fact that
stock markets around the world are amazingly diverse in their organization
[Barro et al., 1989].

In local currency units, the minimum
decline was in Austria ($-11.4\%$) and the maximum was in Hong Kong
($-45.8\%$). Out of $23$ major industrial countries 
(Autralia, Austria, Belgium, Canada, Denmark, France,
Germany, Hong Kong, Ireland, Italy, Japan, Malaysia, Mexico, Netherland, New
Zealand, Norway, Singapore, South Africa, Spain, Sweden, Swizerland, United
Kingdom, United States), $19$ had a
decline greater than $20\%$. Contrary to a common belief, the US was not the
first to decline sharply. Non-Japanese Asian markets began a severe decline on
October $19$, 1987, their time, and this decline was echoed first on a number of
European markets, then in North American, and finally in Japan. However, most of
the same markets had experienced significant but less severe declines in
the latter part of the previous week. With the exception of the US and Canada,
other markets continued downward through the end of October, and some of these
declines were as large as the great crash on October $19$.

A lot of work has been carried out to unravel the origin(s) of the crash,
notably in the properties of trading and the structure of markets; however, no
clear cause has been singled out. It is noteworthy that the strong market
decline during October 1987
followed what for many countries had been an unprecedented market increase
during the first nine months of the year and even before. In the US market for
instance, stock prices advanced $31.4 \%$ over those nine months. Some
commentators have suggested that the real cause of October's decline was that
over-inflated prices generated a speculative bubble during the earlier period.

The main explanations people have come up with are the following.
\begin{enumerate}
\item {\bf Computer Trading}. In computer trading,
also known as program trading, 
computers were programmed to automatically order large stock trades when certain
market trends prevailed, in particular sell orders after losses.
However, during the 1987 U.S. Crash, other stock markets which 
did not use program trading also crashed, some with
losses even more severe than the U.S. market. 

\item {\bf Derivative Securities}. 
Index futures and derivative securities 
have been claimed to increase the variability, risk and uncertainty 
of the U.S. stock markets.
Nevertheless, none of these techniques or practices existed in previous large and
sudden market declines in 1914, 1929, and 1962. 

\item {\bf Illiquidity}. During the crash, the large flow of sell orders
could not be digested by the trading mechanisms 
of existing financial markets. Many common stocks in the New
York Stock Exchange were not traded until late in the morning of October 19
because the specialists could not find enough buyers to purchase the amount of
stocks that sellers wanted to get rid of at certain prices. 
This insufficient liquidity may have had
a significant effect on the size of the price drop, since investors had
overestimated the amount of liquidity. However, negative news about
the liquidity of stock markets cannot explain why so many
people decided to sell stock at the same time.

\item {\bf Trade and budget deficits}. The third quarter of 1987 had the largest
U.S. trade deficit since 1960, which together with the budget deficit, 
led investors into thinking
that these deficits would cause a fall of the U.S. stocks compared with foreign
securities. However, if the
large U.S. budget deficit was the cause, why did stock markets in other countries
crash as well? Presumably, if unexpected changes in the trade deficit are bad
news for one country, it should be good news for its trading partner.

\item {\bf Overvaluation}. Many analysts agree that stock prices were overvalued in
September, 1987. While Price/Earning
 ratio and Price/Dividend ratios 
were at historically high levels, similar Price/Earning and Price/Dividends
values had been seen for most of the 1960-72 period over which no sudden crash occurred.
Overvaluation does not seem to trigger crashes every time.
\end{enumerate}

Other cited potential causes involve the auction system itself, the presence or absence
of limits on price movements, regulated margin requirements, off-market and
off-hours trading (continuous auction and automated quotations), the presence or
absence of floor brokers who conduct trades but are not permitted to invest on
their own account, the extent of trading in the cash market versus the forward
market, the identity of traders (i.e., institutions
such as banks or specialized trading firms), the significance of transaction
taxes...    
    
More rigorous and systematic analyses on univariate associations and multiple
regressions of these various factors  
 conclude that it is not clear at all what was the origin of the crash 
 [Barro et al., 1989; Roll, 1988]. The
most precise statement, albeit somewhat self-referencing, is that the most
statistically significant explanatory variable in the October crash can be
ascribed to the normal response of each country's stock market to a worldwide
market motion. A world market index was thus constructed [Barro et al., 1989; Roll, 1988]
by equally weighting the local currency indexes
of the $23$ major industrial countries mentioned above and normalized to $100$ on
september $30$. It fell to $73.6$ by October 30. The important result is that
it was found to be
statistically related to monthly returns in every country during the period
from the beginning of $1981$ until the month before the crash, albeit with a
wildly varying magnitude of the responses across countries [Barro et al., 1989; Roll, 1988].
This correlation was
found to swamp the influence of the institutional market characteristics.
This signals the possible existence of a subtle but nonetheless present
world-wide cooperativity at times preceding crashes.

\vskip-0.3cm
\subsection{How? Historical crashes}

In the financial world, risk, reward and catastrophe come in
irregular cycles witnessed by every
generation. Greed, hubris and systemic fluctuations have given us the 
Tulip Mania, the South Sea bubble, the land booms in the 1920s and 1980s, 
the US stock market and
great crash in 1929, the Oct. 1987 crash,
to name just a few of the hundreds of ready examples [White, 1996]. 

\vskip-0.3cm
\subsubsection{The Tulip mania}

The years of tulip speculation fell within a period of great prosperity
in the republic of the Netherlands. Between 1585 and 1650, Amsterdam
became the chief commercial emporium, the center of the trade of the 
northwestern part of Europe, owing to the growing commercial activity
in newly discovered America. The tulip as a cultivated flower was
imported into Western Europe from Turkey and it is first mentioned
around 1554. The scarcity of tulips and their beautiful colors made
them valuable and a must for members of the upper society.

During the build-up of the tulip market, the participants were not making money
through the actual process of production. Tulips acted as the medium of
speculation and its price determined the wealth of participants in the tulip
business. It is not clear whether the build-up attracted new investment or new
investment fueled the build-up, or both. What is known is that, as the build-up continued
more and more, people were roped in to invest their hard won earnings. The
price of the tulip lost all correlation to its comparative value with other goods
or services.

What we now call the ``tulip mania'' of the seventeenth century was the ``sure
thing'' investment during the period from mid-1500s to 1636. Before its
devastating end in 1637, those who bought tulips rarely lost money. People became
too confident that this ``sure thing'' would always make them money and, at its
peak, the participants mortgaged their houses and businesses to trade tulips. The
craze was so overwhelming that some tulip bulbs of a rare variety sold for the
equivalent of a few tens of thousand dollars. Before the crash, any suggestion
that the price of tulips was irrational was dismissed by all the participants.

The conditions now generally associated with the first period of a boom were
all present: an increasing currency, a new economy with novel colonial 
possibilities, an increasingly prosperous country,
all together had created the optimistic atmosphere in which booms
are said to grow. 

The crisis came unexpectedly. On february 4th, 1637, the possibility
of the tulips becoming definitely unsalable was mentioned for the first time.
From then to the end of May 1637, all attempts of coordination between
florists, bulbgrowers as well as by the States of Holland were met with failure.
Bulbs worth tens of thousand of US dollars (in present value) in early 1637
became valueless a few months later. This remarkable event is often 
discussed in present days and parallels are drawn with modern
speculation mania and the question is asked:
does the tulip market's build-up and its subsequent crash has any relevance for
today's times?

\vskip-0.3cm
\subsubsection{The South Sea bubble}

The South Sea Bubble is the name given to the enthusiatic speculative
fervor ending in the first great stock
market crash in England in 1720 [White, 1996]. 
The South
Sea Bubble is a fascinating story of mass hysteria, political corruption, and
public upheaval. It is really a collection of thousands of stories, tracing the
personal fortunes of countless individuals who rode the wave of stock speculation
for a furious six months in 1720. The ``Bubble year'' as it is referred to,
actually involves several individual ``bubbles'' as all kinds of fraudulent
joint-stock companies sought to take advantage of the mania for speculation.
The following account borrows from 
(The) Bubble Project.

In 1711, the South Sea Company was given a
monopoly of all trade to the south seas. The real prize was the anticipated
trade that would open up with the rich Spanish colonies in South America.
In return for this monopoly, the South Sea Company would
assume a portion of the national debt that England had incurred during the 
 War of the Spanish Succession. When
Britain and Spain officially went to war again in 1718, the immediate prospects
for any benefits from trade to South America were nil. What mattered to
speculators, however, were future prospects, and here it could always be argued
that incredible prosperity lay ahead and would be realized when open hostilities
came to an end.

The early 1700s was also a time of international finance. By 1719 the South Sea
directors wished, in a sense, to imitate the manipulation of public credit that
John Law had achieved in France with the Mississippi Company, 
which was
given a monopoly of French trade to North America; Law had connived to drive the
price of its stock up, and the South Sea directors hoped to do the same. 
In 1719 the South Sea directors made a proposal to assume the entire
public debt of the British government. On April 12, 1720 this offer was accepted. The
Company immediately starts to drive the price of the stock up through artificial
means; these largely took the form of new subscriptions combined with the
circulation of pro-trade-with-Spain stories designed to give the impression that
the stock could only go higher. Not only did capital stay in England, but many Dutch
investors bought South Sea stock, thus increasing the inflationary pressure.

South Sea stock rose steadily from January through to the spring. And as every
apparent success would soon attract its imitators, all kinds of joint-stock
companies suddenly appeared, hoping to cash in on the speculation mania. Some of
these companies were legitimate but the bulk were bogus schemes designed to take
advantage of the credulity of the people. Several of the bubbles, both large and
small, had some overseas trade or ``New World'' aspect. In addition to the South
Sea and Mississippi ventures, there was a project for improving the Greenland
fishery, another for importing walnut trees from Virginia. Raising capital sums
by selling stock in these enterprises was apparently easy work. The projects mentioned so
far all have a tangible specificity at least on paper if not in practice; others
were rather vague on details but big on promise. The most remarkable was ``A
company for carrying on an undertaking of great advantage, but nobody to know
what it is''.
The prospectus stated that
 ``the required capital was half a million, in five
thousand shares of 100 pounds each, deposit 2 pounds per share. Each subscriber,
paying his [or her] desposit, was entitled to 100 pounds per annum per share. How
this immense profit was to be obtained, [the proposer] did not condescend to
inform [the buyers] at that time.'' As
T.J. Dunning [!860] wrote: ``Capital eschews no profit, or very
small profit.... With adequate profit, capital is very bold. A certain 1\% percent
will ensure its employment anywhere; 20 percent certain will produce eagerness;
50 percent, positive audacity; 100 percent will make it ready to trample on all
human laws; 300 percent and there is not a crime at which it will scruple, nor a
risk it will not run, even to the chance of its owner being hanged.''
Next morning, at nine o'clock, this great man opened an office in
Cornhill. Crowds of people beset his door, and when he shut up at three o'cock,
he found that no less than one thousand shares had been subscribed for, and the
deposits paid. He was thus, in five hours, the winner of 2000 pounds. He was
philosophical enough to be contented with his venture, and set off the same
evening for the Continent. He was never heard of again. 

Such scams were bad for the speculation business and so largely through the
pressure of the South Sea directors, the so-called ``Bubble Act'' was passed on
June 11, 1720 requiring all joint-stock companies to have a royal charter. For a
moment, the confidence of the people was given an extra boost, and they responded
accordingly. South Sea stock had been at 175 pounds at the end of February, 380
at the end of March, and around 520 by May 29. It peaked at the end of June at
over 1000 pounds (a psychological barrier in that four-digit number).

With credulity now stretched to the limit and rumors of more and more people
(including the directors themselves) selling off, the bubble then burst
according to a slow, very slow at first, but steady
deflation (not unlike the 60\% drop of the Japanese Nikkei index
after its all time peak at the end of Dec. 1990). 
By mid August, the bankruptcy listings in the London Gazette reached an
all-time high, an indication of how people bought on credit or margin. Thousands
of fortunes were lost, both large and small. The directors attempted to pump-up more
speculation. They failed. The full collapse came by the end of September when the
stock stood at 135 pounds. The crash remained in the consciouness of the Western
world for the rest of the eighteenth century, not unlike our cultural memory of
the 1929 Wall Street Crash.

\vskip-0.3cm
\subsubsection{The Great crash of Oct. 1929}

The Roaring 1920s -- a time of growth and prosperity on Wall Street and Main
Street -- ended with the Great Crash of October 1929 (for
the most thorough and authoritative account and analysis, see [Galbraith, 1997]). 
Two thousand investment firms went under.
And the American banking industry underwent the biggest structural changes of its
history, as a new era of government regulation began.
Roosevelt's New Deal politics would follow.
The Great Depression that followed put
13 million Americans out of work (that the crash of 
Oct. 1929 caused the Great Depression is a part of financial folklore, but nevertheless 
probably not fully accurate.
For instance, using a regime switching framework, Coe [2002] finds that
a prolonged period of crisis began not with the 1929 stock market crash but
with the first banking panic of October 1930).

The Oct. 1929 crash is a remarkable illustration of several remarkable features
often associated with crashes.
First, stock market crashes are often unforeseen for most people, especially economists.
``In a few months, I expect to see the stock market much higher than today.'' Those
words were pronounced by Irving Fisher, America's distinguished and famous
economist, Professor of Economics at Yale University, 14 days before Wall Street
crashed on Black Tuesday, October 29, 1929.

``A severe depression such as 1920-21 is outside the range of probability. We are
not facing a protracted liquidation.'' This was the analysis offered days after
the crash by the Harvard Economic Society to its subscribers. After continuous
and erroneous optimistic forecasts, the Society closed its doors in 1932. Thus,
the two most renowned economic forecasting institutes in America at the time
failed to predict that a crash and a depression were forthcoming, and continued with
their optimistic views, even as the Great Depression took hold of America.
The reason is simple: predictions of trend-reversals
constitutes by far the most
difficult challenge posed to forecasters and is very 
unreliable especially within the linear framework of standard
(auto-regressive) economic models.

A second general feature exemplified by the Oct. 1929 event 
is that a financial collapse has never happened when things look bad. 
On the contrary, macroeconomic flows look good before crashes.
Before every collapse, economists say the economy is in the best of all worlds.
Everything looks rosy, stock markets go up and up, and macroeconomic flows
(output, employment, and so on.) appear to be improving further and further. This
explains why a crash catches most people, especially economists, totally by
surprise. The good times are invariably extrapolated linearly into the future. Is
it not perceived as senseless by most people in today's euphoria to talk about
crash and depression?

\begin{figure}
\begin{center}
\epsfig{file=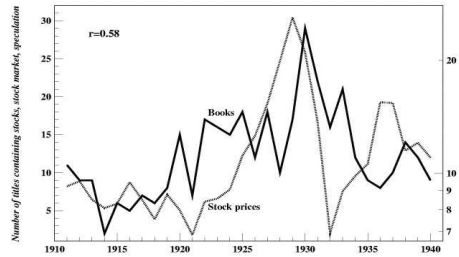,width=8cm,height=6cm}
\caption{\protect\label{bookthermo} Comparison between the number of yearly published books
about stock market speculation and the level of stock prices (1911-1940).
Black line: books at
Harvard library whose titles contain one of the words ``stocks'',
``stock market'' or ``speculation'';
grey line: Standard and Poor index of common stocks.
The curve of published books lags
behind the price curve with a time-lag of about 1.5 years, which
can be explained by the time needed for a book to get published.
Source: The stock price index is taken from the Historical Abstract
of the United States. Reproduced from [Roehner and Sornette, 2000]. }
\end{center}
\end{figure}

During the
build-up phase of a bubble such as the one preceding the Oct. 1929 crash,
there is a growing interest in the public for the
commodity in question, whether it consists in stocks, diamonds or coins.
That interest can be estimated through different indicators: increase in
the number of books published on the topic (see figure \ref{bookthermo}), 
and increase in the subscriptions
to specialized journals. Moreover, the well-known empirical rule according to
which the volume of sales is growing during a bull market finds a natural
interpretation: sales increases in fact reveal and
pinpoint the progress of the bubble's diffusion throughout society.
These features has been recently re-examined for evidence
of a bubble, a `fad' or `herding' behavior, by studying individual stock returns
[White and Rappoport, 1995].
One story often advanced for the boom of 1928 and 1929 is that it was driven by
the entry into the market of largely uninformed investors, who followed the
fortunes of and invested in `favorite' stocks. 
The result of this behavior would be a tendency for the favorite
stocks' prices to move together more than would be predicted by their shared
fundamental economic values. The comovement indeed increased significantly during the
boom and was a signal characteristic of the tumultuous market of the early 1930s.
These results are thus consistent with the possibility that a fad or crowd
psychology played a role in the rise of the market, its crash and subsequent
volatility [White and Rappoport, 1995].

The political mood before the Oct. 1929 crash
was also optimistic. In November 1928, Herbert Hoover was
elected President of the United States in a landslide, and his election set off
the greatest increase in stock buying to that date. Less than a year after the
election, Wall Street crashed.

\vskip-0.3cm
\subsection{Why? Extreme events in complex systems}

Financial markets are not the only systems with extreme events.
Financial markets constitute one among many other systems
exhibiting a complex organization and dynamics with similar behavior.
Systems with a large number of mutually interacting parts, often
open to their environment, self-organize their internal structure and 
their dynamics with novel and sometimes surprising macroscopic
(``emergent'') properties. The complex system approach, which 
involves ``seeing'' inter-connections and relationships, i.e., the whole
picture as well as the component parts, is nowadays pervasive
in modern control of engineering devices and business management.
It is also plays an increasing role in most of the scientific disciplines, including
biology (biological networks, ecology, evolution, origin of life, immunology, neurobiology,
molecular biology, and so on), geology (plate-tectonics, earthquakes and volcanoes, erosion and
landscapes, climate and weather, environment, and so on.), 
economy and social sciences (including cognition, distributed learning,
interacting agents, and so on.). There is a growing recognition that progress in
most of these disciplines, in many of the pressing issues for our future welfare
as well as for the management of our everyday life, will need
such a systemic complex system and multidisciplinary approach. This view
tends to replace the previous reductionist approach, consisting of
decomposing a system in components, such that the detailed understand of
each component was believed to bring understanding in the functioning of the whole.

A central property of a complex system is the possible occurrence of
coherent large-scale collective behaviors with a very rich structure, resulting from the 
repeated non-linear interactions among its constituents: the whole turns out 
to be much more than the sum of its parts. A part of the scientific community
holds that
most complex systems are not amenable to mathematical, analytic descriptions
and can only be explored by means of ``numerical experiments''
(see for instance [Wolfram, 2002] from an extreme implementation of this view and
[Kadanoff, 2002] for a enlightening criticism). In the context
of the mathematics of algorithmic complexity
[Chaitin, 1987], many complex systems are 
said to be computationally irreducible,
i.e. the only way to decide about their evolution is to actually let them evolve in
time. Accordingly, the ``dynamical'' future time evolution of 
complex systems would be inherently unpredictable. This unpredictability 
refers to the frustration to satisfy the quest for the knowledge of
what tomorrow will be made of, often filled by the vision of 
 ``prophets'' who have historically inspired or terrified the masses.
 
The view that complex systems are unpredictable has
recently been defended persuasively in concrete prediction applications, such as 
the socially important issue of earthquake prediction
[Geller et al., 1997] (see the 
contributions in [Nature debate, 1999] for arguments put forward
by leading seismologists and geophysicts either defending
or fighting this view). In addition to the persistent 
failures at reaching a reliable earthquake predictive scheme, this view is rooted 
theoretically in the analogy between earthquakes and 
self-organized criticality [Bak, 1996].
In this ``fractal'' framework, there is no characteristic scale and the power law
distribution of earthquake sizes reflects the fact that 
the large earthquakes are nothing but small earthquakes that did not stop. They
are thus unpredictable because their nucleation is not different from that of
the multitude of small earthquakes which obviously cannot be all predicted.

Does this really hold for all features of complex systems? Take our personal life.
We are not really interested in knowing in 
advance at what time we will go to a given store or drive to a highway. We are much more 
interested in forecasting the major bifurcations ahead of us, involving the 
few important things, like health, love and work that count for our happiness.
Similarly, predicting the detailed evolution of complex systems has no real value and the
fact that we are taught that it is out of reach from a fundamental point of view does
not exclude the more interesting possibility of predicting phases of evolutions of 
complex systems that really count, like the extreme events. 

It turns out that 
most complex systems in natural and social sciences do
exhibit rare and sudden transitions, that occur over
time intervals that are short compared to the
characteristic time scales of their posterior evolution.
Such extreme events express more than anything else the underlying ``forces'' usually
hidden by almost perfect balance and thus provide the potential for a better
scientific understanding of complex systems. 

These crises have fundamental societal
impacts and range from large natural catastrophes such as earthquakes, volcanic
eruptions, hurricanes and tornadoes, landslides,avalanches, lightning strikes,
meteorite/asteroid impacts, catastrophic events of environmental degradation, to
the failure of engineering structures, crashes in the stock market, social unrest
leading to large-scale strikes and upheaval, economic drawdowns on national and
global scales, regional power blackouts, traffic gridlock, diseases and
epidemics, and so on. It is essential to realize that
the long-term behavior of these complex systems is
often controlled in large part by these rare catastrophic events: the universe
was probably born during an extreme explosion (the ``big-bang''); the nucleosynthesis of
all important heavy atomic elements constituting our matter results from the colossal
explosion of supernovae (these stars more heavy than our sun whose
internal nuclear combustion diverges at the end of their life); the largest
earthquake in California repeating about once every two centuries accounts for
a significant fraction of the total tectonic deformation; landscapes are more
shaped by the ``millenium'' flood that moves large boulders rather than the action
of all other eroding agents; the largest volcanic eruptions lead to major 
topographic changes as well as severe climatic disruptions; 
according to some contemporary views,  evolution is 
probably characterized by phases
of quasi-stasis interrupted by episodic bursts of activity and destruction
[Gould and Eldredge, 1993]; 
financial crashes, which can destroy in an instant trillions
of dollars, loom over and shape the psychological state of investors;
political crises and revolutions shape the long-term geopolitical 
landscape; even our personal life is shaped on the long run by a few key
decisions or happenings. 

The outstanding scientific question is thus how such large-scale patterns
of catastrophic nature might evolve from a series of interactions on the smallest
and increasingly larger scales. In complex systems, it has been found that the
organization of spatial and temporal correlations do not stem, in general, from a
nucleation phase diffusing across the system. It results rather from a
progressive and more global cooperative process occurring over the whole system
by repetitive interactions. For instance, 
scientific and technical discoveries are often quasi-simultaneous in several 
laboratories in different parts of the world,
signaling the global nature of the maturing process. 

Standard models and simulations of scenarios of 
extreme events are subject to numerous sources of error, 
each of which may have a negative impact on the validity of the predictions 
[Karplus, 1992]. 
Some of the uncertainties are under control in the modeling
process; they usually involve trade-offs between a more faithful description
and manageable calculations. Other sources of errors are beyond control as they are
inherent in the modeling methodology of the specific disciplines. The two known 
strategies for modeling are both limited in this respect\,: analytical theoretical
predictions are still out of reach for many complex problems even if
notable counter-examples exist (see for instance 
[Barra et al., 2002; Arad et al., 2002; Falkovich et al., 2001]).
Brute force numerical
resolution of the equations (when they are known) or of scenarios is reliable in the
``center of the distribution'', i.e., in the regime far from the extremes where
good statistics can be accumulated. Crises are extreme events that occur rarely,
albeit with extraordinary impact, and are thus completely under-sampled and thus 
poorly constrained. Even the introduction of teraflop (or even petaflops in the future)
supercomputers does not change qualitatively this fundamental limitation.

Notwithstanding these limitations, we believe that
the progress of science and of its multidisciplinary enterprises make the time
ripe for a full-fledge effort towards the prediction of complex systems.
In particular, novel approaches are possible for modeling and
predicting certain catastrophic events, or ``ruptures'', that is, sudden
transitions from a quiescent state to a crisis or catastrophic event
[Sornette, 1999].
Such ruptures involve interactions between structures at many
different scales. In the present review, we apply these ideas to one
of the most dramatic events in social sciences, financial crashes.
The approach described here combines
ideas and tools from mathematics, physics, engineering and social sciences
to identify and classify possible universal structures that occur at
different scales, and to develop application-specific methodologies to
use these structures for prediction of the financial ``crises''.
Of special interest will be the study of the premonitory processes before
financial crashes or ``bubble'' corrections in the stock market.

For this, we will describe a new set of computational
methods which are capable of searching and comparing patterns, simultaneously
and iteratively, at multiple scales in hierarchical systems.
We will use these patterns to improve the understanding
of the dynamical state before and after a financial crash and to enhance the
statistical modeling of social hierarchical systems with the goal of
developing reliable forecasting skills for these large-scale financial crashes.

\vskip-0.3cm
\subsection{When? Is prediction possible? A working hypothesis}

Our hypothesis is that stock market 
crashes are caused by the slow buildup
of long-range correlations leading to a global cooperative behavior of the market
eventually ending into a collapse in a short
critical time interval. The use of the word ``critical''
is not purely literary
here: in mathematical terms, complex dynamical systems can go through
``critical'' points, defined as the explosion to infinity of a normally well-behaved
quantity. As a matter of fact, as far as nonlinear dynamical systems go, the
existence of
critical points is more the rule than the exception. Given the puzzling and
violent nature of stock market  crashes, it is worth investigating whether
there could possibly be a link between stock market crashes and critical
points.

\begin{itemize}

\item Our key assumption is
that a crash may be caused by {\em local} self-reinforcing imitation
between traders.
This self-reinforcing imitation process leads to the blossoming of a bubble. If
the tendency for traders to ``imitate'' their ``friends'' increases
up to a certain point called the ``critical'' point, many traders may
place the same order (sell) at the same time, thus causing a crash. The
interplay between the progressive strengthening of imitation and the ubiquity
of noise requires a probabilistic description\,: a crash is {\it not} a certain
outcome of the bubble but
can be characterised by its hazard rate, {\it i.e.}, the probability per
unit time that the crash will happen in the next instant provided it has not
happened yet.

\item Since the crash is not a certain deterministic outcome of the bubble, it
remains rational for investors to remain in the market provided they are
compensated by a higher rate of growth of the bubble for taking the risk of
a crash, because there is a finite probability of  ``landing smoothly'',
{\it i.e.}, of attaining the end of the bubble without crash. 

\end{itemize}

In a series of research articles, 
we have shown extensive evidence that
the build-up of bubbles manifests itself as an
over-all power law acceleration in the price decorated by ``log-periodic''
precursors, a concept related to fractals as will be become clear later. 
This article is to tell this story, to explain why and how these precursors occur, what
do they mean? What do they imply with respect to prediction? 

We claim that there is a degree of predictive skill associated with these
patterns. This has already been used in practice and is investigated by our 
co-workers and us as well 
as several others, academics and most-of-all practitioners (see
[Sornette and Johansen, 2001] for a recent review and assessment
and [Zhou and Sornette, 2002] for non-parametric tests using a
generalization of the so-called $q$-derivative).

The evidence we shall discuss include 
\begin{itemize}
\item the Wall street Oct.~1929, the World Oct.~1987, the Hong-Kong Oct.~1987,
the World Aug. 1998, the Nasdaq April 2000 crashes,
\item the 1985 foreign exchange event on the US dollar,
the correction of the US dollar
against the Canadian dollar and the Japanese Yen starting in Aug. 1998,
\item the bubble on the Russian market
and its ensuing collapse in 1997-98,
\item twenty-two significant bubbles followed by large crashes or by severe
corrections in
the Argentinian, Brazilian, Chilean, Mexican, Peruvian, Venezuelan, Hong-Kong,
Indonesian, Korean, Malaysian, Philippine and Thai stock markets.
\end{itemize}
In all these cases, it has been found that log-periodic power laws adequately
describe speculative bubbles on the western as well
as on the emerging markets with very few exceptions. 

Notwithstanding the drastic differences in epochs and contexts, we 
shall show that these
financial crashes share a common underlying background as well as structure. 
The rationale for this rather surprising result is probably rooted in the fact 
that humans are endowed with basically the same emotional and
rational qualities in the 21st century as they were in the 17th century (or at
any other epoch).
Humans are still essentially driven by at least a grain of greed and fear in their 
quest for a better well-being. The ``universal'' structures we are
going to uncover may be understood as the robust
emergent properties
of the market resulting from some characteristic ``rules'' of interaction 
between investors. These interactions can change in details due, for instance, to 
computers and electronic communications. They have not changed at a qualitative level.
As we shall see, complex system theory allows us to account for this robustness.

\section{Financial crashes are ``outliers''}

In the spirit of Bacon in Novum Organum about 
400 years ago, ``Errors of Nature, Sports and Monsters correct the 
understanding in regard to ordinary things, and reveal general forms. For 
whoever knows the ways of Nature will more easily notice her deviations; 
and, on the other hand, whoever knows her deviations will more accurately 
describe her ways,'' we document in this section the evidences showing that large
market drops are ``outliers'' and that they reveal 
fundamental properties of the stock market.

\vskip-0.3cm
\subsection{What are ``abnormal'' returns?}

Stock markets can exhibit very large motions, such as rallies and crashes. 
Should we expect these
extreme variations? Or should we consider them as anomalous?
\begin{figure}
\begin{center}
\caption{\protect\label{DJ-Nad90-00} Distribution of daily returns 
for the DJIA and the Nasdaq index
for the period Jan. 2nd, 1990 till Sept. 29, 2000. The lines corresponds to fits of the
data by an exponential law. The branches of negative returns have been folded back
onto the positive returns for comparison.}
\end{center}
\end{figure}

Figure \ref{DJ-Nad90-00} shows the distribution of daily returns of the DJIA
and of the Nasdaq index for the period Jan. 2nd, 1990 till Sept. 29, 2000.
For instance, we read on the figure \ref{DJ-Nad90-00} that five
negative and five positive
daily DJIA market returns larger or equal to $4\%$ have occurred.
In comparison, 15 negative and 20 positive returns
larger or equal to $4\%$ have occurred for the Nasdaq index.
The larger fluctuations of returns
of the Nasdaq compared to the DJIA are also quantified by the
so-called volatility (standard deviation of returns), 
equal to $1.6\%$ (respectively $1.4\%$) for positive
(respectively. negative) returns of the DJIA, and equal to
$2.5\%$ (respectively $2.0\%$) for positive
(respectively negative) returns of the Nasdaq index.
The lines shown in figure \ref{DJ-Nad90-00} correspond to
represent the data by an exponential function.
The upward convexity of the trajectories defined by the symbols for the Nasdaq
qualifies a stretched exponential model
[Laherr\`ere and Sornette, 1998] which embodies the fact that 
the tail of the distribution is ``fatter'', i.e., there
are larger risks of large drops (as well as ups) in the Nasdaq
compared to the DJIA.

Let us use the exponential model and calculate the probability to observe
a return amplitude larger than, say, $10$ standard deviations ($10\%$ in our 
example). The result is $0.000045$, which corresponds to $1$ event in $22,026$ days,
or in $88$ years. The drop of $22.6\%$ of Oct. 19, 1987 would correspond to one 
event in $520$ million years, which qualifies it as an ``outlier''.
Thus, according to the exponential model, a $10\%$ return amplidude does not qualify 
as an ``outlier'', in a clear-cut and undisputable manner. In addition,
the discrimination between normal and abnormal returns depends on our choice
for the frequency distribution. Qualifying what is the correct description of the 
frequency distribution, especially for large positive and negative returns, is a 
delicate problem that is still a hot domain for research. Due to the lack of 
certainty on the best choice for the frequency distribution, this approach does
not seem the most adequate for characterizing anomalous events.
We now introduce another diagnostic that allows us to characterize
abnormal market phases in a much more precise and non-parametric way, i.e., 
without refering to a specific mathematical representation of the frequency
distribution.

\vskip-0.3cm
\subsection{Drawdowns (runs)}

Extreme value theory (EVT) provides an alternative approach, still based on the 
distribution of returns estimated at a fixed time scale. Its most practical 
implementation is based on the so-called ``peak-over-threshold'' distributions
[Embrechts et al., 1997; Bassi et al., 1998], which is founded
on a limit theorem known as the Gnedenko-Pickands-Balkema-de Haan theorem which 
gives a natural limit law for peak-over-threshold values in the form of the 
Generalized Pareto Distribution (GPD), a family of distributions with two parameters
based on the Gumbel, Weibull and Frechet extreme value distributions. 
The GPD is either an exponential or has a power law tail. Peak-over-threshold
distributions put the emphasis on the characterization of the tails of 
distribution of returns and have thus been scrutinized for their potential
for risk assessment and management of large and extreme events
(see for instance [Phoa, 1999; McNeil, 1999]). In particular,
extreme value theory provides a general foundation for the estimation of 
the value-at-risk for very low-probability ``extreme'' events. There
are however severe pitfalls [Diebold et al., 2001] in the use of extreme
value distributions for risk management because of its reliance on
the (unstable) estimation of tail probabilities. In addition, the EVT
literature assumes independent returns, 
which implies that the degree of fatness in the tails decreases as 
the holding horizon lengthens (for the values
of the exponents found empirically).  Here, we show that this is not the case:
returns exhibit strong correlations at special times precisely 
characterized by the occurrence of extreme events, the regime that EVT aims
to describe. This suggests to re-examine EVT and extend it to variable
time scales, for instance by analyzing the EVT of the distribution of 
drawdowns and drawups. 

A drawdown is defined as a
persistent decrease in the price over consecutive days. A drawdown
is thus the cumulative loss from the
last maximum to the next minimum of the price. Drawdowns embody
a rather subtle dependence
since they are constructed from runs of the same sign variations. Their
distribution thus captures the way successive drops can influence 
each other and
construct in this way a persistent process.
This persistence is not measured by the
distribution of returns because, by its very definition, it forgets about
the relative positions of the returns as they unravel themselves as a function
of time by only counting their frequency. This is not detected either by
the two-point correlation function, which measures an {\it average}
linear dependence over the whole time series, while the dependence may only
appear at special times, for instance for
very large runs, a feature that
will be washed out by the global averaging procedure.

To demonstrate the information contained in drawdowns and contrast it
with the fixed time-scale returns, let us consider the hypothetical
situation of a crash of $30\%$ occurring over three days with three 
successive losses of exactly $10\%$. 
The crash is thus defined as the total loss or drawdown
of $30\%$. Rather than looking at drawdowns, let us now follow
the common approach and examine the daily data, in particular the 
daily distribution of returns. The $30\%$ drawdown is now seen as
three daily losses of $10\%$. The essential point to realize is that the 
construction of the distribution of returns amounts to count the
number of days over which a given return has been observed. The crash
will thus contribute three days of $10\%$ loss, {\it without} the
information that the three losses occurred sequentially!
To see what this loss of information entails, we consider a market in 
which a $10\%$ daily loss occurs typically once every $4$ years (this 
is not an unreasonable number for the Nasdaq composite index at
present times of high volatility). Counting
approximately $250$ trading days per year, $4$ years correspond to $1,000$ trading
days and $1$ event in 1000 days thus corresponds to a probability $1/1,000=0.001$
for a daily loss of $10\%$. The crash of $30\%$ has been dissected as three events which 
are not very remarkable (each with a relatively short average
recurrence time of four years). The plot thickens
when we ask what is, according to this description,
the probability for three successive daily losses of $10\%$? 
Elementary probability tells us that it is the probability of one daily loss of
$10\%$ times the probability of one daily loss of
$10\%$ times the probability of one daily loss of
$10\%$, giving $10^{-9}$.
This corresponds to a $1$ event in $1$ billion trading days! 
We should thus wait typically 
$4$ millions years to witness such an event! 

What has gone wrong? Simply, looking at daily returns and at their
distributions has destroyed the information that
the daily returns may be correlated, at special times! This crash is like
a mammoth which has been dissected in pieces without memory of the connection
between the parts and we are left with what look as mouses (bear
with the slight exageration)! Our estimation 
that three successive losses of $10\%$ are utterly impossible relied on the 
incorrect hypothesis that these three events are independent. Independence
between successive returns is remarkably well-verified most of the time.
However, it may be that large drops may not be independent. In other words,
there may be ``burst of dependence'', i.e., 
``pockets of predictability''.

It is clear that drawdowns will keep precisely the information 
relevant to identify the possible burst of local dependence leading
to possibly extraordinary large cumulative losses. 
Our emphasis on drawdowns is thus motivated by two considerations: 1) 
drawdowns are important measures of risks used by practitioners because they represent 
their cumulative loss since the last estimation of their wealth. It
is indeed a common psychological trait of people to estimate a loss
by comparison with the latest maximum wealth; 2) drawdowns automatically capture an
important part of the time dependence of price returns, similarly to the 
 run-statistics often used in statistical testing [Knuth, 1969]
and econometrics [Campbell et al., 1997; Barber and Lyon, 1997].
As previously showed [Johansen and Sornette, 1998, 2002], 
the distribution of drawdowns 
contains an information which is quite 
different from the distribution of returns over a fixed time scale. 
In particular, a drawndown embodies the interplay 
between a series of losses and hence measures a ``memory''  of the 
market. Drawdowns examplify the effect of correlations in price 
variations when they appear, which must be taken into account for a 
correct characterisation of market price variations. They are direct
measures of a possible amplification or ``flight of fear'' where previous 
losses lead to further selling, strengthening the downward trend,
occasionally ending in a crash. We stress
that drawdowns, by the ``elastic'' time-scale used to define them,
are effectively function of several higher order correlations at the same time.

Johansen and Sornette [2002] have shown that
the distribution of drawdowns for independent
price increments $x$ is asymptotically an exponential 
(while the body of the distribution is Gaussian [Mood, 1940]), when the distribution
of $x$ does not decay more slowly than an exponential, i.e., belong to the class
of exponential or super-exponential distributions. In contrast, for sub-exponentials
(such as stable L\'evy laws, power laws and stretched exponentials), the 
tail of the distribution of drawdowns is asymptotically the same as the 
distribution of the individual price variations. 
Since stretched exponentials have been found to offer 
an accurate quantification of price variations 
[Lah\'errere and Sornette, 1998; Sornette et al., 2000; Andersen and Sornette, 2001]
thus capturing a possible sub-exponential behavior
and since they contain the exponential law as a special case,
the stretched exponential law is a good null hypothesis.

The cumulative stretched distribution is defined by
\be \label{stretched}
N_c\lp x\rp = A ~\exp \left( - (|x|/\chi)^z \right) ~,
\ee
where $x$ is either a drawdown or a drawup. 
When $z<1$ (resp. $z>1$), $N_c\lp x\rp$ is a stretched exponential
or sub-exponential (resp. super-exponential). The special case $z=1$
corresponds to a pure exponential. In this case, $\chi$ is nothing but
the standard deviation of $|x|$.

Johansen and Sornette [2002] have analyzed the major financial
indices, the major currencies, gold, the twenty 
largest U.S. companies in terms of capitalisation as well as nine others
chosen randomly. They find that approximately $98\%$ of the distributions of drawdowns
is well-represented by an exponential or a stretched exponential, while the largest to the few ten
largest drawdowns are occurring with a significantly larger rate
than predicted by the exponential. This is confirmed by 
extensive testing on surrogate data. Very large drawdowns thus
belong to a different class of their own and call for a specific amplification
mechanism. Drawups (gain from the last local minimum to the next local
maximum) exhibit a similar behavior in only about half the markets examined.
We now present some of the most significant results.

\vskip-0.3cm
\subsection{Testing outliers}

Testing for ``outliers'' or more generally for a change of
population in a distribution is a subtle problem:
the evidence for outliers and extreme events does not require
and is not even synonymous in general with the existence of
a break in the distribution of the drawdowns. Let us illustrate this
pictorially by borrowing from another domain of active scientific
investigation, namely the search for the understanding of the complexity
of eddies and vortices in turbulent hydrodynamic flows, such as
in mountain rivers or in the weather. Since
solving the exact equations of these flows does not provide
much insight as the results are forbidding,
a useful line of attack has been to simplify the problem
by studying simple toy models, such as
``shell'' models of turbulence, that are believed to capture the
essential ingredient of these flows, while being amenable
to analysis. Such ``shell'' models replace the three-dimensional
spatial domain by a series of uniform onion-like spherical layers
with radii increasing as a geometrical series $1, 2, 4, 8, ..., 2^n$
and communicating with each other typically with nearest and next-nearest neighbors.

As for financial returns, a quantity of great interest
is the distribution of velocity variations between two instants at
the same position or between two points simultaneously. Such a
distribution for the square of the velocity variations has been
calculated  [L'vov et al., 2001] and exhibits an approximate
exponential drop-off as well as a co-existence with larger fluctuations, 
quite reminiscent of our findings in finance 
[Johansen and Sornette, 1998; 2002].
Usually, such large fluctuations are not considered to be
statistically significant and do not provide any specific insight.
Here, it turns out that it can be shown that these large
fluctuations of the fluid velocity correspond to intensive
peaks propagating coherently over several shell layers with a
characteristic bell-like shape, approximately independent of
their amplitude and duration (up to a re-scaling of their size
and duration). When extending these observations to very long
times so that the anomalous fluctuations can be sampled much better, one gets
a continuous distribution [L'vov et al., 2001]. Naively, one would 
expect that the
same physics apply in each shell layer (each scale) and, as a consequence,
the distributions in each shell should be the same, up to a change of 
unit reflecting
the different scale embodied by each layer. It turns out that the three curves
for three different shells
can indeed by nicely collapsed, but only for the small velocity 
fluctuations, while
the large fluctuations are described by very different heavy tails.
Alternatively, when one tries to collapse the curves in the region of the large
velocity fluctuations, then the portions of the curves close to the origin
are not collapsed at all and are very different.
The remarkable conclusion is that the distributions
of velocity increment seem to be composed of two regions, a region of 
``normal scaling'' and a domain of extreme events.
The theoretical analysis of L'vov et al. [2001] further substantiate
the fact that the largest fluctuations result from a different mechanism.

Here is the message that comes out of this discussion: the concept of 
outliers and of
extreme events does not rest on the
requirement that the distribution should not be smooth. Noise and
the very process of constructing the distribution
will almost always smooth out the curves. What is found by L'vov et al. [2001]
is that the distribution is made of two different populations, the body and
the tail, which have different physics, different
scaling and different properties. This is a clear demonstration that this model
of turbulence exhibits outliers in the sense that there is a well-defined
population of very large and quite rare events that punctuate the dynamics and
which cannot be seen as scale-up versions of the small fluctuations.

As a consequence, the fact that the distribution of small events might show
up some curvature or continuous behavior does not tell anything against
the outlier hypothesis. It is essential to keep this point in mind when
looking at the evidence presented below for the drawdowns.

Other groups have recently presented supporting evidence that 
crash and rally days significantly differ in their statistical properties from the 
typical market days.
For instance, Lillo and Mantegna [2000] investigated the return distributions of an ensemble 
of stocks simultaneously traded in the New York Stock Exchange (NYSE) during
market days of extreme crash or rally 
in the period from January 1987 to December 1998. Out of
two hundred distributions of returns, one for each 
of two hundred trading days where the ensemble of returns is constructed over the whole
set of stocks traded on the NYSE, anomalous large widths
and fat tails are observed specifically
on the day of the crash of Oct. 19 1987, as well as during a few other
turbulent days. Lillo and Mantegna document another 
remarkable behavior associated with crashes
and rallies, namely that the distortion of the distributions of returns
are not only strong in the tails describing large moves but also in their center.
Specifically, they show that the overall shape of the distributions
is modified in crash and rally days. 
Closer to our claim that markets develop precursory signatures of bubbles
of long time scales, Mansilla (2001) has also shown, using a measure of
relative complexity, that time sequences corresponding to ``critical''
periods before large market corrections or crashes
have more novel informations with respect to the whole price time series than those
sequences corresponding to periods where nothing happened. The conclusion is that, in the
intervals where no financial turbulence is observed, that is, where the markets works fine,
the informational contents of the (binary-coded) price time series is small. 
In contrast, there seems to be significant information in the price time series
associated with bubbles. This finding is consistent with the appearance of 
a collective herding behavior modifying the texture of the price time series compared
to normal times.

\vskip-0.3cm
\subsection{The Dow Jones Industrial Average}

\begin{figure}
\begin{center}
\epsfig{file=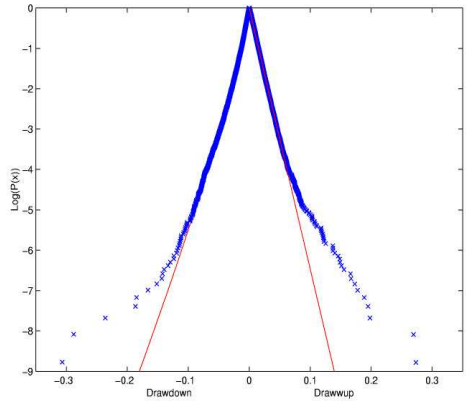,width=8cm,height=7cm}
\caption{\protect\label{djdd} Normalized natural logarithm of the 
cumulative distribution of drawdowns and of the complementary cumulative 
distribution of drawups for the Dow Jones Industrial Average index (US stock market). 
The two continuous lines show the fits of these
two distributions with the stretched exponential distribution. 
Negative values such as $-0.20$ and $-0.10$ correspond
to drawdowns of amplitude respectively equal to $20\%$ and $10\%$. Similarly, 
positive values corresponds to drawups with,
for instance, a number $0.2$ meaning a drawup of $+20\%$. 
Reproduced from [Johansen and Sornette, 2001c]}
\end{center}
\end{figure}

Figure \ref{djdd} shows the distribution of drawdowns for the returns of the 
DJIA over this century. 

The (stretched) exponential distribution 
has been derived on the assumption that
successive price variations are independent. There is a large body of 
evidence for the correctness of this assumption for most 
trading days [Campbell et al., 1997]. However, consider, for instance, 
the $14$ largest drawdowns that have occurred in the Dow Jones Industrial 
Average in this century. Their characteristics are presented in table
\ref{largedddj}. Only $3$ lasted one or two days, whereas $9$ lasted four
days or more. 
Let us examine in particular the largest drawdown. It started on Oct. 
14, 1987 (1987.786 in decimal years), lasted four days and led to a 
total loss of $-30.7\%$. This crash is thus a run of four consecutive 
losses: first day the index is down by $3.8\%$, second day by $6.1\%$, 
third day by $10.4\%$ and fourth by $30.7\%$. In terms of consecutive 
losses this correspond to $3.8\%$, $2.4\%$, $4.6\%$ and with $22.6\%$ on 
what is known as the Black Monday of Oct. 1987.

\begin{table}[]
\begin{center}
\begin{tabular}{|c|c|c|c|c|c|c|} \hline
rank & starting time & index value & duration (days) & loss  \\ \hline
1 & $1987.786$ & $2508.16$ & 4 & $-30.7\%$ \\ \hline
2 & $1914.579$ & $76.7$  & 2 & $-28.8\%$ \\ \hline
3 & $1929.818$ & $301.22$  & 3 & $-23.6\%$ \\ \hline
4 & $1933.549$ & $108.67$ & 4 & $-18.6\%$ \\ \hline
5 & $1932.249$ & $77.15$ & 8 & $-18.5\%$ \\ \hline
6 & $1929.852$ & $238.19$ & 4 & $-16.6\%$ \\ \hline
7 & $1929.835$ & $273.51$ & 2 & $-16.6\%$ \\ \hline
8 & $1932.630$ & $67.5$ & 1 & $-14.8\%$ \\ \hline
9 & $1931.93$ & $90.14$ & 7 & $-14.3\%$ \\ \hline
10 & $1932.694$ & $76.54$ & 3 & $-13.9\%$ \\ \hline
11 & $1974.719$ & $674.05$ & 11 & $-13.3\%$ \\ \hline
12 & $1930.444$ & $239.69$ & 4 & $-12.4\%$ \\ \hline
13 & $1931.735$ & $109.86$ & 5 & $-12.9\%$ \\ \hline
14 & $1998.649$ & $8602.65$ & 4 & $-12.4\%$ \\ \hline
\end{tabular}
\vspace{5mm}
\caption[]{Characteristics of the 14 largest drawdowns of
the Dow Jones Industrial Average in this century. The starting 
dates are given in decimal years. Reproduced from [Johansen and Sornette, 2001c]
}
\label{largedddj}
\end{center}
\end{table}

The observation of large successive drops is suggestive of the existence
of a transient correlation as we already pointed out. 
For the Dow Jones, this reasoning can be adapted as follows. 
We use a simple functional form for the distribution of daily losses, 
namely an exponential distribution with decay rate $1/0.63\%$ obtained
by a fit to the distribution of drawdowns shown in figure \ref{djdd}.
The quality of the exponential model
is confirmed by the direct calculation of the average loss amplitude equal
to $0.67\%$ and of its standard deviation equal to $0.61\%$ (recall
that an exact exponential would give the three values 
exactly equal: 1/decay$=$average$=$standard deviation).
Using these numerical values, the probability for a drop equal to or
larger than $3.8\%$ is $\exp(-3.8/0.63)= 2.4~10^{-3}$ (an event occurring
about once every two years);  the probability for a drop equal to or
larger than $2.4\%$ is $\exp(-2.4/0.63)= 2.2~10^{-2}$ (an event occurring
about once every two months);  the probability for a drop equal to or
larger than $4.6\%$ is $\exp(-4.6/0.63)= 6.7~10^{-4}$ (an event occurring
about once every six years);  the probability for a drop equal to or
larger than $22.6\%$ is $\exp(-22.6/0.63)= 2.6~10^{-16}$ (an event occurring
about once every $10^{14}$ years). All together, under the hypothesis that 
daily losses are uncorrelated from 
one day to the next, the sequence of four drops making the largest drawdown
occurs with a probability $10^{-23}$, i.e., once in about $4$ thousands
of billions of billions years. This exceedingly negligible value  $10^{-23}$
suggests that the hypothesis of uncorrelated daily 
returns is to be rejected: drawdowns 
and especially the large ones may exhibit intermittent correlations 
in the asset price time series.

\vskip-0.3cm
\subsection{The Nasdaq composite index}

\begin{figure}
\begin{center}
\epsfig{file=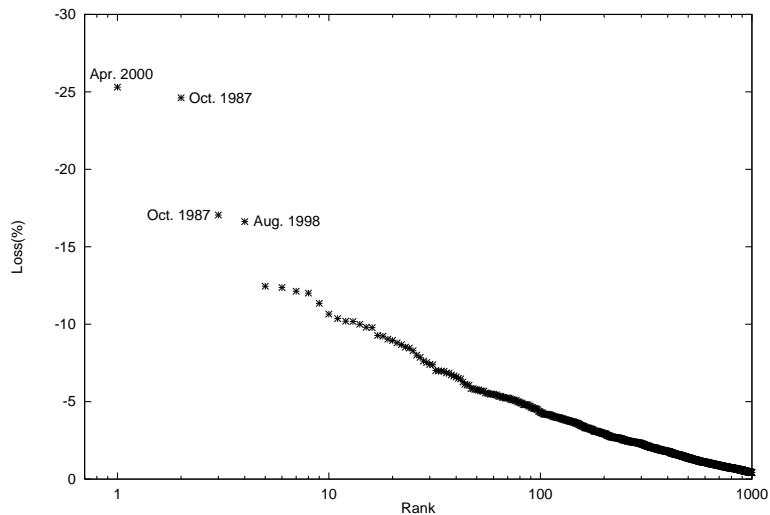,height=7cm}
\caption{\protect\label{nascumu}  Rank ordering of drawdowns
in the Nasdaq Composite since its establishment in 1971 until 18 April 2000.
Rank 1 is the largest drawdown. Rank 2 is the second largest, and so on.
Reproduced from [Johansen and Sornette, 2000a]}
\end{center}
\end{figure}

In figure
\ref{nascumu}, we see the rank ordering plot of drawdowns for the Nasdaq composite index, 
since its establishment in 1971 until
18 April 2000. The rank ordering plot, which is the same as the
(complementary) cumulative distribution with axis interchanged, puts
emphasis on the largest events. The four largest events are not situated
on a continuation of the distribution of smaller events: the jump between
rank 4 and 5 in relative value is larger than $33\%$ whereas the corresponding jump between
rank 5 and 6 is less than $1\%$ and this remains true for higher ranks. This means
that, for drawdowns less than $12.5\%$, we have a more or less ``smooth''
curve
and then a larger than $ 33\%$ gap to rank 3 and 4. The four events are according
to rank the crash of April 2000, the crash of Oct. 1987, a
larger than $17\%$ ``after-shock'' related to the crash of Oct. 1987 and a larger than $16\%$
drop related to the ``slow crash'' of Aug. 1998, that we shall discuss later on.

To further establish the statistical confidence with which we can conclude
that the four largest events are outliers, the daily
returns have been reshuffled 1000 times generating 1000 synthetic data sets. This
procedure means that the synthetic data sets will have exactly the same
distribution of daily returns. However, higher order correlations and
dependence that may be present
in the largest drawdowns are destroyed by the reshuffling. This
 ``surrogate'' data analysis of the distribution of drawdowns has the advantage
of being {\it non-parametric}, {\it i.e.}, independent of the quality of fits
with a model such as the exponential or any other model. We will
now compare the
distribution of drawdowns both for the real data and the synthetic data.
With respect to the synthetic data, this can be done in two complementary ways.

\begin{figure}
\begin{center}
\epsfig{file=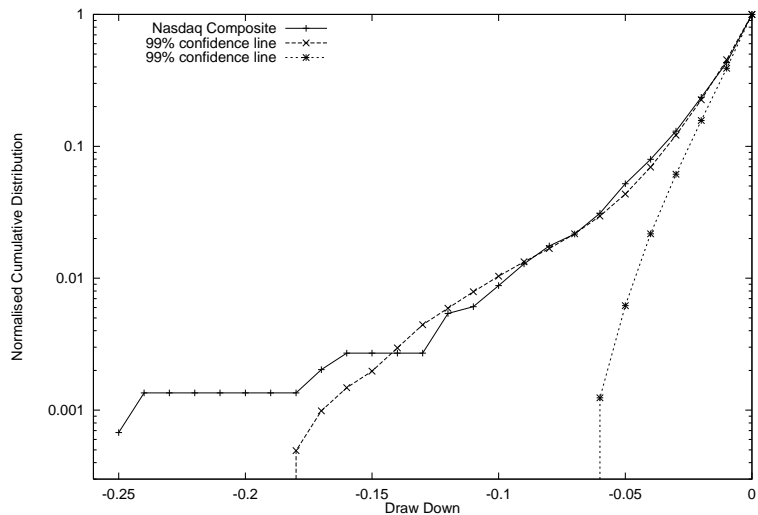,height=7cm}
\caption{\protect\label{confidence}  Normalised cumulative distribution of drawdowns 
in the Nasdaq Composite since its establishment in 1971 until 18 April
2000. The $99\%$ confidence lines are estimated from synthetic tests obtained
by generating surrogate financial time series constructed by reshuffling the 
daily returns at random.
Reproduced from [Johansen and Sornette, 2000a]}
\end{center}
\end{figure}

In figure \ref{confidence}, we see the distribution of drawdowns in the
Nasdaq Composite compared with the two lines constructed at the $99\%$
confidence level for the entire
{\it ensemble} of synthetic drawdowns, {\it i.e.} by considering the
individual drawdowns
as independent: for any given drawdown, the upper (resp. lower)
confidence line is such that $5$ of the synthetic distributions are above (below) it;
as a consequence, 990 synthetic times series out of the 1000 are within the
two confidence lines for any drawdown value which define the typical interval
within which we expect to find the empirical distribution.

The most striking feature apparent on figure \ref{confidence} is that
the distribution of the true data
breaks away from the $99\%$ confidence intervals at approximately $15\%$, showing
that the four largest events are indeed ``outliers''. In other words,
chance alone cannot reproduce these largest drawdowns. We are thus forced to
explore the possibility that an amplification mechanism and dependence across daily 
returns might appear at special and rare times to create these outliers.

A more sophisticated analysis is to consider each synthetic data set
{\it separately} and calculate the {\it conditional probability} of
observing a given drawdown given some prior observation of drawdowns.
This gives a more precise estimation of the statistical significance
of the outliers, because the previously defined confidence lines neglect
the correlations created by the ordering process
which is explicit in the construction of a cumulative distribution.

Out of 10,000 synthetic data sets that were generated, 
we find that 776 had a single drawdown larger
than $16.5\%$, 13 had two drawdowns larger than $16.5\%$, 1 had three drawdowns 
larger than $16.5\%$ and none had 4 (or more) drawdowns larger than
$16.5\%$ as in the real data. This means that, given the distribution of
returns, by chance we have a $8\%$ probability of observing a drawdowns 
larger than $16.5\%$, a $0.1\%$ probability of observing two
drawdowns larger than $16.5\%$ and for all practical purposes zero
probability of observing three or more drawdowns larger than $16.5\%$.
Hence, we can reject the hypothesis that the four largest drawdowns observed on the Nasdaq 
composite index could result from chance alone with a probability or confidence
better than $99.99\%$, i.e., essentially with certainty.
As a consequence, we are lead again to conclude that the largest market events are
characterised by a stronger dependence than
is observed during ``normal'' times.

This analysis confirms the conclusion from the analysis of the DJIA
shown in figure \ref{djdd}, that drawdowns larger than about $ 15$\% are to be considered as
outliers with high probability. It is interesting that the same amplitude
of approximately $15$\% is found for both markets considering the much larger
daily volatility of the Nasdaq Composite. This may result from the fact that,
as we have shown, very large drawdowns are more controlled by transient
correlations leading to runs of losses lasting a few days than by the amplitude of 
a single daily return.

The statistical analysis of the Dow Jones Average and the Nasdaq
Composite suggests that large crashes {\it are} special. In following sections,
we shall show that there are other specific indications associated with 
these ``outliers'', such as precursory
patterns decorating the speculative bubbles ending in crashes.

\vskip-0.3cm
\subsection{The presence of ``Outliers'' is a general phenomenon} 

\begin{figure}
\begin{center}
\epsfig{file=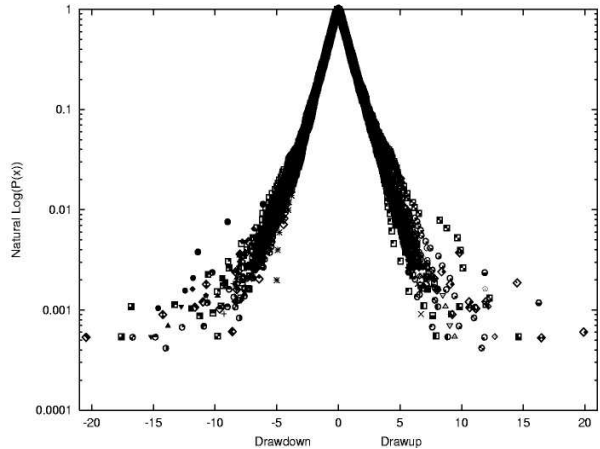,width=8cm,height=6cm}
\caption{\protect\label{rescaleall-1} Cumulative distribution of 
drawdowns and complementary cumulative distribution of drawups for
$29$ companies, which include the $20$ largest USA companies in terms of capitalisation
according to Forbes at the beginning of the year 2000, and in addition
Coca Cola (Forbes number 25), Qualcomm (number 30), 
Appl. Materials (number 35), Procter \& Gamble (number 38)
JDS Uniphase (number 39), General Motors (number 43), Am. Home Prod. (number 46),
Medtronic (number 50) and Ford (number 64). This figure plots each
distribution $N_c$ normalized by its corresponding factor $A$ 
as a function of the variable $y \equiv {\rm sign}(x)~(|x|/\chi)^z$, where 
 $\chi$ and $z$ are specific to each distribution and obtained
 from the fit to the stretched exponential model.
 Reproduced from [Johansen and Sornette, 2001c].
}
\end{center}
\end{figure}

To avoid a tedious repetition of many figures, we group the cumulative
distributions of drawdowns and complementary cumulative distributions
of several stocks in the same figure \ref{rescaleall-1}. In order to 
construct this figure, we have fitted the stretched
exponential model (\ref{stretched}) to each distribution 
and obtained the corresponding parameters
$A$, $\chi$ and $z$ given in [Johansen and Sornette, 2001c]. 
We then construct the
normalized distributions 
\be
N_C^{(n)}(x) = N_c\lp (|x|/\chi)^z \rp /A
\label{ghbwhsz}
\ee
using the triplet $A$, $\chi$ and $z$ which is specific to each distribution.
Figure \ref{rescaleall-1} plots the expression (\ref{ghbwhsz}) for each
distribution, i.e., $N_c/A$ as a function of 
$y \equiv {\rm sign}(x)~(x/\chi)^z$. If the stretched
exponential model (\ref{stretched}) held true for all the drawdowns and all the drawups,
all the normalized distributions should collapse exactly onto the 
``universal'' functions
$e^y$ for the drawdowns and $e^{-y}$ for the drawups. We observe that this is
the case for values of $|y|$ up to about $5$, i.e., up to typically $5$ standard
deviations (since most exponents $z$ are close to $1$), beyond which there is a clear upward
departure observed both for drawdowns and for drawups. Comparing with the
extrapolation of the normalized stretched exponential model $e^{-|y|}$,
the empirical normalized distributions give about $10$ times too many drawdowns
and drawups larger than $|y|=10$ standard deviations and more the $10^4$
too many drawdowns and drawups larger than $|y|=20$ standard deviations.
Note that for AT\&T, a crash of $\approx 73\%$ occurred which lies
beyond the range shown in figure \ref{rescaleall-1}.

The results obtained in [Johansen and Sornette, 1998; 2000a; 2001c] can
be summarized as follows.
\begin{enumerate}
\item Approximately $1-2\%$ of the largest drawdowns are not at all explained
by the exponential null-hypothesis or its extension 
in terms of the stretched exponential (\ref{stretched}).
Large drawdowns up to three times larger than expected from the null-hypothesis
are found to be ubiquitous occurrences of essentially all the times series
that we have investigated, the only noticeable exception being the French index
CAC40. We term ``outliers'' these anomalous drawdowns.
\item About half of the time series show outliers for the drawups. The drawups 
are thus different statistically from the drawdowns and constitute a less
conspicuous structure of financial markets.
\item For companies, large drawups of more than $15\%$ occur approximately twice as
often as large drawdowns of similar amplitudes.
\item The bulk ($98\%$) of the drawdowns and drawups are very well-fitted
by the exponential null-hypothesis (based on the assumption of independent
price variations) or by the stretched exponential model. 
\end{enumerate}

The most important result is the demonstration that the very largest drawdowns
are outliers. This is true notwithstanding the fact that the very largest
daily drops are {\it not} outliers, except for the exceptional and unique daily drop 
on Oct. 29, 1987. Therefore, the anomalously
large amplitude of the drawdowns can only be explained
by invoking the emergence of rare but sudden persistences of 
successive daily drops, with in addition correlated amplification of the
drops. Why such successions of correlated daily moves occur
is a very important question with 
consequences for portfolio management and systemic risk, to cite
only two applications, that we are going to investigate in the following sections.

\vskip-0.3cm
\subsection{Implications for safety regulations of stock markets}

The realization that large drawdowns and crashes in particular may result from
a run of losses over several successive days is not without consequences for 
the regulation of stock markets. Following the market crash of Oct. 1987,
in an attempt to head off future one-day stock market tumbles of historic proportions,
the Securities and Exchange Commission and the three major
U.S. stock exchanges agreed to install so-called circuit breakers.
Circuit breakers are designed to gradually inhibit
trading during market declines, first curbing New York Stock Exchange program
trades and eventually halting all U.S. equity, options and futures activity.
Similar circuit breakers are operating in the other world stock markets 
with different specific definitions.

The argument is that the halt triggered by a circuit breaker will provide
time for brokers and dealers to contact their clients when there are large price
movements and to get new instructions or additional margin. They also limit
credit risk and loss of financial confidence by providing a ``time-out'' to settle
up and to ensure that everyone is solvent. This inactive period is of further use
for investors to pause, evaluate and inhibit panic. Finally, circuit breakers
clarify the illusion of market liquidity by spelling out the economic fact of life
that markets have limited capacity to absorb massive unbalanced volumes. They thus
force large investors, such as pension portfolio managers and mutual funds, to 
take even more account of the impact of their ``size order'', thus possibly cushioning
large market movements. Others argue that a
trading halt can increase risk by inducing trading in anticipation of a trading
halt. Another disadvantage is that they prevent some traders from liquidating their positions,
thus creating market distorsion by preventing price discovery [Harris, 1997]. 

For the Oct. 1987 crash, 
countries that had stringent circuit breakers, such as 
France, Switzerland  and Israel, had also some of the largest cumulative losses. According to 
the evidence presented here
that large drops are created by transient and rare dependent losses occurring
over several days, we should be cautious in considering
circuit breakers as reliable crash killers.

\section{Positive feedbacks} 

Since it is the 
actions of investors whose buy and sell decisions move prices up and down,
any deviation from a random walk in the stock
market price trajectory has ultimately to be traced back to the behavior
of investors. We are in particular interested in mechanisms that may lead
to positive feedbacks on prices, i.e., to the fact that, conditioned
on the observation that the market has recently moved 
up (respectively down), this makes it more probable to keep it moving up 
(respectively down), so that a large cumulative move ensues. 
The concept of ``positive feedbacks'' has a long
history in economics and is related to the idea of 
``increasing returns''-- which says that goods become cheaper the more of them
 are produced (and the closely related idea that some products, like fax
 machines, become more useful the more people use them). 
 ``Positive feedback'' is the opposite of ``negative feedback'', 
 a concept well-known for instance in population dynamics: the larger the population
 of rabbits in a valley, the less they have grass per rabbit. If the population
 grows too much, they will eventually starve, slowing down their reproduction rate
 which thus reduces their population at a later time.
 Thus negative feedback means that the higher the population, the 
 slower the growth rate, leading to a spontaneous regulation of 
 the population size; negative feedbacks thus tend to regulate growth
 towards an equilibrium.
 In contrast, positive feedback asserts that the 
 higher the price or the price return in the recent past, the higher will be
 the price growth in the future. Positive feedbacks, when unchecked, can produce
 runaways until the deviation from equilibrium is so large that other 
 effects can be abruptly triggered and lead to rupture or crashes.
 Youssefmir et al. [1998] have stressed the importance of
 positive feedback in a dynamical theory of 
 asset price bubbles that exhibits the appearance
of bubbles and their subsequent crashes. The positive feedback
leads to speculative trends which may
dominate over fundamental beliefs and which make the system
increasingly susceptible to any exogenous shock, thus eventually precipitating a
crash. 
 
There are many mechanisms in the stock
market and in the behavior of investors which may lead to positive feedbacks.
We describe a general mechanism for positive feedback, which
is now known as the  ``herd'' or ``crowd''
effect, based on imitation processes. We present a simple model of the best 
investment strategy that an investor can develop based on interactions with
and information taken from other investors. We show how the repetition of
these interactions may lead to a remarkable cooperative phenomenon in which
the market can suddenly ``solidify'' a global opinion, leading to large price
variations.

\vskip-0.3cm
\subsection{Herding}
              
There are growing empirical evidences of the existence of herd or ``crowd'' 
behavior in speculative markets (see  [Shiller, 2000] and references therein). 
Herd behavior is often said to occur when many people take the same action,
because some mimic the actions of others. The term ``herd'' obviously
refers to similar behavior observed in animal groups. Other terms such 
as ``flocks'' or ``schools'' describe the collective
coherent motion of large numbers of self-propelled organisms, such 
as migrating birds and gnus, lemmings and ants. In recent years,
much of the observed
herd behavior in animals has been shown to result from the action of simple
laws of interactions between animals. 
With respect to humans, there is a long history of analogies
between human groups and organized matter [Callen and Shapero, 1974; 
Montroll and Badger, 1974]. 
More recently, extreme crowd motions such as under panic have been remarkably
well quantified by models that treat the crowd as a collection of 
individuals interacting as a granular medium with friction such as the familiar
sand of beaches [Helbing et al, 2000].

Herding
has been linked to many economic activities, such as investment recommendations
[Scharfstein and Stein, 1990; Graham, 1999; Welch, 2000s], price behavior of IPO's 
(Initial Public Offering) [Welch, 1992], fads and
customs [Bikhchandani et al., 1992], earnings forecasts
[Trueman, 1994], corporate conservatism [Zwiebel, 1995] and delegated
portfolio management [Maug and Naik, 1995].
Researchers are investigating the
incentives investment advisors face when deciding whether to herd and, in particular,
whether economic conditions and agents' individual
characteristics affect their likelihood of herding. 
Although herding behavior appears inefficient from a social
standpoint, it can be rational from the perspective of managers who are
concerned about their reputations in the labor market,
Such behavior can be rational 
and may occur as an information cascade [Welch, 1992; Bikhchandani et al., 1992;
Devenow and Welch, 1996], 
a situation in which every subsequent actor, 
based on the observations of others, makes the same choice independent of
his/her private signal.
Herding among investment newsletters, for instance, is found to decrease
with the precision of private information [Graham, 1999]: the less information you 
have, the more important is your incentive to follow the consensus.

Research on herding in finance can be subdivided in the following 
non-mutually exclusive manner [Devenow and Welch, 1996; Graham, 1999].
\begin{enumerate}
\item {\bf Informational cascades} occur
when individuals choose to ignore or downplay their private information
and instead jump on the bandwagon by mimicking the actions of individuals
who acted previously. 
Informational cascades occur when the existing
aggregate information becomes so overwhelming that an individual's
single piece of private information is not strong enough to reverse the
decision of the crowd. Therefore, the individual chooses to mimic the action
of the crowd, rather than act on his private information. If this scenario
holds for one individual, then it likely also holds for anyone acting after
this person. This domino-like effect is often referred to as a cascade. 
The two crucial ingredients for an informational cascade to develop are: [1]
sequential decisions with subsequent actors
observing decisions (not information) of previous actors; and [2] 
a limited action space.

\item {\bf Reputational herding}, like cascades, takes place when an agent chooses to
ignore his or her private information and mimic the action of another agent who
has acted previously. However, reputational herding models have an additional
layer of mimicking, resulting from positive reputational properties
that can be obtained by acting as part of a group or choosing a certain
project. Evidence has been found that 
a forecaster's age is positively related to
the absolute first difference between his forecast and the group mean. This has
been interpreted as evidence that as a forecaster ages, evaluators develop
tighter prior beliefs about the forecasterÕs ability, and hence the forecaster
has less incentive to herd with the group. On the other hand, the incentive
for a second-mover to discard his private information and instead mimick the market
leader increases with his initial reputation, as he strives to protect his 
current status and level of pay [Graham, 1999].

\item {\bf Investigative herding} occurs when an analyst chooses to investigate a piece
of information he or she believes others also will examine. The analyst would like
to be the first to discover the information but can only profit from an investment
if other investors follow suit and push the price of the asset in the
direction anticipated by the first analyst. Otherwise, the first analyst may
be stuck holding an asset that he or she cannot profitably sell. 

\item {\bf Empirical herding} refers to observations by many researchers of 
``herding'' without reference to a specific model or explanation. 
There is indeed evidence of herding and clustering 
among pension funds, mutual funds, and institutional
investors when a disproportionate share of investors engage in buying,
or at other times selling, the same stock. These works
suggest that clustering can result from momentum-following also
called ``positive feedback investment,'' e.g., buying past winners or perhaps
from repeating the predominant buy or sell pattern from the previous period.

\end{enumerate}

There are many reported case of herding. One of the most dramatic and clearest in 
recent times is the observation [Huberman and Regevon, 2001] of
a contagious speculation associated with a non-event in the following sense.
A sunday New York Times article on a potential development of a new 
cancer-curing drugs caused the biotech company EntreMed's stock to rise from 12.063 at the
Friday May 1, 1998 close to open at 85 on Monday May 4, close near 52 
on the same day and remain above 39 in the
three following weeks. The enthusiasm spilled over to other biotechnology stocks.
It turns out that the potential breakthrough in cancer research already had
been reported in one of the leading scientific journal `Nature' and in various
popular newspaper (including the Times) more than five months earlier. At that time, 
market reactions were essentially nil.
Thus the enthusiastic public attention induced a long-term rise in share prices, even
though no genuinely new information had been presented. The very prominent 
and exceptionally optimistic Sunday New York Times article of May 3, 1998 led
to a rush on EntreMed's stock and other biotechnology companies' stocks,
which is reminiscent of similar rushs leading to bubbles in historical times 
previously discussed. It is to be expected that information technology, the internet
and biotechnology are among the leading new frontiers on which sensational
stories will lead to enthusiasm, contagion, herding and speculative bubbles.

\vskip-0.3cm
\subsection{It is optimal to imitate when lacking information \label{ratioimit}}

All the traders in the world are organized
into a network of family, friends, colleagues, contacts, and so on, which are
 sources of opinion and they influence each
other {\em locally} through this network 
[Boissevain and Mitchell, 1973]. We call ``neighbors'' of agent Anne on this
world-wide graph the set of people in direct contact with Anne.
Other sources of influence also involve
newspapers, web sites, TV stations, and so on.
Specifically, if Anne is directly
connected with $k$ ``neighbors'' in the worldwide graph of connections, 
then there are only two forces that
influence Anne's opinion: (a) the opinions of these $k$ people together with 
the influence of the media; and (b) an
idiosyncratic signal that she alone receives (or generates). According to the concept of
herding and imitation, the assumption is that
agents tend to {\em imitate} the opinions of their ``neighbors'', not
contradict them. It is easy to see that force (a) will tend to create
order, while
force (b) will tend to create disorder, or in other words, heterogeneity. The main story here
is the fight between order and disorder and the question we are now
going to investigate is: what behavior can result from this fight? 
Can the system go through unstable regimes, such as crashes? 
Are crashes predictable? We show that the science of self-organizing systems
(sometimes also refered to as ``complex systems'') bears very significantly
on these questions: the stock market and the web of traders' connections can 
be understood in large part from the science of critical phenomena, in a sense
that we are going to examine in some depth in the following sections, 
from which important consequences can be
derived.
    
To make progress, we formalize a bit the problem and 
consider a network of investors: each one can be named by an integer
$i=1,\dots,I$, and $N(i)$ denotes the set of the agents who are directly
connected to agent $i$ according to the world-wide graph of
acquaintances. If we isolate one trader, Anne, $N({\rm Anne})$ is the number of
traders in direct contact with her and who can exchange direct information with her and
exert a direct influence on her.
For simplicity, we assume
that any investor such as Anne can be in only one of several possible states. In the simplest
version, we can consider only
two possible states: $s_{\rm Anne} = -1$ or $s_{\rm Anne}=+1$.
We could interpret these states as ``buy'' and ``sell'', ``bullish'' and
``bearish'', ``optimistic'' and ``pessimistic'', and so on. 
In the next paragraph, we show that, based only on the information of
the actions $s_j(t-1)$ performed yesterday (at time $t-1$) by her $N({\rm Anne})$
``neighbors'', Anne maximizes her return by having taken yesterday the 
decision $s_{\rm Anne}(t-1)$ given by the sign of the sum of the actions of 
all her ``neighbors''. In other words, the optimal decision of Anne, based on
the local polling of her ``neighbors'' who she hopes represents a sufficiently
faithful representation of the market mood, is to imitate the 
majority of her neighbors. This is of course up to some possible deviations 
when she decides to follow her own idiosynchratic ``intuition'' rather than
being influenced by her ``neighbors''. Such an idiosynchratic move can be captured
in this model by a stochastic component independent of the decisions of the 
neighbors or of any other agent. Intuitively, the reason why it is in general optimal 
for Anne to follow the opinion of the majority is simply because
prices move in that direction, forced by the law of suppy and demand. 
This apparently 
innocuous evolution law produces remarkable self-organizing patterns.

Consider $N$ traders in a network, whose links represent the communication
channels through which the traders exchange information. The graph describes
the chain of intermediate
acquaintances between any two people in the world.
We denote $N(i)$ the number of traders directly connected to a given trader $i$ on the graph.
The traders buy or sell one asset at price $p(t)$ which evolves as a
function of
time assumed to be discrete and measured in units
of the time step $\Delta t$. In the simplest version of the model, each
agent can
either buy or sell only one unit of the asset. This is quantified by
the buy state $s_i=+1$ or the sell state $s_i=-1$. Each agent can trade at
time
$t-1$ at the price $p(t-1)$ based on all previous information including
that at $t-1$.
The asset price variation is taken simply proportional to 
the aggregate sum $\sum_{i=1}^N s_i(t-1)$ of all traders' actions:
indeed, if this sum is zero, there are as many buyers as they are sellers 
and the price does not change since there is a perfect balance between supply
and demand. If, on the other hand, the sum is positive, there are more buy orders
than sell orders, the price has to increase to balance the supply and the demand, 
as the asset is too rare to satisfy all the demand. There are many other 
influences impacting the price change from one day to the other, and this can
usually be accounted for in a simple way by adding a stochastic component
to the price variation. This term alone would give the usual log-normal
random walk process [Cootner, 1967] while the balance between supply and 
demand together with imitation leads to some organization as we show below.

At time $t-1$, just when the price $p(t-1)$ has been announced, the trader $i$
defines her strategy $s_i(t-1)$ that she will hold from $t-1$ to $t$, thus
realizing the profit (or loss) equal to the price difference
$(p(t)-p(t-1))$ times her position $s_i(t-1)$. 
To define her optimal strategy $s_i(t-1)$, the trader should
calculate her expected profit $P_E$, given the past information and her
position,
and then choose $s_i(t-1)$ such that $P_E$ is maximum. Since the price
moves with the general opinion $\sum_{i=1}^N s_i(t-1)$, the best 
strategy is to buy if it is positive and sell if it is negative. 
The difficulty
is that a given trader cannot poll the positions
$s_j$ that will take all other traders which will determine the price drift
according to the balance between supply and demand. The next best thing that trader $i$ can
do is to poll her $N(i)$ ``neighbors'' and construct her prediction for the
price drift from this information. The trader needs an additional
information, namely
the a priori probability $P_+$ and $P_-$ for each trader to buy or sell.
The probabilities $P_+$ and $P_-$ are the only information that she can
use for all the
traders that she does not poll directly. From this, she can form
her expectation of the price change. The simplest case corresponds to a
market without drift where $P_+=P_-=1/2$. 

Based on the previously stated rule that the price variation is
proportional to the sum of actions of traders, the best guess of 
trader $i$ is that the future price change will be
proportional to the sum of the actions of
her neighbors that she has been able to poll, hoping that this provides
a sufficiently reliable sample of the total population. Traders are indeed
constantly sharing information, calling each other to ``take the temperature'',
effectively polling each other before taking actions. 
It is then clear that the strategy that maximizes her expected profit is
such that her position is of the sign given by the sum of the actions of 
all her ``neighbors''. This is exactly the meaning of 
expression (\ref{eq:state}) 
\be
\label{eq:state}
s_{i}(t-1)  =  {\rm sign}\left(K\sum_{j\in N_i} s_j+  \varepsilon_i\right)
\ee
such that this position $s_{i}(t-1)$
gives her the maximum payoff based on her best prediction of 
the price variation $p(t)-p(t-1)$ from yesterday to today.
The function ${\rm sign}(x)$ is defined by being equal to $+1$ (to $-1$) for positive
(negative) argument $x$, $K$ is a positive constant of proportionality between the price
change and the aggregate buy-sell orders. It is inversely proportional to the ``market
depth'': the larger the market, the smaller is the relative impact of
a given unbalance between buy and sell orders, hence the smaller is the price change.
$\varepsilon_i$ is a noise
and $N(i)$ is the number of neighbors with whom trader $i$ interacts significantly.
In simple terms, this law (\ref{eq:state}) states that the best investment 
decision for a given trader is to take that of the majority of her neighbors, 
up to some uncertainly (noise) capturing the
possibility that the majority of her neighbors might give an incorrect prediction of the
behavior of the total market.

Expression (\ref{eq:state}) can be thought of as a mathematical formulation 
of Keynes' beauty contest. Keynes [1936] argued that stock prices
are not only determined by the firm's fundamental value, but, in addition, mass psychology
and investors' expectations influence financial
markets signifcantly. It was his opinion that professional investors
prefer to devote their energy, not to estimating fundamental values 
but rather, to analyzing how
the crowd of investors is likely to behave in the future. As a result, he said, 
most persons are largely concerned, not with making superior long-term 
forecasts of the probable yield of an investment over its whole life but,
with foreseeing changes in the conventional basis 
of valuation a short time ahead of the general public. Keynes uses
his famous beauty contest as a parable for stock markets. In order to predict
the winner of beauty contest, objective beauty is not much important, but knowledge
or prediction of others'prediction of beauty is much more relevant. In Keynes'view,
the optimal strategy is not to pick those faces the player thinks the prettiest,
but those the other players are likely to think the average opinion will be, or
those the other players will think the others will think the average opinion will be,
or even further along this iterative loop. Expression (\ref{eq:state}) precisely
captures this concept: the opinion $s_{i}$ at time $t$ 
of an agent $i$ is a function of all the
opinions of the other ``neighboring'' agents at the previous time $t-1$, which 
themselves depend on the opinion of the agent $i$ at time $t-2$, and so on. 
In the stationary equilibrium situation in which all agents
finally form an opinion after many such iterative feedbacks 
have had time to develop, the solution of (\ref{eq:state})  is precisely the one
taking into account all the opinions in a completely self-consistent way
compatible with the infinitely iterative loop.
Similarly, Orl\'ean [1984; 1986; 1989; 1991; 1995] has captured
the paradox of combining rational and imitative behavior under the name
``mimetic rationality'' ({\it rationalit\'e mim\'etique}).
He has developed models of mimetic contagion 
of investors in the stock markets that are based on irreversible processes of 
opinion forming. See also [Krawiecki et al., 2002] for a recent
generalization with time-varying coupling strength $K$ leading to on-off
intermittency and attractor bubbling.

\vskip-0.3cm
\subsection{Cooperative behaviors resulting from imitation \label{coop}}

The imitative behavior discussed in section \ref{ratioimit}
and captured by the expression (\ref{eq:state}) belongs 
to a very general class of stochastic dynamical models 
developed to describe interacting elements, particles, agents in a large
variety of contexts, in particular in physics and biology [Liggett, 1985; 1997].
The tendency or force
towards imitation is governed by the coupling strength $K$;
the tendency towards idiosyncratic (or noisy) behavior is governed by the 
amplitude $\sigma$ of the noise term. Thus
the value of $K$ relative to $\sigma$ determines the outcome of the battle
between order and disorder, and eventually the structure of the market prices. More
generally, the coupling strength $K$ could be heterogeneous across pairs of
neighbors, and it would not substantially affect  the properties of the model.
Some of the $K_{ij}$'s could even be negative, as long as the average of all
$K_{ij}$'s was strictly positive.

\begin{figure}
\parbox[l]{7.5cm}{
\epsfig{file=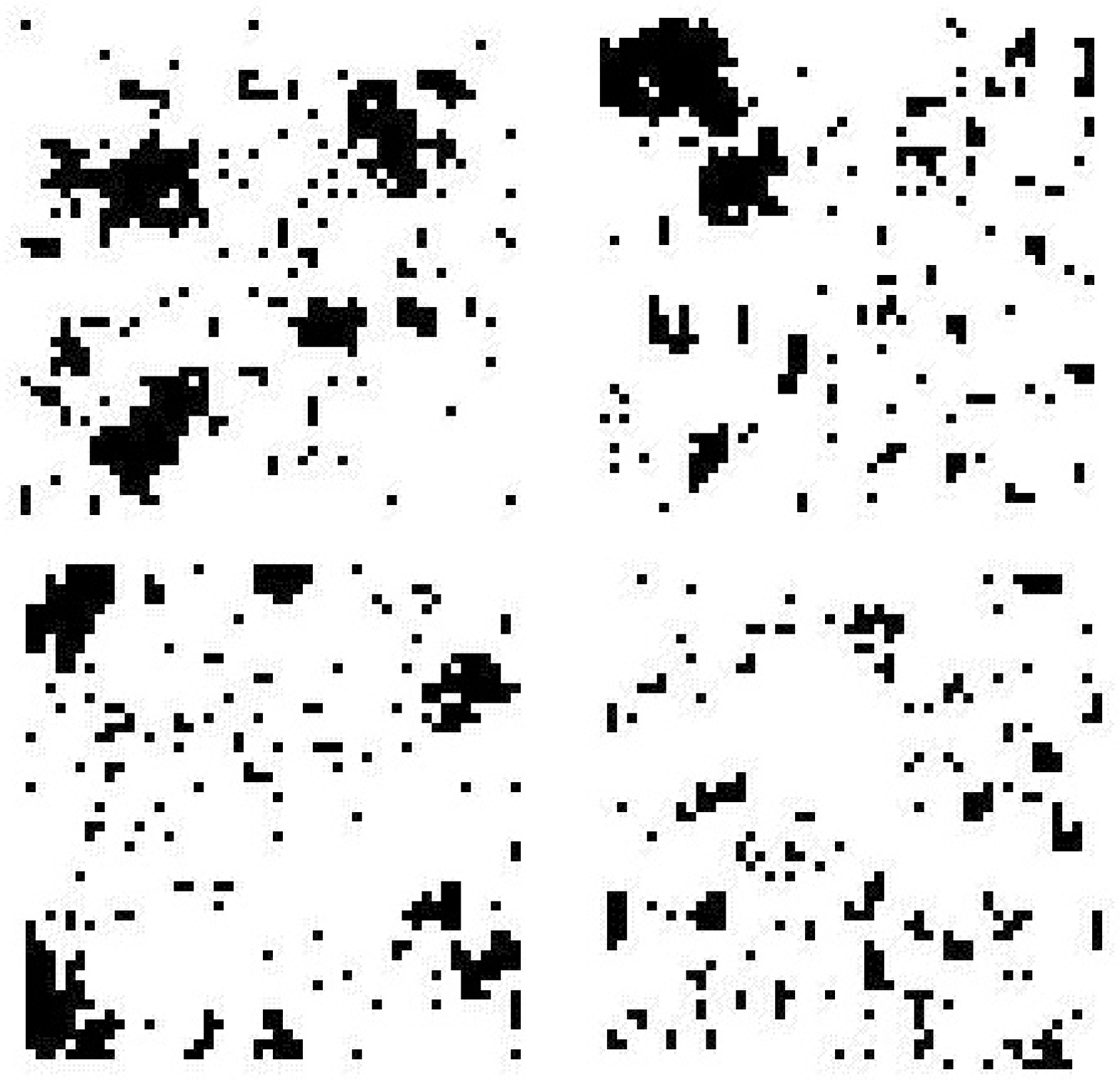,height=7cm,width=7cm}
\caption{\protect\label{IsingEvol} Four snapshots at four successive times of the state of 
a planar system of $64 \times 64$ agents put on a regular square lattice.
Each agent placed within a small square interacts
with her four nearest neighbors according to the imitative rule (\ref{eq:state}).
White (resp. black) squares correspond to ``bull'' (resp. ``bear''). The four cases
shown here correspond to the existence of a majority of buy orders as white is the 
predominant color. }}
\hspace{5mm}
\parbox[r]{7.5cm}{
\epsfig{file=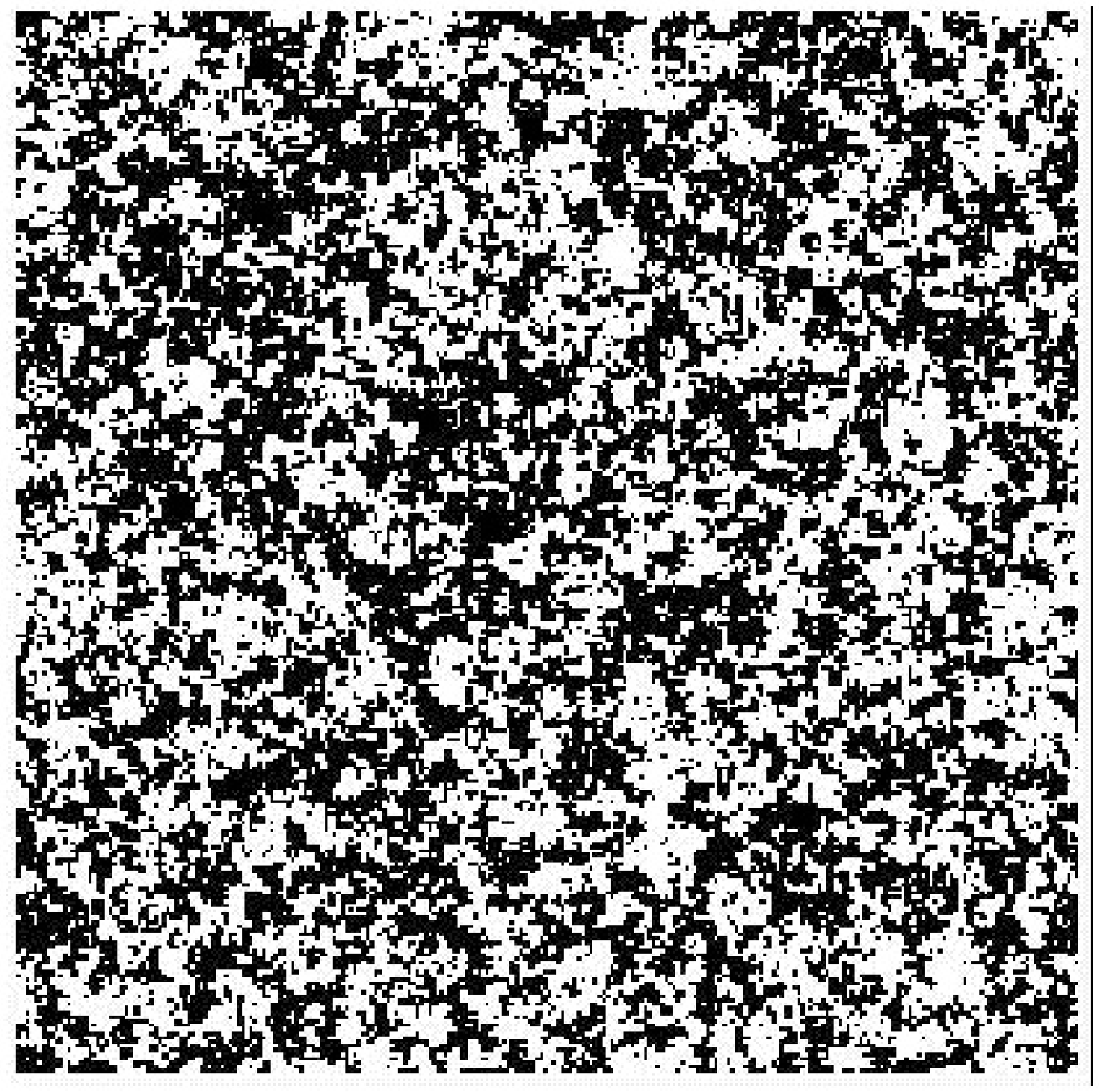,height=7cm,width=7cm}
\caption{\protect\label{FigIsingAboveTc} $K<K_c$: buy 
(white squares) and sell (back squares) configuration 
in a two-dimensional
Manhattan-like planar network of $256 \times 256$ agents interacting 
with their four nearest neighbors. There are approximately the same number
of white and black sells, i.e., the market has no consensus. The size of 
largest local
clusters quantifies the correlation length, i.e., the distance over which
the local imitations between neighbors propagate before being significantly
distorded by the ``noise'' in the transmission process resulting from
the idiosynchratic signals of each agent.}}
\end{figure}

The expression (\ref{eq:state}) only describes the state of an agent at
a given time. In the next instant, new $\varepsilon_i$'s are realized, new
influences propagate themselves to neighbors, and agents can change their decision.
The system is thus constantly changing and reorganizing as shown in figure \ref{IsingEvol}.
The model does {\it not} assume instantaneous opinion
interactions between neighbours. In real markets, opinions tend indeed not to be instantaneous
but are formed over a period of time by a process involving family, friends,
colleagues, newspapers, web sites, TV stations, and so on.
Decisions about trading activity of a given agent may occur when the consensus
from all these sources reaches a trigger level. This is precisely this feature
of a threshold reached by a consensus that expression (\ref{eq:state}) captures\,:
the consensus is quantified by the sum over the $N(i)$ agents connected to agent $i$
and the threshold is provided by the sign function. 
The delay in the formation of the opinion of a given 
trader as a function of other traders' opinion is captured
by the progressive spreading of information 
during successive updating steps (see for instance [Liggett, 1985; 1997]).

The simplest possible network is a
two-dimensional grid in the Euclidean plane. Each agent has four nearest
neighbors: one to the North, one
to the South, the East and the West. The tendency $K$ towards imitation is balanced by
the tendency $\sigma$ towards idiosyncratic behavior. In the context of the alignment
of atomic spins to create magnetisation (magnets), this model is identical to the
 two-dimensional Ising model which has been solved explicitly by
Onsager [1944]. Only its formulation is different from what is usually
found in textbooks [Goldenfeld, 1992], as we emphasize a dynamical view point. 

In the Ising model, there exists a critical point $K_c$ that determines the
properties of the system. When $K<K_c$ (see figure \ref{FigIsingAboveTc}), 
disorder reigns: the
sensitivity to a small global influence is small, the clusters of agents
who are in agreement remain of small size, and imitation only propagates
between
close neighbors. In this case, the susceptibility $\chi$ of the system to 
external news is small as many clusters of different opinion react 
incoherently, thus more or less cancelling out their response.

\begin{figure}
\parbox[l]{7.5cm}{
\epsfig{file=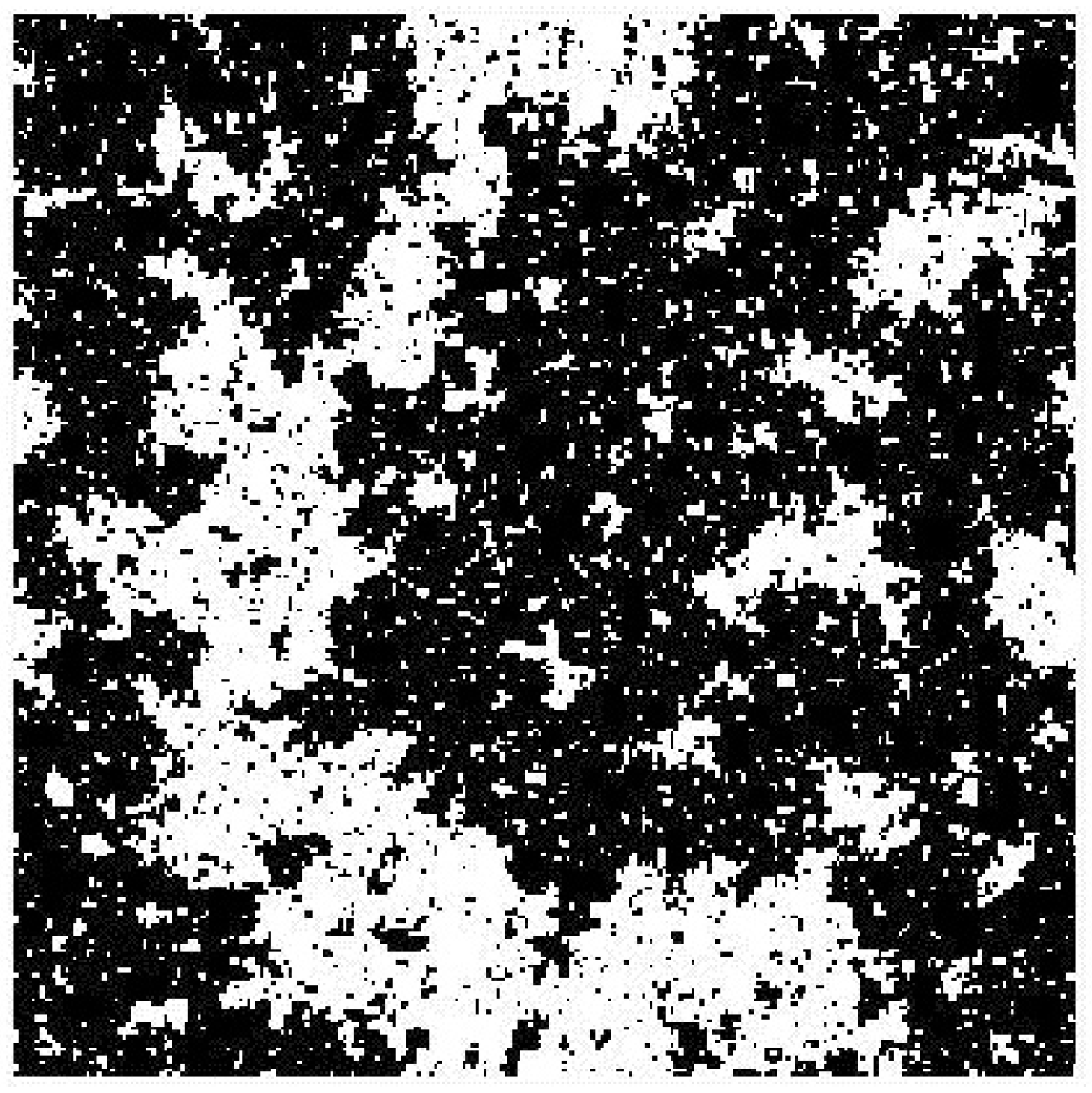,height=7cm,width=7cm}
\caption{\protect\label{FigIsingatTc} Same as figure \ref{FigIsingAboveTc} 
for $K$ close to $K_c$. 
There are still approximately the same number
of white and black sells, i.e., the market has no consensus. However,
the size of the largest local clusters has grown to become comparable to the
total system size. In addition, holes and clusters of all sizes can be observed.
The ``scale-invariance'' or ``fractal'' looking structure is the hallmark
of a ``critical state'' for which the correlation length and the 
susceptibility become infinite (or simply bounded by the size of the system).}}
\hspace{5mm}
\parbox[r]{7.5cm}{
\epsfig{file=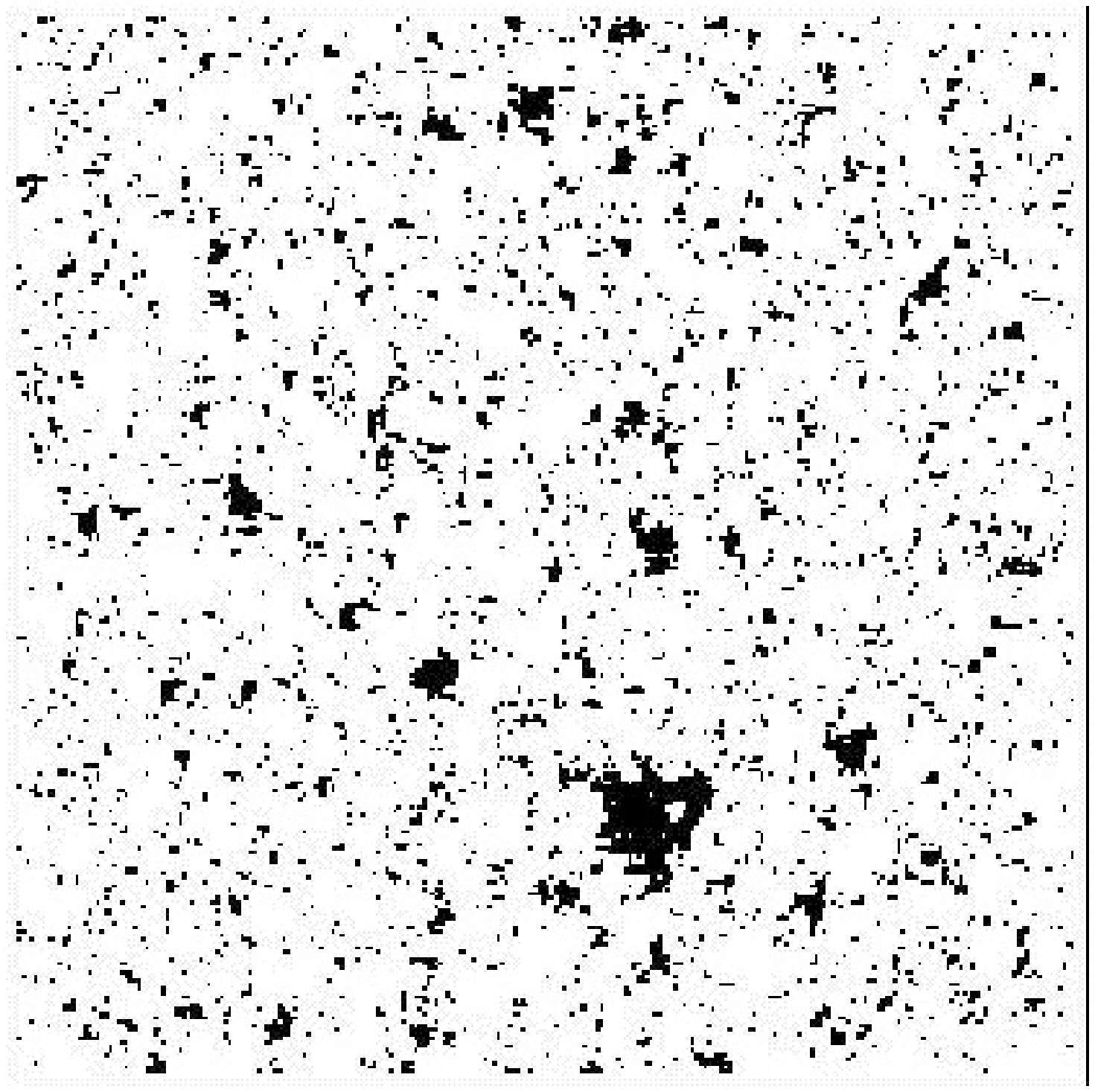,height=7cm,width=7cm}
\caption{\protect\label{FigIsingbelowTc} Same as figure \ref{FigIsingAboveTc} for $K>K_c$. 
The imitation is so strong that the network of agents spontaneously
break the symmetry between the two decisions and one of them predominates.
Here, we show the case where the ``buy'' state has been selected. 
Interestingly, the collapse onto one of the two states is essentially random
and results from the combined effect of a slight initial bias and of 
fluctuations during the imitation process. Only small and isolated islands of 
``bears'' remain in an ocean of buyers. This state would correspond to 
a bubble, a strong bullish market.}}
\end{figure}

When the imitation strength
$K$ increases and gets close to $K_c$ (see figure \ref{FigIsingatTc}), order
starts to appear: the system becomes extremely sensitive to a small global
perturbation, agents who agree with each other form large clusters, and
imitation propagates over long distances. In the Natural Sciences, these are
the characteristics of {\em critical} phenomena. Formally, in this
case the susceptibility $\chi$ of the system goes to infinity.  The
hallmark of
criticality is the {\em power law}, and indeed the susceptibility goes to
infinity according to a power law $\chi\approx A(K_c-K)^{-\gamma}$,
where $A$ is a positive constant and $\gamma>0$ is called the {\em critical
exponent} of the susceptibility (equal to $7/4$ for the 2-d Ising model).
This kind of critical behavior is found in many other models of interacting
elements [Liggett, 1985; 1997] (see also 
[Moss de Oliveira et al., 1999] for applications to finance
among others). The large susceptibility means that the system is unstable:
a small external perturbation may lead to a large collective reaction of the
traders who may revise drastically their decision, which may abruptly produce
a sudden unbalance between supply and demand, thus triggering a crash or a rally.
This specific mechanism will be shown to lead to crashes in the model described
in the next section.

For even stronger imitation strength $K>K_c$, 
the imitation is so strong that the idiosynchratic signals become
negligible and the traders self-organize into a strong imitative
behavior as shown in figure \ref{FigIsingbelowTc}. The selection
of one of the two possible states is determined from small and subtle
initial biases as well as from the fluctuations during the evolutionary
dynamics.

These behaviors apply more generically to other network topologies. Indeed,
the stock market constitutes an ensemble of interacting investors who differ in
size by many orders of magnitudes ranging from individuals to gigantic
professional investors, such as pension funds. Furthermore, structures at
even higher levels, such as currency influence spheres (US\$, DM, YEN ...),
exist and with the current globalization and de-regulation of the market
one may argue that structures on the largest possible scale, {\it i.e.},
the world economy, are beginning to form. 
This observation and the network of connections between traders show
that the two-dimensional lattice representation used in the figures
\ref{IsingEvol}, \ref{FigIsingAboveTc}, \ref{FigIsingatTc}
and \ref{FigIsingbelowTc} is too naive. A better representation of
the structure
of the financial markets is that of hierarchical
systems with ``traders'' on all levels of the market. Of course, this
does not imply that any strict hierarchical structure of the stock market
exists, but there are numerous examples of qualitatively hierarchical
structures in society. In fact, one may say that horizontal organisations
of individuals are rather rare. This means that the plane network used in
our previous discussion may very well represent a gross over-simplification.

Even though the predictions of these models are quite detailed, 
they are very robust to model
misspecification. We indeed claim that models that combine the following
features would display the same characteristics, in particular apparent coordinate
buying and selling periods, leading eventually to several financial crashes. These
features are:
\begin{enumerate}
\item A system of traders who are influenced by their ``neighbors'';
\item Local imitation propagating spontaneously into global cooperation;
\item Global cooperation among noise traders causing collective behavior;
\item Prices related to the properties of this system;
\item System parameters evolving slowly through time;
\end{enumerate}
As we shall show in the following sections, 
a crash is most likely when the locally imitative system
goes through a {\em critical} point.

In Physics, critical points are widely considered to be one of the most
interesting
properties of complex systems. A system goes critical when local influences
propagate over long distances and the average state of the system becomes
exquisitely sensitive to a small perturbation, {\it i.e.} different parts of
the system become highly correlated. Another characteristic is that
critical systems are self-similar across scales: in figure \ref{FigIsingatTc}, at the
critical point, an ocean of traders who are mostly bearish may have within it several
continents of traders who are mostly bullish, each of which in turns surrounds
seas of bearish traders with islands of bullish traders; the progression
continues all the way down to the smallest possible scale: a single trader [Wilson, 1979].
Intuitively speaking, critical self-similarity is why
local imitation cascades through the scales into global coordination.
Critical points are described in mathematical parlance as singularities
associated with bifurcation and catastrophe theory. 

The previous Ising model is one of the simplest possible description of 
cooperative behaviors resulting from repetitive interactions between
agents. Many other models have recently been developed in order to capture
more realistic properties of people and of their economic interactions.
These multi-agent models, often explored by computer simulations, support
the hypothesis that the observed characteristics 
of financial prices, such as non-Gaussian ``fat'' 
tails of distributions of returns, mostly unpredictable returns,
clustered and excess volatility,
may result endogenously from the interaction
between agents. 

Several works have modelled the epidemics of opinion and speculative
bubbles in financial markets from an adaptative agent point-of-view
[Kirman, 1991; Lux, 1995; 1998; Lux and Marchesi, 1999; 2000].
The main mechanism for bubbles is that above
average returns are reflected in a generally more optimistic attitude that
fosters the disposition to overtake others' bullish beliefs and vice
versa. The adaptive nature of agents is reflected in the alternatives
available to agents to choose between several class of strategies, for
instance to invest according to fundamental economic valuation or
by using technical analysis of past price trajectories. Other 
relevant works put more emphasis on the heterogeneity and threshold
nature of decision making which lead in general to irregular cycles
[Takayasu et al., 1992; Youssefmir et al., 1998; Levy et al., 1995; 
Sato and Takayasu, 1998; Levy et al., 2000; Gaunersdorfer, 2000].

\section{Modeling financial bubbles and market crashes}

In this section, we describe three complementary models that we have
developed to describe bubbles and crashes. The first two models
are extensions of the
rational expectation model of bubbles and crashes of
Blanchard [1979] and Blanchard and Watson [1982]. They
originally introduced the model of
rational expectations (RE) bubbles to account for the possibility, often
discussed in the empirical literature and by practitioners, that observed
prices may deviate significantly and over extended time intervals
from fundamental prices.
While allowing for deviations from fundamental prices, rational bubbles
keep a fundamental anchor point of economic modelling, namely that bubbles
must obey the condition of rational expectations. In contrast,
recent works stress that investors are not fully rational, or have at most bound
rationality, and that behavioral and psychological
mechanisms, such as herding, may be important in the shaping of market
prices [Thaler, 1993; Shefrin, 2000; Shleifer, 2000].
However, for fluid assets, dynamic investment strategies rarely perform
over simple buy-and-hold strategies
[Malkiel, 1999], in other words, the market is not far from being efficient
and little arbitrage opportunities exist as a result of the constant
search for gains by sophisticated investors. For the first
two models, we shall work within
the conditions of rational expectations and of no-arbitrage condition,
taken as useful approximations. Indeed, the rationality of both expectations
and behavior often
does not imply that the price of an asset be equal to its fundamental
value. In other words, there can be rational deviations of the price from
this value, called rational bubbles. A rational bubble can arise when
the actual
market price depends positively on its own expected rate of change, as
sometimes occurs in asset markets, which is the mechanism underlying the
models of Blanchard [1979] and Blanchard and Watson [1982].
The third model proposes to complement the modeling of bubbles and crashes 
by studying the effects of interactions between 
the two typical opposite attitudes of investors in stock markets, namely
imitative and contrarian behaviors. 

\vskip-0.3cm
\subsection{The risk-driven model \label{riskdriven}}

This first model contains the following ingredients [Johansen et al., 1999a,b; 2000a]:
\begin{enumerate}
\item A system of traders who are influenced by their ``neighbors''.
\item Local imitation propagating spontaneously into global cooperation.
\item Global cooperation among traders causing crash.
\item Prices related to the properties of this system.
\end{enumerate}

The interplay between the progressive strengthening of imitation
controlled by the three first ingredients and the ubiquity of
noise requires a stochastic description. A crash is not certain but can
be characterized by its hazard rate $h(t)$, i.e., the probability per unit
time that the crash will happen in the next instant if it has not happened yet.

The crash hazard rate $h(t)$ embodies subtle uncertainties of the market\,:
when will the traders realize with sufficient clarity that the market is
over-valued? When will a significant fraction of them believe that the
bullish trend
is not sustainable? When will they feel that other traders think that a
crash is coming?
Nowhere is Keynes's beauty contest analogy more relevant than in the
characterization of the crash hazard rate, because the survival of the
bubble rests on the
overall confidence of investors in the market bullish trend.

A crash happens when a large group
of agents place sell orders simultaneously. This group of agents must create
enough of an imbalance in the order book for market makers to be unable to
absorb the other side without lowering prices substantially. A notable fact
is that the agents in this group typically do not know each other. They did
not convene a meeting and decide to provoke a crash. Nor do they take orders
from a leader. In fact, most of the time, these agents disagree with one
another, and submit roughly as many buy orders as sell orders (these are all
the times when a crash {\em does not} happen). The key question is to
determine by what
mechanism did they suddenly manage to organize a coordinated sell-off?

We propose the following answer [Johansen et al., 1999a,b] already
outline above: all the traders in the
world are organized
into a network (of family, friends, colleagues, and so on) and they influence each
other {\em locally} through this network\,: for instance, an active
trader is constantly on the phone exchanging information and opinions with a
set of selected colleagues. In addition, there are indirect interactions
mediated
for instance by the media. Specifically, if I am directly
connected with $k$ other traders, then there are only two forces that
influence my opinion: (a) the opinions of these $k$ people and of the global
information network; and (b) an
idiosyncratic signal that I alone generate. Our working assumption here is that
agents tend to {\em imitate} the opinions of their connections.
The force (a) will tend to create order, while
force (b) will tend to create disorder. The main story here
is a fight between order and disorder. As far as asset prices are
concerned, a crash happens when order wins (everybody has the same
opinion: selling), and normal times are when disorder wins (buyers and
sellers disagree with each other and roughly balance each other out). We
must stress that this is exactly the opposite of the popular
characterization of crashes as times of chaos. Disorder, or a balanced
and varied opinion spectrum, is what keeps the market liquid in normal times.
This mechanism does not require an overarching
coordination mechanism since macro-level coordination can
arise from micro-level imitation and it relies on a
realistic model of how agents form opinions by constant interactions.

\vskip-0.3cm
\subsubsection{Finite-time singularity in the crash hazard rate \label{sectcrashhaim}}

In the spirit of ``mean field'' theory of collective systems
[Goldenfeld, 1992],
the simplest way to describe an imitation
process is to assume that the hazard rate $h(t)$ evolves according to the
following equation\,:
\be
{dh \over dt} = C~h^{\delta}~,~~~~~~~{\rm with}~\delta > 1~,
\label{azzer}
\ee
where $C$ is a positive constant.
Mean field theory amounts to embody the diversity of trader actions by a
single effective representative behavior determined from an average
interaction between the
traders. In this sense, $h(t)$ is the collective result of the interactions
between
traders. The term $h^{\delta}$ in the r.h.s. of (\ref{azzer}) accounts for
the fact that the
hazard rate will increase or decrease due to the presence of {\em
interactions} between the traders. The exponent $\delta > 1$ quantifies the
effective number
equal to $\delta - 1$ of interactions felt by a typical trader. The condition
$\delta > 1$ is crucial to model interactions and is, as we now show, essential
to obtain a singularity (critical point) in finite time.
Indeed, integrating (\ref{azzer}), we get
\be
h(t) = {B \over (t_c - t)^{\alpha}}~,~~~~~~~~{\rm with}~\alpha \equiv {1
\over \delta - 1}~.
\label{cjdjlk}
\ee
The critical time $t_c$ is determined by the initial conditions at some
origin of time.
The exponent $\alpha$ must lie between zero and one for an economic
reason\,: otherwise, as we shall see,
the price would go to infinity when approaching $t_c$ (if the bubble has
not crashed in the mean time). This condition translates into $2 < \delta <
+\infty$\,: a typical trader must be
connected to more than one other trader. There is a large body of literature in
Physics, Biology and Mathematics on the microscopic modeling of systems of
stochastic dynamical
interacting agents that lead to critical behaviors of the type
(\ref{cjdjlk}) [Liggett, 1985, 1997]. The macroscopic
model (\ref{azzer}) can thus be substantiated by specific microscopic
models [Johansen et al., 2000].

Before continuing, let us provide an intuitive explanation for 
the creation of a finite-time singularity at $t_c$. 
The faster-than-exponential growth of the return and of the crash
hazard rate correspond to non-constant growth
rates, which increase with the return and with the hazard rate. The following
reasoning allows us to understand intuitively the origin of the appearance
of an infinite slope or infinite value in a finite time at $t_c$, called
a finite-time singularity.
Suppose for instance that the growth rate of the hazard rate doubles when the
hazard rate doubles. For simplicity, we consider discrete time intervals as
follows. Starting with a hazard rate of $1$, we assume it grows at a constant
rate of $1\%$ per day until it doubles. We estimate the doubling time
as proportional to the inverse of the growth rate, i.e.,
approximately $1/1\%=1/0.01=$ one hundred days. There is a multiplicative correction term
equal to $\ln 2=  0.69$ such that the doubling time is $\ln 2 /1\%= 69$ days. 
But we factor out this proportionality factor $\ln 2=  0.69$
for the sake of pedagogy and simplicity. Including
it multiplies all time intervals below by $0.69$ without changing
the conclusions.

When
the hazard rate turns $2$, we assume that the growth rate doubles to $2\%$ and stays
fixed until the hazard rate doubles again to reach $4$. This new doubling time is only 
approximately $1/0.02=50$ days
at this $2\%$ growth rate. When the hazard rate reaches $4$, its growth rate is
doubled to $4\%$. The doubling time of the hazard rate is therefore approximately
halved to $25$ days and the scenario continues with a doubling of the growth rate
every time the hazard rate doubles. Since the doubling time is approximately
halved at each step, we have the following sequence (time=$0$, hazard rate=$1$,
growth rate=$1\%$), (time=$100$, hazard rate=$2$, growth rate=$2\%$), (time=$150$,
hazard rate=$4$, growth rate=$4\%$), (time=175, hazard rate=$8$, growth rate=$8\%$) and
so on. We observe that the time interval needed for the hazard rate to double is
shrinking very rapidly by a factor of two at each step. In the same way that 
\be
{1 \over 2} + {1 \over 4} + {1 \over 8} + {1 \over 16} +... = 1~,
\ee
which was immortalized by the Ancient Greeks as Zeno's
paradox, the infinite sequence of doubling thus takes a finite time and the
hazard rate reaches infinity at a finite ``critical time'' approximately equal to
$100+50+25+...=200$ (a rigorous mathematical treatment requires a continuous time
formulation, which does not change the qualitative content of the example).  A
spontaneous singularity has been created by the increasing growth rate! This
process is quite general and applies as soon as the growth rate possesses the
property of being multiplied by some factor larger than 1 when the hazard rate or
any other observable is
multiplied by some constant larger than 1.

\vskip-0.3cm
\subsubsection{Derivation from the microscoping Ising model}

The phenomenological equations (\ref{azzer}) and (\ref{cjdjlk}) can be
derived from the microscopic model of agent interactions described by
equation (\ref{eq:state}). For this,
let us assume that the imitation strength $K$ changes smoothly with time,
as a result for instance of the varying confidence level of investors, the 
economic outlook, and so on. The simplest assumption, which does not change the
nature of the argument, is that $K$ is proportional to time. Initially, 
$K$ is small and only small clusters of investors self-organize, as shown in 
figure \ref{FigIsingAboveTc}. As $K$ increases, the typical size of the clusters
increases as shown in figure \ref{FigIsingatTc}. These kinds of systems exhibiting
cooperative behavior are characterized by a broad distribution of cluster sizes $s$
(the size of the black islands for instance) up to a maximum $s^*$ which itself increases
in an accelerating fashion up to the critical value $K_c$. Right at $K=K_c$,
the geography of clusters of a given kind becomes self-similar with a continuous
hierarchy of sizes from the smallest (the individual investor) to the largest
(the total system). Within this phenomenology, the probability for a crash to occur is 
constructed as follows.

First, a crash corresponds to a coordinated sell-off of a large number of investors.
In our simple model, this will happen as soon as a single cluster of connected investors, which
is sufficiently large to set the market off-balance, decides to sell-off. Recall indeed that
``clusters'' are defined by the condition that all investors in the same cluster move in 
concert. When a very large cluster of investors sells, this creates a sudden unbalance 
which triggers an abrupt drop of the price, hence a crash. To be concrete, we assume
that a crash occurs when the size (number of investors) $s$ of the active cluster 
is larger than some minimum value $s_m$. The specific value $s_m$ is not important,
only the fact that $s_m$ is much larger than $1$ so that a crash can only occur
as a result of a cooperative action of many traders who destabilize the 
market. At this stage, we do not
specify the amplitude of the crash, only its triggering as an instability.
For this explanation to make sense, investors change opinion and send market orders
only rarely. Therefore, we should expect only one or few large clusters to be 
simultaneously active and able to trigger a crash.

For a crash to occur, we thus need (1) to find at least one cluster of size larger than 
$s_m$ and (2) to verify that this cluster is indeed actively selling-off. Since these
two events are independent,
the probability for a crash to occur is thus the product of the probability to find 
such a cluster of 
size larger than the threshold $s_m$ by the probability that such a cluster begins
to sell-off collectively.
The probability to find a cluster of size $s$ is a well-known characteristic of
critical phenomena [Goldenfeld, 1992; Stauffer and Aharony, 1994]:
it is a power law distribution 
truncated at a maximum $s^*$; this 
maximum increases without bound (except for the total system size) on the approach
to the critical value $K_c$ of the imitation strength.

If the decision to sell off by an investor belonging to a given cluster of size $s$ was
independent of the decisions of all the other investors in the same cluster, 
then the probability per unit time that such a cluster of size $s$ becomes active 
would be simply proportional to the number $s$ of investors in that cluster.
However, by the very definition of a cluster, 
investors belonging to a given cluster do interact with each other.
Therefore, the decision of an investor to sell off is probably quite strongly 
coupled with those of the other investors in the same cluster. Hence, 
the probability per unit time that a specific cluster of $s$ investors
becomes active is a function of the number $s$ of investors belonging
to that cluster and of all the interactions between these investors.
Clearly, the maximum number of interactions within a
cluster is $s \times (s-1)/2$, that is, for large $s$,
it becomes proportional to the square of the number of investors in that cluster.
This occurs when each of the $s$ investors speaks to each of his or her $s-1$ colleagues. The
factor $1/2$ accounts for the fact that if investor Anne speaks to investor
Paul then in general Paul also speaks to Anne and their two-ways interactions must
be counted only once. Of course, one can imagine more complex situations in which Paul listen
to Anne but Anne does not reciprocate but this does not change the results. 
Nothwithstanding these complications, one sees that the probability $h(t) \Delta t$ per 
unit time $\Delta t$ that a specific cluster of $s$ investors
becomes active must be a function growing with the cluster size $s$ faster than $s$
but probably slower than the maximum number of interactions (proportional to $s^2$). 
A simple parameterization is to take 
$h(t) \Delta t$ proportional to the cluster size $s$ elevated to some power $\alpha$
larger than $1$ but smaller than $2$. This exponent $\alpha$ captures the collective
organization within a cluster of size $s$ due to the multiple interactions between
its investors. It is related to the concept of fractal dimensions.

The probability for a crash to occur, which is the same as the probability of finding
at least one active cluster of size larger than the minimum destabilizing size $s_m$,
is therefore the sum over all sizes $s$ larger than $s_m$ of all the products
of probabilities $n_s$ to find a cluster of a specific size $s$ by their probability
per unit time to become active (itself proportional to $s^{\alpha}$ as we have argued).
With mild technical conditions, it can then be shown that the crash hazard rate exhibits
a power law acceleration with a singular behavior. Intuitively, this result stems from
the interplay between the existence of larger and larger clusters as the interaction 
parameter $K$ approached its critical value $K_c$ and from the nonlinear accelerating
probability per unit time for a cluster to become active as its typical
size $s^*$ grows with the approach of $K$ to $K_c$

The diverging acceleration of the crash probability 
implies a remarkable prediction for the crash hazard rate: indeed, the crash hazard
rate is nothing but the rate of change of the probability of a crash as a function of time
(conditioned on it not having happened yet). The crash
hazard rate thus increases without bounds as $K$ goes to $K_c$. The risk of a crash
per unit time, knowing that the crash has not yet occurred, increases dramatically
when the interaction between investors becomes strong enough so that the network
of interactions between traders self-organized into a hierarchy containing a few
large spontaneously formed groups acting collectively. 

We stress that $K_c$ is not the value of the 
imitation strength at which the crash occurs, because the crash
could happen for any value before $K_c$, even though this is not very likely.
$K_c$ is the most probable value of the imitation strength for which
the crash occurs. To translate these results as a function of time,
it is natural to expect that the imitation strength $K$ is changing slowly with time as 
a result of several factors influencing the tendancy of investors to herd.
A typical trajectory $K(t)$ of the imitation strength as a function of time $t$ 
is erratic and smooth. The critical time $t_c$ is defined as
the time at which the critical imitation strength $K_c$ is reached for the first time
starting from some initial value. $t_c$ is not the time of the crash, it is
the end of the bubble. It is the most
probable time of the crash because the hazard rate is largest at that time. 
Due to its probabilistic nature, the crash can occur at any other time, with a 
likelihood changing with time following the crash hazard rate. 

The critical time $t_c$ (or $K_c$) signals the death of the speculative bubble.
We stress that $t_c$ is not {\em the} time of the crash because the crash
could happen at any time before $t_c$, even though this is not very likely.
$t_c$ is simply the most probable time of the crash. There
exists a finite probability
\be
1- \int_{t_0}^{t_c} h(t) dt >0
\ee
of  ``landing'' smoothly, i.e., of
 attaining the end of the bubble without crash. This residual probability is
crucial for the coherence of the model, because otherwise agents would
anticipate the crash
and would exit from the market.

\vskip-0.3cm
\subsubsection{Dynamics of prices from the rational expectation condition}

Assume for simplicity that, during a crash, the price drops
by a fixed percentage $\kappa\in(0,1)$, say between $20$ and $30\%$ of the
price increase above a reference value $p_1$.
Then, the dynamics of the asset price before the crash are given by:
\be
\label{eq:crash}
dp = \mu(t)\,p(t)\,dt-\kappa [p(t)-p_1] dj~,
\ee
where $j$ denotes a jump process whose value is zero before the crash and one
afterwards. In this simplified model,
we neglect interest rate, risk aversion, information asymmetry,
and the market-clearing condition.

As a first-order approximation of the market organization, we assume that
traders do their best and price the asset so that a fair game condition holds.
Mathematically, this stylized rational expectation model
is equivalent to the familiar martingale hypothesis:
\be
\label{eq:martingale}
\forall t'>t \qquad\  E_t[p(t')] = p(t)
\ee
where $p(t)$ denotes the price of the asset at time $t$ and $E_t[\cdot]$
denotes the expectation conditional on information revealed up to time $t$.
If we do not allow the asset price to fluctuate under the impact of noise,
the solution to equation (\ref{eq:martingale}) is a constant: $p(t) = p(t_0)$,
where $t_0$ denotes some initial time. $p(t)$ can be interpreted as the
price in
excess of the fundamental value of the asset. This rational 
expectation bubble model can be extended to general and arbitrary risk-aversion within
the general stochastic discount factor theory [Sornette and Johansen, 2001]. 

Putting (\ref{eq:crash}) in (\ref{eq:martingale}) leads to
\be
\mu(t) p(t) = \kappa [p(t)-p_1] h(t)~.
\label{hfjqklq}
\ee
In words, if
the crash hazard rate $h(t)$ increases, the return $\mu$ increases to
compensate the traders
for the increasing risk. Plugging (\ref{hfjqklq}) into
(\ref{eq:crash}), we obtain a ordinary differential equation. For
$p(t) - p(t_0) < p(t_0) - p_1$, its solution is
\be
\label{eq:price}
p(t) \approx p(t_0) +  \kappa [p(t_0) - p_1]~\int_{t_0}^t h(t') dt'
\qquad\mbox{before the crash}.
\ee

If instead the price drops by a fixed percentage $\kappa\in(0,1)$ of the
price, the dynamics of the asset price before the crash is given by
\be \label{eq:crash1}
dp = \mu(t)\,p(t)\,dt-\kappa p(t)dj~.
\ee
We then get
\be
{\rm E}_t[dp] = \mu(t)p(t)dt-\kappa p(t)h(t)dt = 0~,
\ee
which yields\,:
\be
\mu(t) = \kappa h(t)~.
\label{hkllmqmlqm}
\ee
and the corresponding equation for the price is\,:
\be
\label{eq:price1}
\log\left[\frac{p(t)}{p(t_0)}\right] = \kappa\int_{t_0}^t  h(t')dt'
\qquad\mbox{before the crash}.
\ee
This gives the logarithm of the price as the relevant observable. These
two different scenarios for the price drops raises a rather
interesting question. If the first scenario is the correct one, then crashes
are nothing but (a partial) depletion of preceding bubbles and hence signals 
the markets return towards equilibrium. Hence, it may as such be taken as a 
sign of economical health, as also suggested by [Barro et al., 1989] in relation to
the crash of Oct. 1987. 
On the other hand, if the second scenario is true, this
suggest that bubbles and crashes are instabilities which are built-in or inherent 
in the market structure and that they are signatures of a market constantly
out-of-balance, signaling fundamental systemic instabilities. 
We will return to this question in the conclusion. Johansen and Sornette [2001b] have
shown that the first scenario is slightly more warranted according to the data.

The higher the probability of a crash, the faster the price must increase
(conditional on having no crash) in order to satisfy the martingale (no
free lunch) condition.
Intuitively, investors must be compensated by the chance of a higher return
in order to be induced to hold an asset that might crash. This effect
may go against the naive preconception that price is adversely affected by
the probability of the crash, but our result is the only one consistent with
rational expectations. Complementarily, from a behavioral and
dynamical point of view of the financial market, a faster rising price decreases
the probability that it can be sustained much longer and may announce an instable
phase in the mind of investors. We thus face a kind of ``chicken and egg'' problem.

Plugging (\ref{cjdjlk}) into (\ref{eq:price}) gives the
following price law:
\be
\label{eq:solution}
p(t) \approx p_c -\frac{\kappa B}{z}\times(t_c-t)^{z}
\qquad\mbox{before the crash}.
\ee
where $z = 1-\alpha\in(0,1)$ and $p_c$ is the price at the critical
 time (conditioned on no crash having been triggered). The price before the
crash thus
follows a power law with a finite upper bound $p_c$.
The trend of the price becomes unbounded as we
approach the critical date. This is to compensate for an unbounded
crash rate in the next instant.

The last ingredient of the model is to recognize that
the stock market is made of actors which differs in
size by many orders of magnitudes ranging from individuals to gigantic
professional investors, such as pension funds. Furthermore, structures at
even higher levels, such as currency influence spheres (US\$, Euro, YEN ...),
exist and with the current globalization and de-regulation of the market
one may argue that structures on the largest possible scale, i.e.,
the world economy, are beginning to form. This means that the structure
of the financial markets have features which resembles that of hierarchical
systems with ``traders'' on all levels of the market. Of course, this
does not imply that any strict hierarchical structure of the stock market
exists, but there are numerous examples of qualitatively hierarchical
structures in society. Models of imitative interactions on
hierarchical structures recover the power law behavior (\ref{eq:solution})
[Sornette and Johansen, 1998; Johansen et al., 2000]. But in
addition, they predict that the critical exponent $\alpha$ can be a
complex number!
The first order expansion of the general solution for the hazard rate is then
\be
\label{eq:hazard3}
h(t)\approx B_0(t_c-t)^{-\alpha}
+B_1(t_c-t)^{-\alpha}\cos[\omega\log(t_c-t)-\psi ].
\ee
Once again, the crash hazard rate explodes near the critical date. In
addition,
it now displays log-periodic oscillations. The evolution of the
price  before the crash and before the critical date is given by:
\be
\label{eq:complex}
p(t) \approx p_c -\frac{\kappa}{z}\left\{
B_0(t_c-t)^{z}
+B_1(t_c-t)^{z}\cos[\omega\log(t_c-t)-\phi]\right\}
\ee
where $\phi$ is another phase constant. The key feature is that oscillations
appear in the price of the asset before the critical date. 
This means that the local maxima of the function are 
separated by time intervals that tend to zero at the
critical date, and do so in geometric progression, i.e., the ratio of
consecutive time
intervals between maxima is a constant
\be
\lambda \equiv e^{2 \pi \over \omega}~.
\ee
This is very useful from an empirical point
of view because such oscillations are much more strikingly visible in actual
data than a simple power law\,: a fit can ``lock-in'' on the oscillations which
contain information about the critical date $t_c$.
Note that complex exponents and log-periodic oscillations do
not necessitate a pre-existing hierarchical structure as mentioned above, but
may emerge spontaneously from the non-linear
complex dynamics of markets [Sornette, 1998].

To sum up, we have constructed a model in which the stock market price is
driven by the risk of a crash, quantified by its hazard rate. In turn, imitation and
herding forces drive the crash hazard rate. When the imitation strength becomes
close to a critical value, the crash hazard rate diverges with a characteristic
power law behavior. This leads to a specific power law acceleration of the market price,
providing our first predictive precursory pattern anticipating a crash.

\vskip-0.3cm
\subsection{The price-driven model \label{pricedriven}}

The price-driven model inverts the logic of the previous risk-driven model: here, 
again as a result of the action of rational investors,
the price is driving the crash hazard rate rather than the reverse. The price itself
is driven up by the imitation and herding behavior of the ``noisy'' investors.

As before, a stochastic description is required to capture
the interplay between the progressive strengthening of imitation
controlled by the connections and interactions between traders
and the ubiquity of
idiosyncratic behavior as well as the influence of many other factors
that are impossible to model in details. As a consequence, the price
dynamics are stochastic and the occurrence of a
crash is not certain but can
be characterized by its hazard rate $h(t)$, defined as the probability per unit
time that the crash will happen in the next instant if it has not happened yet.

Keeping a basic tenet of economic theory, rational expectations, the model 
developed in [Sornette and Andersen, 2002]
captures the nonlinear positive feedback between agents in the stock market
as an interplay between nonlinearity and multiplicative noise. 
The derived
hyperbolic stochastic finite-time singularity formula transforms a Gaussian
white noise into a rich time series possessing 
all the stylized facts of empirical prices, as well as 
accelerated speculative bubbles preceding crashes. 

Let us give the premise of the model and some preliminary results.
We start from the geometric Brownian model of the 
bubble price $B(t)$,
$d B = \mu B dt + \sigma B dW_t$,
where $\mu$ is the instantaneous return rate, $\sigma$ is the volatility
and $dW_t$ is the infinitesimal increment of the random walk with unit variance
(Wiener process).  We generalize this expression into
\be
d B(t) = \mu(B(t)) B(t) dt + \sigma(B(t)) B(t) dW_t - \kappa(t) B(t) dj~,  \label{nfkaaak}
\ee
allowing $\mu(B(t))$ and $\sigma(B(t))$ to depend arbitrarily 
and nonlinearly on the instantaneous realization of the price.
A jump term has been added to describe a correction or a crash
of return amplitude $\kappa$, which can be a stochastic variable taken from 
an a priori arbitrary distribution. 
Immediately after the last crash which becomes the new origin of time $0$,
$dj$ is reset to $0$ and will eventually jump to $1$ with a
hazard rate $h(t)$, defined such that the probability that a crash occurs between $t$ 
and $t+dt$ conditioned
on not having occurred since time $0$ is $h(t) dt$. 

Following [Blanchard , 1979; Blanchard and Watson, 1982], $B(t)$ is a 
rational expectations bubble which accounts for the possibility, often
discussed in the empirical literature and by practitioners, that observed
prices may deviate significantly and over extended time intervals
from fundamental prices.
While allowing for deviations from fundamental prices, rational bubbles
keep a fundamental anchor point of economic modelling, namely that bubbles
must obey the condition of rational expectations. This translates essentially
into the no-arbitrage condition with risk-neutrality, which states that
the expectation of $d B(t)$ conditioned on the past
up to time $t$ is zero. This allows us to determine 
the crash hazard rate $h(t)$ as a function of $B(t)$.
Using the definition of the hazard rate $h(t) dt = \langle dj \rangle$,
where the bracket denotes the expectation over all possible outcomes since the last crash, 
this leads to $\mu(B(t)) B(t)   - \langle \kappa \rangle B(t) h(t) = 0$,
which provides the hazard rate as a function of price:
\be
h(t) = {\mu(B(t)) \over \langle \kappa \rangle}~.  \label{bvfjuaj}
\ee
Expression (\ref{bvfjuaj}) quantifies the fact that the theory of rational expectations
with risk-neutrality
associates a risk to any price: for example, if the bubble price explodes, so will the crash
hazard rate, so that the risk-return trade-off is always obeyed. 
We note that it is easy to incorporate risk-aversion by introducing 
a risk-premium rate or by amplifying the risk of a crash perceived by traders.

The dependence of $\mu(B(t))$ and $\sigma(B(t))$ is chosen so as
to capture the possible appearance of
positive feedbacks on prices. There are many mechanisms in the stock
market and in the behavior of investors which may lead to positive feedbacks.
First, investment strategies with ``portfolio insurance'' are such that
sell orders are issued whenever a loss threshold (or stop loss) is passed. It is clear that
by increasing the volume of sell order, this may 
lead to further price decreases.  Some commentators have indeed attributed the 
crash of Oct. 1987 to a cascade of sell orders.
Second, there is a growing empirical evidence of the existence of herd or ``crowd'' 
behavior in speculative markets [Shiller, 2000], 
in fund behaviors [Scharfstein and Stein, 1990; Grinblatt et al., 1995] and 
in the forecasts made by financial analysts [Trueman, 1991]. 
Although this behavior is inefficient from a social
standpoint, it can be rational from the perspective of managers who are
concerned about their reputations in the labor market.
As we have already mentioned,
such behavior can be rational
and may occur as an information cascade, a situation in which every subsequent actor, 
based on the observations of others, makes the same choice independent of
his/her private signal [Bikhchandani et al., 1992]. 
Herding leads to positive nonlinear feedback. Another mechanism for positive
feedbacks is the so-called ``wealth'' effect: a rise of the stock market
increases the wealth of investors who spend more, adding to the earnings
of companies, and thus increasing the value of their stock.

The evidence for nonlinearity has a strong empirical support: for instance, 
the coexistence of
the absence of correlation of price changes and the strong autocorrelation of their 
absolute values can not be explained by any linear model [Hsieh, 1985].
Comparing additively nonlinear processes and multiplicatively nonlinear 
models, the later class of models are found consistent
with empirical price changes and with options' implied volatilities.
With the additional insight that hedging strategies of general
Black-Scholes option models lead to a positive feedback on the volatility 
[Sircar and Papanicolaou, 1998],
we are led to propose the following
simplistic nonlinear model with multiplicative noise in which
the return rate and the 
volatility are nonlinear increasing power law of $B(t)$ [Sornette and Andersen, 2002]: 
\ba
\mu(B) B &=& {m \over 2B} [B \sigma(B)]^2 + \mu_0 [B(t)/B_0]^m ~,  \label{buyaauqka}   \\
\sigma(B) B &=&  \sigma_0 [B(t)/B_0]^m~,  \label{fjallqaaq}
\ea
where $B_0$, $\mu_0$, $m>0$ and $\sigma_0$ are four parameters of the model, setting respectively
a reference scale, an effective drift and the strength of the nonlinear 
positive feedback. The first term in the r.h.s. (\ref{buyaauqka}) is added as 
a convenient device to simplify the Ito calculation of these stochastic differential
equations. The model can be reformulated in the Stratonovich interpretation
\be
{d B \over dt} = (a \mu_0 + b \eta)~B^m~,  \label{jfja}
\ee
where $a$ and $b$ are two constants and $\eta$ is a delta-correlated Gaussian white noise, 
in physicist's notation such that $\eta dt \equiv dW$. The form (\ref{jfja})
examplifies the fundamental ingredient of the theory developed in
[Sornette and Andersen, 2002]
based on the interplay between nonlinearity and multiplicative noise. The nonlinearity
creates a singularity in finite time 
and the multiplicative noise makes it stochastic. The choice (\ref{buyaauqka},\ref{fjallqaaq})
or (\ref{jfja}) are the simplest generalisation of the standard geometric Brownian model
(\ref{nfkaaak}) recovered for the special case $m=1$. The introduction of the exponent $m$
is a straightforward mathematical trick to account in the simplest and most parsimonious
way for the presence of nonlinearity. Note in particular that, in the limit where $m$ becomes
very large, the nonlinear function $B^m$ tends to a threshold response. The power
$B^m$ can be decomposed as $B^m = B^{m-1} \times B$ stressing the fact that $B^{m-1}$ plays
the role of a growth rate, function of the price itself. The positive feedback effect
is captured by the fact that a larger price $B$ feeds a larger growth rate, which leads
to a larger price and so no. 

The solution of (\ref{nfkaaak}) with (\ref{buyaauqka}) and (\ref{fjallqaaq})
is given by
\be
B(t) = \alpha^{\alpha}~{1 \over \left(\mu_0[t_c - t] - 
{\sigma_0 \over B_0^{m}}~W(t)\right)^{\alpha}}~,  ~~~~{\rm where}~\alpha\equiv {1 \over m-1}
\label{jfkaaakaaa}
\ee
with $t_c= y_0/(m-1)\mu_0$ is a constant determined by the initial condition
with $y_0=1/B(t=0)^{m-1}$. To grasp the meaning of (\ref{jfkaaakaaa}), let
us first consider the deterministic case $\sigma_0=0$, such that the 
return rate $\mu(B) \propto [B(t)]^{m-1}$ is the sole driving term. Then,
(\ref{jfkaaakaaa}) reduces to $B(t) \propto 1/[t_c-t]^{1 \over m-1}$, i.e.,
a positive feedback $m>1$ of the price $B(t)$ on the return rate $\mu$ 
creates a finite-time singularity at a critical time $t_c$ determined by the initial
starting point. This power law
acceleration of the price accounts for the effect of herding resulting from
the positive feedback. It is in agreement with the empirical finding that
price peaks have sharp concave upwards maxima [Roehner and Sornette, 1998].
Reintroducing the stochastic component $\sigma_0 \neq 0$, we see
from (\ref{jfkaaakaaa}) that the finite-time singularity still exists but
its visit is controlled by the first passage of a biased random walk at the 
position $\mu_0 t_c$ such that the denominator
$\mu_0[t_c - t] -  {\sigma_0 \over B_0^{m}}~W(t)$ vanishes.
In practice, a price trajectory will never sample the finite-time
singularity as it is not allowed to approach too close to it due to the jump process
$dj$ defined in (\ref{nfkaaak}). Indeed, from the no-arbitrage condition, the
expression (\ref{bvfjuaj}) for the crash hazard rate ensures that when the price
explodes, so does $h(t)$ so that a crash will occur with larger and larger probability,
ultimately screening the divergence which can never be reached. The endogeneous 
determination (\ref{bvfjuaj}) of the crash probability also ensures that the denominator 
$\mu_0[t_c - t] - {\sigma_0 \over B_0^{m}}~W(t)$ never becomes negative: when it
approaches zero, $B(t)$ blows up and the crash hazard rate increases accordingly.
A crash will occur with probability $1$ before the denominator reaches zero.
Hence, the price $B(t)$ remains always positive and real.
We stress the
remarkably simple and elegant constraint on the dynamics provided by the 
rational expectation condition
that ensures the existence and stationarity of the dynamics at all times, nothwithstanding
the locally nonlinear stochastic explosive dynamics.
When $\mu_0 >0$, the random walk has a positive drift attracting the denominator 
in (\ref{jfkaaakaaa}) to zero
(i.e., attracting the bubble to infinity). However, by the mechanism explained above, 
as $B(t)$ increases, so does the crash hazard rate by the relation (\ref{bvfjuaj}). 
Eventually, a crash occurs that reset the bubble to a lower price. The random walk with drift
goes on, eventually $B(t)$ increases again and reaches ``dangerous waters'', a crash occurs
again, and so on. Note that a crash is not a certain event: an inflated bubble price can
also deflate spontaneously by the random realisation of the random walk $W(t)$ which 
brings back the denominator far from zero.

\begin{figure}
\begin{center}
\epsfig{file=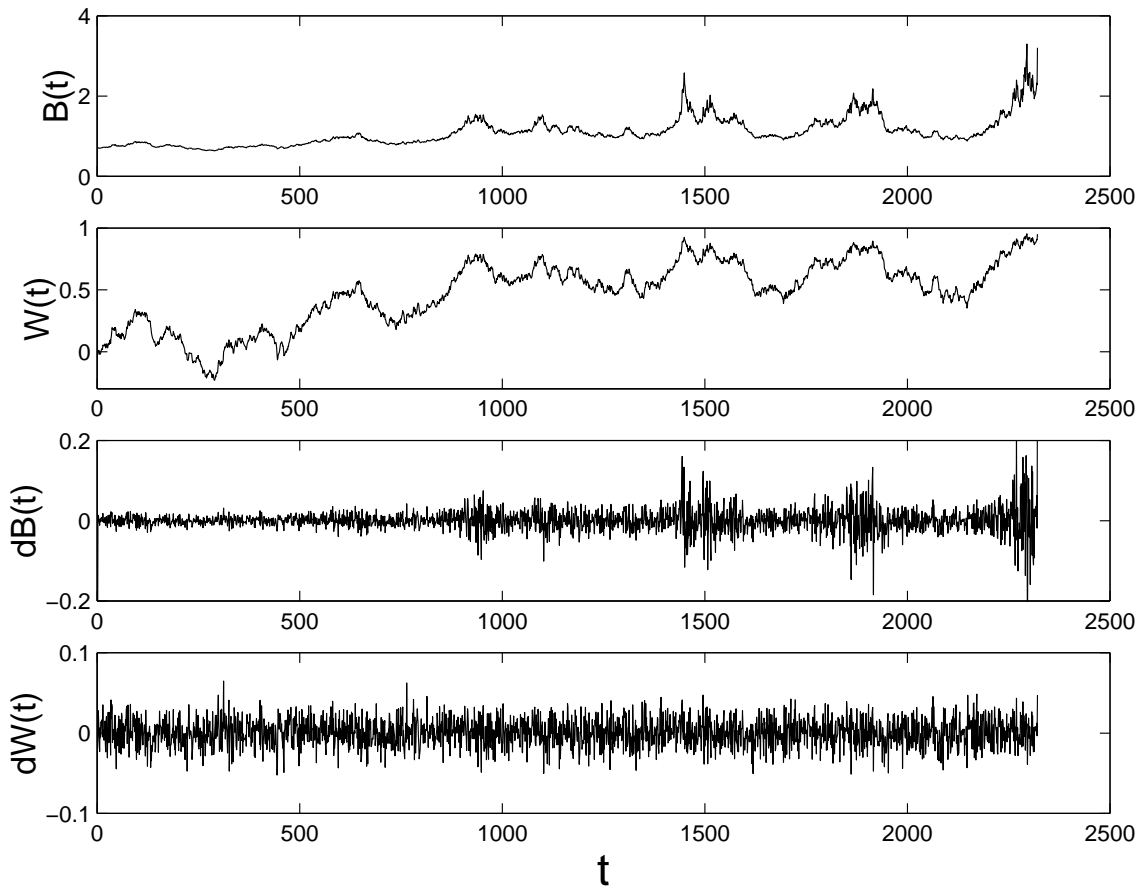,height=12cm}
\caption{\protect\label{hyperbubble1} Top panel: 
realization of a bubble price $B(t)$ as a function of time constructed
from the ``singular inverse random walk''. This corresponds to a specific realization
of the random numbers
used in generating the random walks $W(t)$ represented in the second panel.
The top panel is obtained by 
taking a power of the
inverse of a constant $W_c$ here taken equal to $1$ minus the
 random walk shown in the second panel. In this case, when the random walk approach $1$,
 the bubble diverges. Notice the similarity between the trajectories shown in the top
 ($B(t)$) and second ($W(t)$) panels as long as the random walk $W(t)$ does not approach too much 
 the value $W_c=1$. It is free to wander but when it
 approaches $1$, the bubble price $B(t)$ shows much greater sensitivity and eventually
 diverges as $W(t)$ reaches $1$. Before this happens, $B(t)$ can
 exhibit local peaks, i.e., local bubbles, which come back smoothly. This corresponds
 to realization when
 the random walk approaches $W_c$ without touching it and then spontaneously
 recedes away from it. The third (respectively fourth) panel 
 shows the time series of the increments 
 $dB(t)=B(t)-B(t-1)$ of the bubble (respectively $dW(t) = W(t)-W(t-1)$ of the random walk.
 Notice the intermittent bursts of strong volatility in the bubble compared to the 
 featureless constant level of fluctuations of the random walk.
 (reproduced from [Sornette and Andersen, 2002]).
}
\end{center}
\end{figure}

Figure \ref{hyperbubble1} shows a typical trajectory of the bubble component of the price
generated by the nonlinear positive feedback model of [Sornette and Andersen, 2002], starting
from some initial value up to the time just before the price starts to blow up.
The simplest version of this model consists
in a bubble price $B(t)$ being essentially a power
of the inverse of a random walk $W(t)$ in the following sense.
Starting from $B(0)=W(0)=0$ at the origin of time, when the random walk approaches
some value $W_c$ here taken equal to $1$, $B(t)$ increases
and vice-versa. In particular, when $W(t)$ approaches $1$, $B(t)$ blows up and reaches 
a singularity at the time $t_c$ when the random walk crosses $1$. 
This process generalizes
in the random domain the finite-time singularities described in section \ref{sectcrashhaim},
such that the monotonously increasing process culminating at
a critical time $t_c$ is replaced by the random walk
that wanders up and down before eventually reaching the critical level.
This nonlinear positive
feedback bubble process $B(t)$ can thus be called a ``singular inverse random walk''.
In absence of a crash, the process $B(t)$ can exist only up to a finite time: 
with probability one (i.e., with certainty), 
we know from the study of random walks that $W(t)$ will eventually
reach any level, in particular the value $W_c=1$ in our example at which $B(t)$ diverges. 

The second effect that tampers the possible divergence of the bubble price, by far the most 
important one in the regime of highly overpriced markets, 
is the impact of the price on the crash hazard rate discussed above: 
as the price blows up due to imitation, herding, speculation as well
as randomness, the crash hazard rate increases even faster according to
equation (\ref{bvfjuaj}), so that a crash will occur
and drive the price back closer to its fundamental value. 
The crashes are triggered
in a random way governed by the crash hazard rate which is an increasing function 
of the bubble price. In the present formulation, the
higher the bubble price is, the higher is the probability of a crash. 
In this model, a crash is similar to a purge administered to a patient.

\begin{figure}
\begin{center}
\epsfig{file=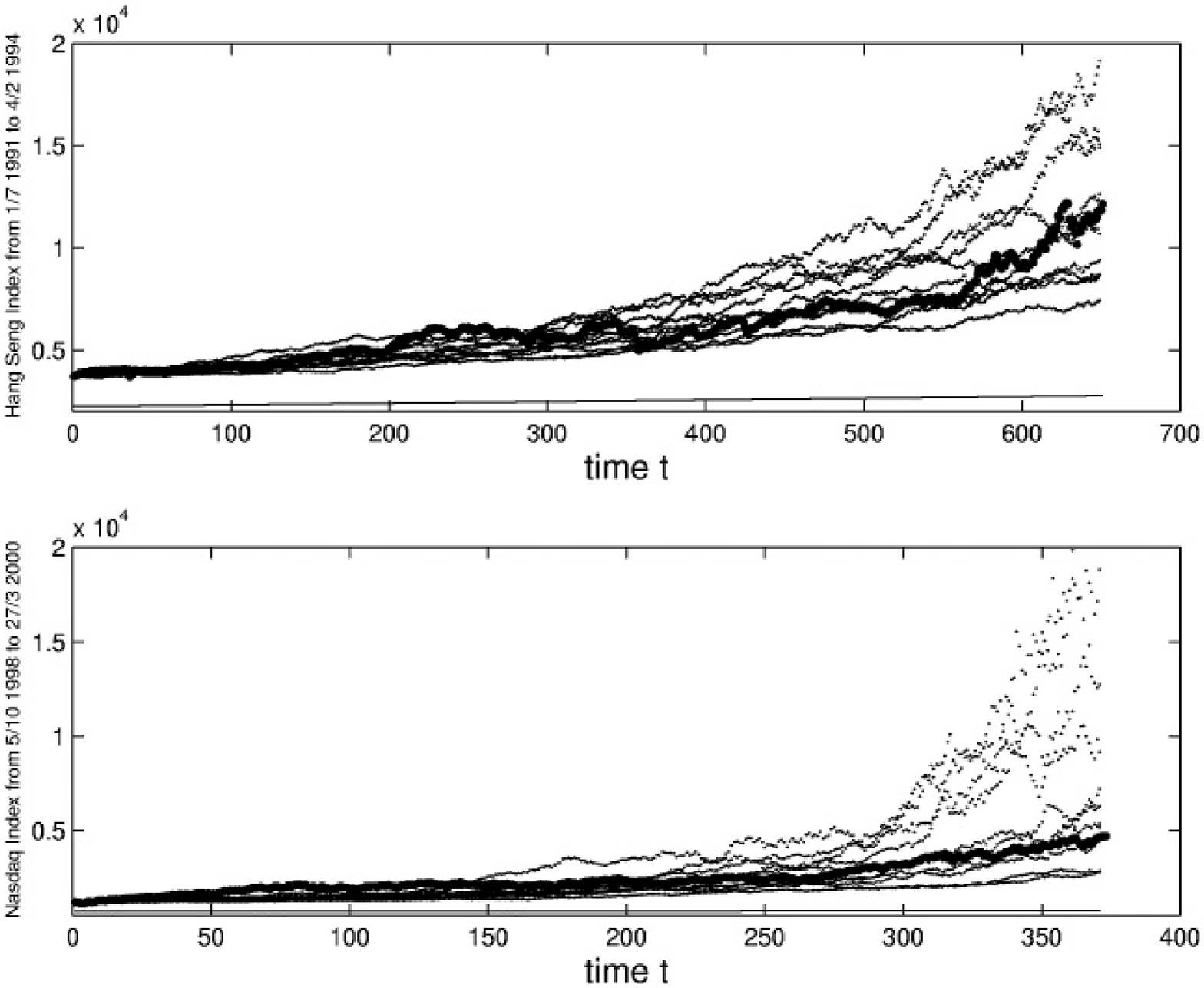,width=14cm,height=12cm}
\caption{\protect\label{Nasdaqandothercomp} Top panel: 
the Hang Seng index from Jul., 1 1991 to Feb. 4 1994
as well as 10 realizations of the 
``singular inverse random walk'' bubble model 
generated by the nonlinear positive feedback model.
Each realization corresponds to an arbitrary random walk whose drift and
variance as been adjusted so as to fit best the distribution of the Heng Seng
index returns.
Bottom panel: the Nasdaq composite index bubble from Oct. 5, 1998 to March 27 2000
as well as 10 realizations of the 
``singular inverse random walk'' bubble model 
generated by the nonlinear positive feedback model.
Each realization corresponds to an arbitrary random walk whose drift and
variance as been adjusted so as to fit best the distribution of the Nasdaq
index returns. 
 (reproduced from [Sornette and Andersen, 2002]).
}
\end{center}
\end{figure}

This model [Sornette and Andersen, 2002]
proposes two scenarios for the end of a bubble: either
a spontaneous deflation or a crash. These two mechanisms are natural features
of the model and have not been artificially added. These two scenarios are
indeed observed in real markets, as will be described later.

This model has an interesting and far-reaching consequence in terms of the repetition
and organization
of crashes in time. Indeed, we see that each time the random walk approaches the 
chosen constant $W_c$, the bubble
price blows up and, according to the no-arbitrage condition together with the 
rational expectations, this implies that the market enters ``dangerous waters'' with a 
crash looming ahead. The random walk model provides a very specific prediction
on the waiting times between successive approaches to the critical value $W_c$, i.e.,
between successive bubbles. The distribution of these waiting times is found to be
a very broad power law distribution, so broad that the average 
waiting time is mathematically infinite [Sornette, 2000a]. 
In practice, this leads to two inter-related
phenomena: clustering (bubbles tend to follow bubbles at short times) and long-term memory (there are
very long waiting times between bubbles once a bubble has deflated for a sufficiently long
time). The ``singular inverse random walk'' bubble model thus predicts very large
intermittent fluctuations in the recurrence time of speculative bubbles.

The solution (\ref{jfkaaakaaa}) can be used to invert real data during periods preceding
financial crashes to obtain the relevant parameters. We present here some tests
using an inversion method based on minimizing the Kolmogorov-Smirnov (KS) distance
between the empirical distribution of returns
and the synthetic one generated by the model,
performed on the Hong Kong market prior to the crash which occurred
in early 1994 and on the Nasdaq composite index prior to the crash of April 2000. 
To construct a meaningful distribution, we propose to
add a constant fundamental price $F$ to the bubble price $B(t)$ as only their sum is
observable in real life:
\be
P(t) = e^{rt} \left[ F + B(t) \right]~.   \label{fundsum}
\ee
We can also include the possibility for a interest rate $r$ or growth of the economy with
rate $r$. We denote $M=\mu_0/\alpha$ and
$\sqrt{V}=\sigma_0 / \alpha B_0^{m}$.
For the Hang Seng index, 
the best fit is with $\alpha=2.5, V=1.1 \cdot 10^{-7}, M=4.23 \cdot 10^{-5},
r=0.00032$ and $F=2267.3$.
corresponding to a KS confidence level of $96.3\%$.
This should be compared with the best Gaussian fit to the empirical price returns giving
a KS confidence level of $11\%$. Thus the model ``gaussianizes'' the data at a
very high significance level: a white-Gaussian noise input is transformed by the
nonlinear multiplicative process
into a realistically looking financial time series. For the 
Nasdaq composite index, we obtain $\alpha=2.0, V=2.1 \cdot 10^{-7}, 
M=-9.29 \cdot 10^{-6}, r=0.00496$
and $F=641.5$, corresponding to a KS confidence level of $85.9\%$. The corresponding
best Gaussian fit to the empirical price gives
a KS confidence level of $73\%$. Here, the improvement is less impressive but 
neverthess present. 

With the parameters of the model that have been obtained by the inversion, we 
can use them to generate many scenarios that are statistically equivalent to the 
real history of the Hang Seng and Nasdaq composite index. Figure 
\ref{Nasdaqandothercomp} shows ten synthetic evolutions of 
the process (\ref{fundsum}) generated with 
the best parameter values for both bubbles. By comparison, the empirical prices
are shown as the thick lines (one time step corresponds approximately to one trading day). 
The smooth continuous line close to the 
horizontal axis is the fundamental price $F e^{rt}$. 

This model together with the
inversion procedure provides a new direct tool for detecting
bubbles, for identifying their starting times and the plausible ends.
Changing the initial time of the time series, the KS probability of the resulting
Gaussian fit of the transformed series $W(t)$ should allow us to determine
the starting date beyond which the model becomes inadequate at a
given statistical level. Furthermore, the exponent
$m$ (or equivalently $\alpha$) provides a direct measure of the 
speculative mood. $m=1$ is the normal regime, while $m>1$ quantifies a positive
self-reinforcing feedback. This opens the possibility for continuously monitoring
it via the inversion procedure and using it as a ``thermometer'' of speculation. 
Furthermore,
the variance $V$ of the multiplicative noise is a measure of volatility,
which is significantly more robust than standard estimators. This is due
to the inversion of the nonlinear formula which removes a large part of the
volatility clustering and of the heavy-tail nature of the distribution of returns.
Its continuous monitoring via the inversion procedure suggests
new ways of looking at dependence between assets.
Preliminary analyses show that most of the stylized facts of 
financial time series are reproduced by this approach
[Sornette and Andersen, 2002]. These stylized facts
concern the absence of two-point correlation between returns, the fat-tail 
structure of distributions of returns, the long-range dependence of the 
two-point correlation of volatility and their persistence, the multifractal
structure of generalized moments of the absolute value of the returns, and so on.
We propose to test them thoroughly to quantify the limitation and predictive
power of the model. Application to shorter time scales covering quarters down to 
months will also be explored to test whether this model and some of its variants
may detect regime of abnormal behavior ($m \neq 1$) in financial time series.

We stress that the proposed
class of nonlinear rational bubble model is fundamentally different from 
bubble models that have been tested previously: all previous models 
assumed exponentially growing bubbles and the results of statistical tests have
not been convincing [Camerer, 1989; Adam and Szafarz, 1992].
In contrast, bubbles may be super-exponential which make
them different in principle from a fundamental price growing at a constant rate. 
By this work, we thus hope to rejuvenate the ``old'' theory of rational bubbles
by extending its universe into the nonlinear stochastic regime. 

An additional layer of refinement can easily be added. Indeed,
following [Hamilton, 1989] which introduced so-called Markov switching techniques for
the analysis of price returns, many scholarly works have documented the empirical evidence
of regime shifts in financial data  sets
[Van Norden and Schaller, 1993; Cai, 1994; Gray, 1996; Van Norden, 1996;
Schaller and van Norden, 1997; Assoe, 1998; Chauvet, 1998; Driffill and Sola, 1998].
For instance, Van Norden and Schaller [1997]
 have proposed a Markov regime switching model
of speculative behavior whose key feature is similar to ours, namely over-valuation
over the fundamental price increases the probability and expected size of
a stock market crash.

This evidence taken together with the fact that bubbles are not expected 
to permeate the dynamics of the price all the time suggests the following natural
extension of the model. In the simplest and most parsimonious extension,
we can assume that only two regimes can occur: bubble and normal.
The bubble regime follows the previous model definition
and is punctuated by crashes occuring with the hazard rate governed by the price level.
The normal regime can be for instance a standard random walk market model with
constant small drift and volatility. 
The regime switches are assumed to be completely random. 
This very simple dynamical model
recovers essentially all the stylized facts of empirical prices, i.e.,
no correlation of returns, long-range correlation of volatilities, fat-tail
of return distributions, apparent fractality and multifractality and sharp peak-flat trough
pattern of price peaks. In addition, the model predicts and we confirm by 
empirical data analysis that times of bubbles are associated with non-stationary
increasing volatility correlations. According to this model,
the apparent long-range correlation of volatility 
is proposed to result from random switching between normal and bubble regimes.
In addition, and maybe most important, the visual appearance of price trajectories are
very reminiscent of real ones, as shown in figure \ref{Nasdaqandothercomp}. The remarkably
simple formulation of the price-driven ``singular inverse random walk'' bubble model is
able to reproduce convincingly the salient properties and appearance of real price
trajectories, with their randomness, bubbles and crashes.

\vskip-0.3cm
\subsection{Risk-driven versus price-driven models}

In common, the risk-driven model of section \ref{riskdriven} and the price-driven model
of section \ref{pricedriven}
describe a system of two populations of traders, the ``rational'' and
the ``noisy'' traders. Occasional imitative and herding behaviors
of the ``noisy'' traders may cause
global cooperation among traders causing a crash. The ``rational'' traders
provide a direct link between the crash risks and the bubble price dynamics.

In the risk-driven model,
the crash hazard rate determined from herding drives the bubble price. In the 
price-driven model, imitation and herding induce positive feedbacks on the price,
which itself creates an increasing risk for a looming yet unrealized financial crash.

We believe that both models capture a part of reality. Studying them independently
is the standard strategy of dividing-to-conquer the complexity of the world.
The price-driven model appears
maybe as the most natural and straightforward as it captures the intuition that
sky-rocketing prices are unsustainable and announce endogeneously a significant correction
or a crash. The risk-driven model captures a most subtle self-organization of 
stock markets, related to the ubiquitous balance between risk and returns. 
Both models embody the notion that the market anticipates the crash in a subtle
self-organized and cooperative fashion, hence releasing precursory
``fingerprints'' observable in the stock market prices. In other words, this
implies that market prices contain information on impending crashes. The next section
explores the origin and nature of these precursory patterns and prepares
the road for a full-fledge analysis of real stock market crashes and their precursors.

\vskip-0.3cm
\subsection{Imitation and contrarian behavior: hyperbolic bubbles, crashes and chaos}

The model of bubbles and crashes that we now discuss complements the 
two previous models of rational expectation (RE)
 bubbles in that it describes a deterministic dynamics
of prices embodying both the bubble phases and the crashes [Corcos et al., 2002]. It is maybe
the simplest analytically tractable model of the interplay
between imitative and contrarian behavior
in a stock market where agents can take at least two states, bullish or bearish.
Each bullish (bearish) agent polls $m$ ``friends'' and changes her opinion to bearish
(bullish)
(1) if at least $m \rho_{hb}$ ($m \rho_{bh}$)  among the $m$ agents inspected are bearish
(bullish) or  (2) if at least  $m \rho_{hh}>m \rho_{hb}$ ($m \rho_{bb}>m \rho_{bh}$) 
among the $m$ agents inspected are bullish (bearish). The condition (1) (resp. (2))
corresponds to imitative (antagonistic) behavior. In the limit where the number 
$N$ of agents is infinite, by using combinatorial techniques, it can be shown that 
the dynamics of the fraction of bullish agents 
is deterministic and exhibits chaotic behavior in a
significant domain of the parameter space $\{\rho_{hb}, \rho_{bh}, \rho_{hh}, \rho_{bb}, m\}$.
The deterministic equation of the price trajectory is found to be of the form
\be
p_{t+1} = F_m(p_t)   ~,
\label{detellaa}
\ee
where the function $F_m(x)$ is a sum of combinatorial factors.
A typical chaotic trajectory can be shown 
to be characterized by intermittent phases of chaos, quasi-periodic
behavior and super-exponentially growing bubbles followed by crashes. A typical bubble
starts initially by growing at an exponential rate and then crosses over to a nonlinear
power law growth rate leading to a finite-time singularity. 
The reinjection mechanism provided by the contrarian behavior introduces a nonlinear
reinjection mechanism rounding off these singularity and leads to chaos. 
This model is one of the rare
agent-based models that give rise to interesting non-periodic complex
dynamics in the limit of an infinite number $N$ of agents. 
A finite number of agents introduces
an endogeneous source of noise superimposed on the chaotic dynamics as shown in
figure \ref{fig:14}. 
One can observe burst of volatility, exploding bubbles and quiescent regimes.

The traditional concept of stock market dynamics envisions a stream of
stochastic ``news'' that may move prices in random directions. This model, in
contrast, demonstrates that certain types of deterministic behavior---mimicry
and contradictory behavior alone---can already lead to chaotic prices.
While the traditional theory of rational anticipations exhibits and
emphasizes
self-re-inforcing mechanisms, without either predicting their inception
nor their collapse, the strength of this model is to justify the occurrence of
speculative bubbles. It allows for their collapse by taking into account the
combination of mimetic and antagonistic behavior in the formation of expectations
about prices.
The specific feature of the model is to combine these two Keynesian aspects of
speculation and enterprise and to derive from them behavioral rules based on
collective opinion: the agents can adopt an imitative and gregarious
behavior, or, on the contrary, anticipate a reversal of tendency, thereby detaching
themselves from the current trend. It is this duality, the continuous
coexistence of these two elements, which is at the origin of the properties
of our model: chaotic behavior and the generation of bubbles.
It is the common wisdom that deterministic chaos leads to a
fundamental limit of predictability because the tiny inevitable fluctuations in
those chaotic systems quickly snowball in unpredictable ways.
This has been investigated in
relation with for instance long-term weather patterns. In our model,
the chaotic dynamics of the returns is not the limiting factor
for predictability, as it contains too much residual correlations.
Endogeneous fluctuations due to finite-size effects
and external news (noise) seem to be needed to retrieve
the observed randomness of stock market prices. 

\begin{figure}
\begin{center}
\caption{\protect\label{fig:14} Time evolution of the price $p_t$  over $10000$ time steps
for $m=60$ polled agents with (a) $N=\infty$, (b) $N=m+1=61$ agents and parameters
$\rho_{hb}=\rho_{bh}=0.72$ and  $\rho_{hh}=\rho_{bb}=0.85$. The panel
(c) represents the noise due to the finite size of the system and is obtained
by substracting the time series in panel (a) from the time series in panel (b).
Reproduced from [Corcos et al., 2002]}
\end{center}
\end{figure}

The model of imitative and contrarian behavior
leads to accelerating bubble prices following finite-time singularity
trajectories aborting into a crash.  The accelerating phase is due
to imitation. The crash is due to the contrarian behavior reinforced later
by the imitation behavior. Quantitatively, the bubble-crash sequence
can be described by studying the logarithm of $p-1/2$ (which is the deviation
from equilibrium where 
the equilibrium is characterized by the equality
between the fraction of bullish agents and the fraction of bearish agents)
as a function of 
linear time. One observes first a linear trend which qualifies
an exponential growth $p-1/2 \propto e^{\kappa t}$ (with the 
factor $\kappa >0$), followed by 
a super-exponential growth accelerating so much as to give the impression
of reaching a singularity in finite-time.

The understanding of this phenomenon comes from the behavior of the ``elasticity''
of $F_m(p)-p$ with respect to $p-1/2$, i.e., the derivative of the logarithm of 
$F_m(p)-p$, where $F_m(p)$ is defined by (\ref{detellaa}), with respect to 
the logarithm of $p-1/2$. 
Two regimes can be observed.
\begin{enumerate}
\item For small $p-1/2$, the elasticity is $1$, i.e., 
\be
F_m(p)-p \simeq \alpha(m) \left(p-\frac{1}{2}\right)~.
\label{linearexp}
\ee
This expression (\ref{linearexp}) explains the exponential growth
observed at early time.

\item For larger $p-1/2$, the elasticity increases above $1$ and stabilizes to a value ${\mu(m)}$
before decreasing again due to the reinjection produced by the contrarian
mechanism. The interval in $p-1/2$ in which the slope is approximately
stabilized at the value ${\mu(m)}$ enables us to write
\be
F_m(p)-p \simeq \beta(m) \left(p-\frac{1}{2}\right)^{\mu(m)}~~~~~~
\mbox{with}~ \mu>1~.
\ee
\end{enumerate}

These two regimes can be collected in the following phenomenological
expression for $F_m(p)$:
\ba
F_m(p)&=&\frac{1}{2} + \left(1 - 2 g_m(1/2)  - g_m'(1/2) \right)
\left(p-\frac{1}{2} \right) + \beta(m) \left(p-\frac{1}{2} \right)^{\mu(m)}~, \\
&=& \frac{1}{2} + \left(p-\frac{1}{2} \right) +\alpha(m) \left(p-\frac{1}{2}
\right) + \beta(m) \left(p-\frac{1}{2} \right)^{\mu(m)}~~~~\mbox{with}~ \mu>1~,
\ea
and
\be
\alpha(m)=  -2 g_m(1/2)  - g_m'(1/2)~.
\ee

Introducing the notation $\epsilon=p-1/2$, the dynamics can be rewritten
\be
\epsilon'-\epsilon = \alpha(m) \epsilon + \beta(m) \epsilon^{\mu(m)},
\ee
which, in the continous time limit, yields
\be
\frac{d \epsilon}{dt}= \alpha(m) \epsilon + \beta(m) \epsilon^{\mu(m)}~.
\ee
Thus, for small $\epsilon$, we obtain an exponential growth rate
\be
\epsilon_t \sim e^{\alpha(m)t}~,
\ee
while for large enough $\epsilon$ 
\be
\epsilon_t \sim (t_c-t)^{-\frac{1}{\mu(m)-1}}~.
\label{finitetimesingm}
\ee

For example, for $m=60$ with $\rho_{hb}=\rho_{bh}=0.72$ and
$\rho_{hh}=\rho_{bb}=0.85$, 
$\mu(m)=3$, which yields for large $\epsilon$
\be
p_t-\frac{1}{2} \sim \frac{1}{\sqrt{t_c-t}}~.
\label{finitetimesingm60}
\ee

The prediction (\ref{finitetimesingm})
implies that the returns $r_t$ should increase in an accelerating 
super-exponential fashion
at the end of a bubble, leading to a price trajectory
\be
\pi_t = \pi_c - C (t_c-t)^{\frac{\mu(m)-2}{\mu(m)-1}}~,
\label{pricetrajbubble}
\ee
where $\pi_c$ is the culminating price of the bubble reached at $t=t_c$
when $\mu(m) >2$, such the finite-time singularity in $r_t$ gives rise
only to an infinite slope of the price trajectory. This behavior
(\ref{pricetrajbubble}) with an exponent $0 < \frac{\mu(m)-2}{\mu(m)-1} < 1$
has been documented in many bubbles [Sornette et al., 1996; Johansen et al., 1999;
2000; Johansen and Sornette, 1999a,b; 2000a; Sornette and Johansen, 2001b,c; Sornette
and Andersen, 2002; Sornette, 2002; 2003].
The case $m=60$ with $\rho_{hb}=\rho_{bh}=0.72$ and
$\rho_{hh}=\rho_{bb}=0.85$ leads to
$\frac{\mu(m)-2}{\mu(m)-1}=1/2$, which is reasonable agreement with 
the values reported previously. 

Interpreted within the present model,
the exponent $\frac{\mu(m)-2}{\mu(m)-1}$ of the price singularity gives
an estimation of the ``connectivity'' number $m$ through the dependence
of $\mu$ on $m$. Such a relationship
has already been argued by Johansen et al. (2000) at a phenomenological level using 
a mean-field equation in which the exponent is directly related to the number
of connections to a given agent.

This model developed recently has strong potential to provide a simple
but powerful approach to modeling financial time series. It can be extended in many
ways, which include (1) introducing at least a third state, called ``neutral'', in addition 
to the ``bullish'' and ``bearish'' states, (2) introducing 
a fundamental price, a population of value investors
and assume that ``noise traders'' follow the imitative-contrarian strategy
previously described, (3) considering the possibility for several stocks to be traded
simultaneously, with in particular the introduction of a riskless asset.

\section{Log-periodic oscillations decorating power laws}

\vskip-0.3cm
\subsection{Status of log-periodicity}

Log-periodicity is an observable signature of the symmetry of
{\em discrete} scale invariance (DSI). DSI is a weaker symmetry than
(continuous) scale invariance [Dubrulle et al., 1997]. 
The latter is the symmetry of a system which 
manifests itself such that an observable ${\cal O}\lp x\rp $ as a function 
of the  ``control'' parameter $x$ is scale invariant under the change 
$x \to \lambda x$ for arbitrary $\lambda$, {\it i.e.}, a number  $\mu\lp 
\lambda\rp$ exists such that
\be
{\cal O} (x) = \mu\lp \lambda\rp {\cal O} (\lambda x) ~~~~.
\label{one}
\ee
The solution of (\ref{one}) is simply a power law ${\cal O}(x) = x^{\alpha}$,
with $\alpha = - {\log \mu \over \log \lambda}$, which can be verified 
directly by  insertion. In DSI, the system or the observable obeys scale 
invariance  (\ref{one}) only for {\em specific} choices of the magnification 
factor $\lambda$, which form in general an infinite but countable set of 
values $\lambda_1, \lambda_2, ...$ that can be written as $\lambda_n = 
\lambda^n$. $\lambda$ is the fundamental scaling ratio determining the 
period of the resulting log-periodicity. This property can
be qualitatively seen to encode a {\it lacunarity} of the fractal structure.
The most general solution of (\ref{one}) with $\lambda$ (and 
therefore $\mu$) is
\be \label{eqdsi}
{\cal O}(x) = x^{\alpha}~P\left({\ln x \over \ln \lambda}\right)
\ee
where $P(y)$ is an arbitrary periodic function of period $1$ in the
argument, hence the name log-periodicity.
Expanding it in Fourier
series $\sum_{n=-\infty}^\infty  c_n \exp\left(2n\pi i{\ln x\over\ln 
\lambda}\right)$,
we see that ${\cal O}(x)$ becomes a sum of power laws with the 
infinitely discrete
spectrum of complex exponents $\alpha_n = \alpha + i 2 \pi n /\ln 
\lambda$, where $n$ is an arbitrary integer. Thus, DSI leads to power laws with 
complex exponents, whose observable signature is log-periodicity. Specifically,
for financial bubbles prior to large crashes, we shall see that a first 
order representation of eq.~(\ref{eqdsi})
\be \label{eqlppow}
I\lp t\rp = A + B\lp t_c -t\rp^\beta + C\lp t_c -t\rp^\beta 
\cos \lp \omega \ln \lp t_c -t\rp
-\phi \rp
\ee
captures well the behaviour of the market price $I\lp t\rp$ prior to a crash or
large correction at a time $\approx t_c$. 

There are many mechanisms known to generate log-periodicity [Sornette, 1998].
The most obvious one is when the system possesses
a pre-existing discrete hierarchical structure. There are however
various dynamical mechanisms generating log-periodicity,
without relying on a pre-existing discrete hierarchical structure.
DSI may be produced dynamically and does not need to be pre-determined 
by e.g., a geometrical network.
This is because there are many ways to break a symmetry, the subtlety here
being to break it only partially. 

\vskip-0.3cm
\subsection{Stock market price dynamics from the interplay between
fundamental value investors and technical analysists \label{sec:market}}

The importance of the interplay of two classes of investors,
fundamental value investors and technical analysts (or trend followers),
has been stressed by several recent works (see for
instance [Lux and Marchesi, 1999] and references therein) to
be essential
in order to retrieve the important stylized facts of stock market 
price statistics.
We build on this insight and construct a simple model of price dynamics,
whose innovation is to put emphasis on the
fundamental {\it nonlinear} behavior of both classes of agents.

\vskip-0.3cm
\subsubsection{Nonlinear value and trend-following strategies}
\label{sec:market-trend}
The price variation of an asset on the stock market is controlled by
supply and demand, in other words by the net order size $\Omega$
through a market impact function [Farmer, 1998].
Assuming that the ratio ${\tilde p}/p$ of the price ${\tilde p}$ at 
which the orders are
executed over the previous quoted price $p$ is solely a function of $\Omega$
and using the condition
that it is impossible to make profits by repeatedly trading through a close
circuit (i.e., buying and selling has to end up with a final net position equal to zero),
Farmer [1998] has shown that the logarithm of the price is given by the 
following equation written in discrete form
\be
\ln p(t+1) - \ln p(t)  = {\Omega(t) \over L} ~.
\label{mqmqmmq}
\ee
The ``market depth'' $L$ is the typical number of 
outstanding stocks traded per unit time
and thus normalizes the impact of a given order size $\Omega(t)$ on the log-price
variations.
The net order size $\Omega$ summed over all traders is changing as a
function of time so as
to reflect the information flow in the market and the evolution of the
traders' opinions and moods. A zero net order size $\Omega=0$ corresponds
to exact balance between supply and demand.
Various derivations
have established a connection between
the price variation or the variation of the logarithm of the price
to factors that control the net order size itself [Farmer, 1998; Bouchaud and Cont, 1998;
Pandey and Stauffer, 2000].

Two basic ingredients of $\Omega(t)$ are thought to be important in determining the price
dynamics: reversal to the fundamental value 
($\Omega_{\rm fund}(t)$) and trend following ($\Omega_{\rm trend}(t)$). Other factors,
such as risk aversion, may also play an important role.

Ide and Sornette [2002] propose to describe the reversal to estimated fundamental value
by the contribution
\be
\Omega_{\rm fund}(t) = -c ~[\ln p(t) - \ln p_f]
~ |\ln p(t) - \ln p_f|^{n-1}~,   \label{fmaaak}
\ee
to the order size,
where $p_f$ is the estimated fundamental value and $n>0$ is an exponent
quantifying the nonlinear nature of reversion to $p_f$.
The strength of the reversion is measured by the coefficient $c>0$, which
reflects that the net order is negative (resp. positive) if the price is
above (resp. below) $p_f$. The nonlinear power law
$[\ln p(t) - \ln p_f]~ |\ln p(t) - \ln p_f|^{n-1}$ of order $n$ is chosen as the
simplest function capturing the following effect. In principle, the 
fundamental value
$p_f$ is determined by the discounted expected future dividends and is thus
dependent upon the forecast of their growth rate and of the risk-less 
interest rate,
both variables being very difficult to predict. The fundamental value is thus
extremely difficult to quantify with high precision and is often estimated
within relatively large bounds:
all of the methods of determining intrinsic value rely on
assumptions that can turn out to be far off the mark.  For instance, 
several academic
studies have disputed the premise that a portfolio of sound, cheaply
bought stocks will, over time, outperform a portfolio selected by any
other method (see for instance [Lamont, 1988]). As a consequence, a trader
trying to track fundamental value has no incentive to react when she feels
that the deviation is small since this deviation is more or less within
the noise. Only when the departure of price from fundamental value
becomes relatively large will the trader act. The relationship (\ref{fmaaak})
with an exponent $n>1$ precisely accounts for this effect: when $n$ 
is significantly
larger than $1$, $|x|^n$ remains small for $|x|<1$ and shoots up rapidly only when
it becomes larger than $1$, mimicking a smoothed threshold behavior.
The nonlinear dependence of $\Omega_{\rm fund}(t)$ on $\ln [p(t)/p_f]=
\ln p(t) - \ln p_f$ shown in 
(\ref{fmaaak}) is the first novel element of our model.
Usually, modelers reduce this term to the linear case $n=1$ while, as 
we shall show, generalizing to larger values $n>1$ will be a crucial feature of the 
price dynamics. In economic language, the exponent 
$n = d \ln \Omega_{\rm fund}/d \ln \left(\ln [p(t)/p_f]\right)$
is called the ``elasticity'' or ``sensitivity''
of the order size $\Omega_{\rm fund}$ with respect
to the (normalized) log-price $\ln [p(t)/p_f]$.

A related ``sensitivity'', that of the money demand to interest rate, has 
has been recently documented to be larger than $1$, similarly to 
the Ide-Sornette [2002] proposal
of taking $n>1$ in (\ref{fmaaak}).
Using a survey of roughly 2,700 households, 
Mulligan and Sala-i-Martin [2000] estimated the interest
elasticity of money demand (the sensitivity or log-derivative of money demand
to interest rate) to be very small at low interest rates. This is due to the fact
that few people decide to invest in interest-producing assets when rates are low,
due to ``shopping'' costs. In contrast, for large interest rates or for
those who own a significant bank account, the interest elasticity of money demand is 
significant. This is a clear-cut example of a threshold-like behavior characterized
by a strong nonlinear response. This can be captured by
$e \equiv d \ln M/ d\ln r = (r/r_{\rm infl})^n$ with $n>1$ such that the elasticity $e$
of money demand $M$ is negligible when the interest $r$ is 
not significantly larger than the inflation
rate $r_{\rm infl}$ and becomes large otherwise.

Trend following (in various elaborated forms)
was (and probably is still) one of the major strategy used by
 technical analysts (see [Andersen et al., 2000] for a review and 
references therein).
More generally, it results naturally when investment strategies are 
positively related to
past price moves.
Trend following can be captured by the following expression of the order size
\be
\Omega_{\rm trend}(t) = a_1[\ln p(t) - \ln p(t-1)] +a_2 [\ln p(t) - \ln p(t-1)]
|\ln p(t) - \ln p(t-1)|^{m-1}~.   \label{jfala}
\ee
This expression corresponds to
driving the price up if the preceding move was up ($a_1>0$ and $a_2>0$).
The linear case $(a_1>0, a_2=0)$ is usually chosen by modelers. Here, we
generalize this model by adding the contribution proportional to $a_2>0$ 
from considerations
similar to those leading to the nonlinear expression
(\ref{fmaaak}) for the reversal term with an exponent $n>1$. We argue
that the dependence of the order size at time $t$ resulting from 
trend-following
strategies is a nonlinear function with exponent $m>1$
of the price change at previous time steps.
Indeed, a small price change from time $t-1$ to time $t$ may not be perceived
as a significant and strong market signal. Since many of the 
investment strategies
are nonlinear, it is natural to consider an average trend-following order size
which increases in an accelerated manner as the price change 
increases in amplitude.
Usually, trend-followers increase
the size of their order faster than just proportionally to the last 
trend. This is
reminiscent of the argument [Andersen et al., 2000] that traders's psychology
is sensitive to a change of trend (acceleration or deceleration) and 
not simply to the
trend (velocity). The fact that trend-following
strategies have an impact on price proportional to the price change 
over the previous
period raised to the power $m>1$ means that trend-following strategies are
not linear when averaged over all of them: they tend to under-react 
for small price
changes and over-react for large ones.
The second term of the right-hand-side of
(\ref{jfala}) with coefficient $a_2$ captures this phenomenology.

\vskip-0.3cm
\subsubsection{Nonlinear dynamical equation for stock market prices}
\label{sec:market-eq}

Introducing the notation
\be
x(t) = \ln [p(t)/p_f]~,  \label{nkkla}
\ee
and the time scale $\delta t$ corresponding to one time step,
and putting all the contributions (\ref{fmaaak}) and (\ref{jfala}) into
(\ref{mqmqmmq}), with $\Omega(t)=\Omega_{\rm fund}(t)+\Omega_{\rm trend}(t)$,
we get
\be
x(t+\delta t) - x(t) = {1 \over L} \left( a_1 ~[x(t) - x(t-\delta t)] + a_2 [x(t) - 
x(t-\delta t)]|x(t) - x(t-\delta t)|^{m-1}
- c ~x(t)  |x(t)|^{n-1} \right)~.   \label{ajkafa}
\ee
Expanding (\ref{ajkafa}) as a Taylor series in powers of $\delta t$, we get
\be
(\delta t)^2 {d^2 x \over dt^2} = - \left[1-{a_1 \over L}\right]~\delta t~ {dx \over dt} +
{a_2 (\delta t)^{m}  \over L} {dx \over dt}
|{dx \over dt}|^{m-1} - {c \over L} x(t)  |x(t)|^{n-1}~+~{\cal O}[(\delta t)^3]~, 
\label{kaklaklq}
\ee
where ${\cal O}[(\delta t)^3]$ represents a term of the order of $(\delta t)^3$.
Note the existence of the second order derivative,
which results from the fact that the price variation from present to
tomorrow is based on analysis of price change between yesterday and
present. Hence the existence of the three time lags leading to
inertia.
A special case of expression (\ref{ajkafa})
with a {\it linear} trend-following term $(a_2=0)$ and a {\it linear} reversal term
$(n=1)$ has been studied in [Bouchaud and Cont, 1998; Farmer, 1998], with the addition
of a risk-aversion term and a
noise term to account for all the other
effects not accounted for by the two terms (\ref{fmaaak}) and (\ref{jfala}). We shall
neglect risk-aversion as well as any other term and focus only on the
reversal and trend-following terms previously discussed to
explore the resulting price behaviors. Grassia [2000] has also studied a similar {\it linear}
second-order differential equation derived from market delay, positive feedback 
and including a mechanism for quenching runaway markets.

Expression (\ref{ajkafa}) is inspired by
the continuous mean-field limit of the model of
Pandey and Stauffer [2000], defined by starting from 
the percolation model of
market price dynamics [Cont and Bouchaud, 2000; Chowdhury and Stauffer, 1999; 
Stauffer and Sornette, 1999] and
developed to account for the dynamics of the Nikkei and Russian market recessions
[Johansen and Sornette, 1999c; 2001b]. The generalization assumes
that trend-following and reversal to fundamental values are two forces that
influence the probability that a trader buys or sells the market. In addition,
Pandey and Stauffer [2000] consider as we do here that the
dependence of the probability to enter the market is a nonlinear function
with exponent $n>1$ of
the deviation between market price and fundamental price. However, they do not
consider the possibility that $m>1$ and stick to the linear
trend-following case.
We shall see that the analytical control offered by our
continuous formulation allows us
to get a clear understanding of the different dynamical phases.

Among the four terms of equation (\ref{kaklaklq}), the first term of
its right-hand-side
is the least interesting. For $a_1<L$, it corresponds to a
damping term which becomes
negligible compared to the second term
in the terminal phase of the growth close to the singularity
when $|dx/dt|$ becomes very large. For $a_1>L$, it corresponds
to a negative viscosity
but the instability it provides is again subdominant for $m>1$. The main
ingredients here are the interplay between the inertia provided by the second
derivative in the left-hand-side, the destabilizing nonlinear
trend-following term
with coefficient $a_2>0$ and the nonlinear reversal term. In order to
simplify the
notation and to simplify the analysis of the different regimes, we
shall neglect
the first term of the right-hand-side of (\ref{kaklaklq}), which amounts
to take the special value $a_1=L$. In a field theoretical
sense, our theory is tuned right at the ``critical point'' with a 
vanishing ``mass'' term.

Equation (\ref{kaklaklq}) can be viewed in two ways. It can be 
seen as a convenient short-hand notation for the intrinsically
discrete equation (\ref{ajkafa}), keeping the time step $\delta t$ small
but finite. In this interpretation, we pose
\ba
\alpha &=& a_2 (\delta t)^{m-2} /L ~,  \label{alphaeq1}    \\
\gamma &=& c /L(\delta t)^2~,   \label{gammaeq2} 
\ea
which depend explicitely on $\delta t$, 
to get
\be
{d^2 x \over dt^2} = \alpha {dx \over dt}
|{dx \over dt}|^{m-1} - \gamma x(t)  |x(t)|^{n-1}~.   \label{kakalaklq}
\ee
A second interpretation is to genuinely take the continuous limit 
$\delta t \to 0$ with the constraints $a_2/L \sim (\delta t)^{2-m}$
and $c/L \sim (\delta t)^2$. This allow us to define the now
$\delta t$-independent coefficients $\alpha$ and $\gamma$ according to 
(\ref{alphaeq1}) and (\ref{kakalaklq}) and obtain the 
truly continuous equation (\ref{kakalaklq}). This equation 
can also be written as
\ba
{d y_1 \over dt} &=& y_2~,  \label{dyn1bis}\\
{d y_2 \over dt} &=& \alpha y_2 |y_2|^{m-1} - \gamma y_1 |y_1|^{n-1}~.
\label{dyn2bis}
\ea
This system leads to a finite-time singularity with accelerating
oscillations for $m>1$ and $n > 1$. The richness of behaviors results
from the competition between these two terms.

\vskip-0.3cm
\subsubsection{Dynamical properties}

The origin $(y_1=0, y_2=0)$ plays
a special role as the unstable (for $m>1$) fixed point around which spiral structures
of trajectories are organized in phase space $(y_1, y_2)$. It is 
particularly interesting that this point plays a special role since
$y_1=0$ means that the observed
price is equal to the fundamental price. If, in addition,
$y_2=0$, there is no trend, i.e., the
market ``does not know'' which direction to take. The fact that this is
the point of instability  around which the price trajectories organize 
themselves provides a fundamental understanding of the cause of
the complexity of market price time series based on the
instability of the fundamental price ``equilibrium''.

\begin{figure}
\parbox[l]{7.5cm}{
\epsfig{file=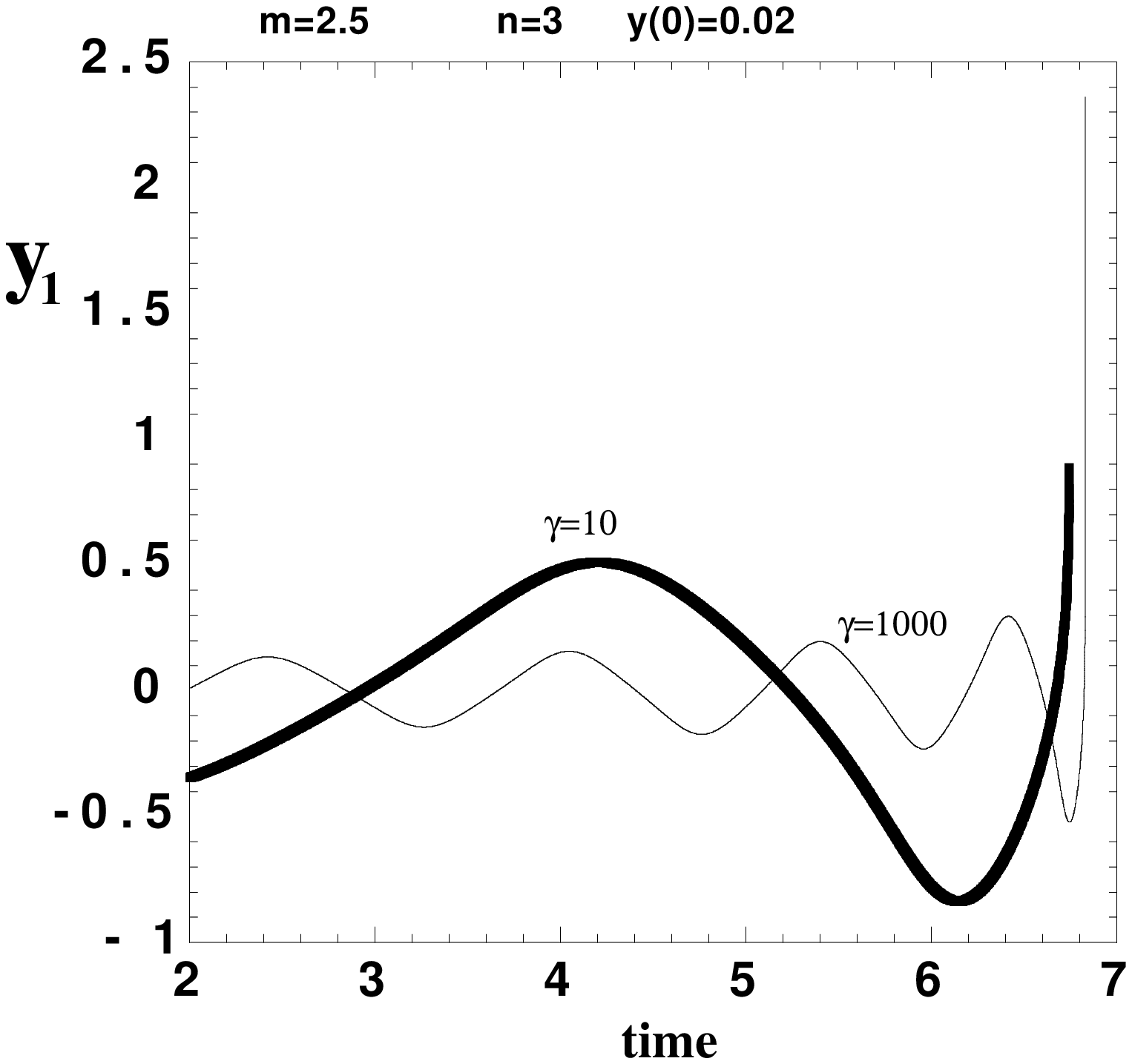,height=5cm,}
\caption{\protect\label{Figm1.5n3}  ``Reduced
price'' as a function of time for a trend-following exponent $m=2.5$ with $n=3$,
$\alpha=1$ and with two
amplitudes $\gamma=10$ and 
$\gamma=1000$ of the  fundamental reversal term. Reproduced from [Ide and Sornette, 2002].}}
\hspace{5mm}
\parbox[r]{7.5cm}{
\epsfig{file=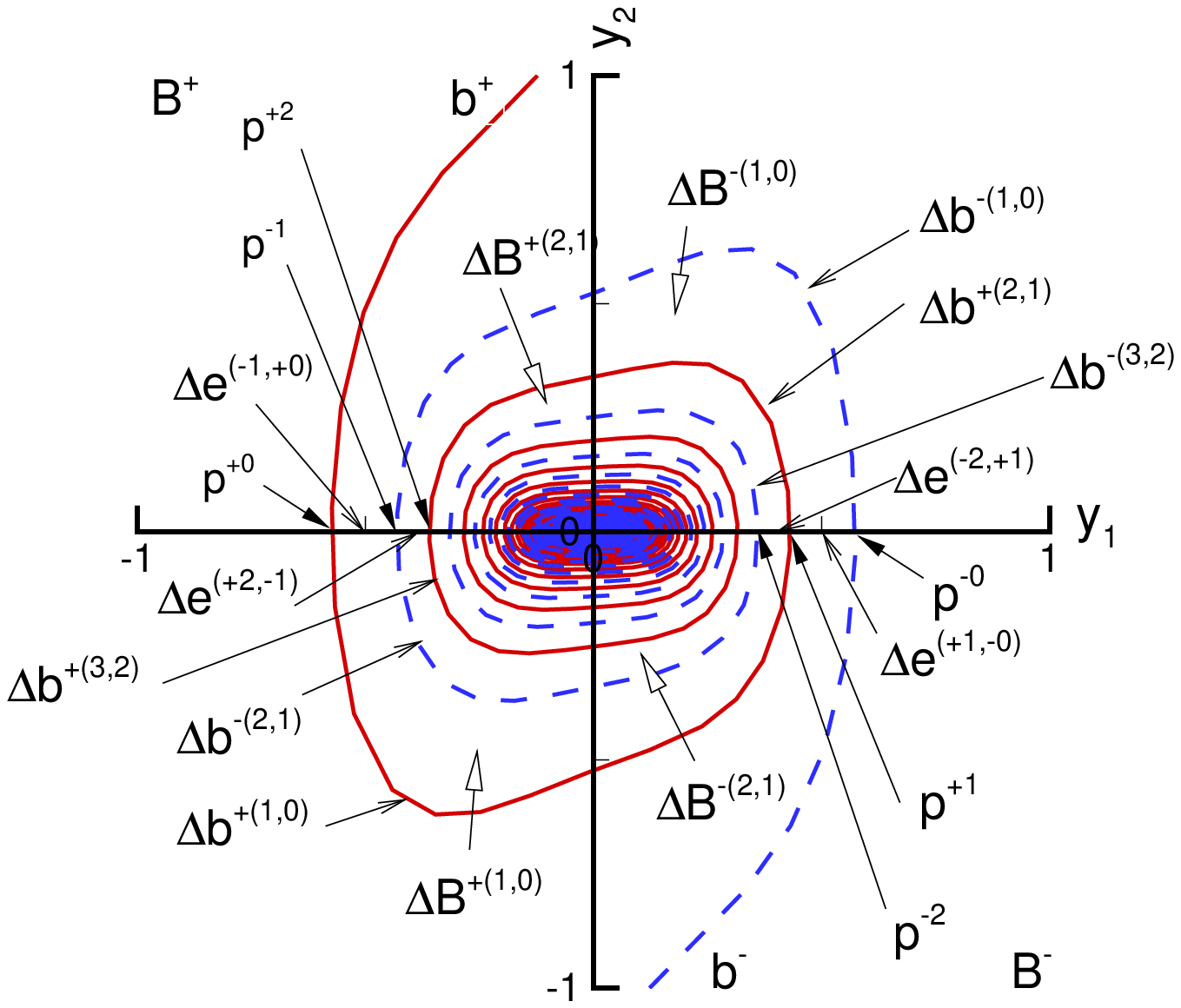,height=5cm,width=7cm}
\caption{\protect\label{fg:basin_def_n3m15g10}  Geometrical spiral showing 
two special trajectories (the continuous and dashed lines)
in the ``reduced price''--``velocity''
plane $(y_1, y_2)$ that connect exactly the origin 
$y_1=0, y_2=0$ to infinity. This spiraling structure, which exhibits
scaling or fractal properties, is at the origin
of the accelerating oscillations decorating the power law behavior close
to the finite-time singularity. Reproduced from [Ide and Sornette, 2002].}}
\end{figure}

Figure \ref{Figm1.5n3} shows the reduced price for the
trend-following exponent $m=2.5$. In this case, the reduced price goes to 
a constant at $t_c$ with an infinite slope (the singularity is thus
on its derivative, or ``velocity''). We can also observe accelerating 
oscillations, reminiscent of log-periodicity. The novel feature
is that the oscillations are only transient, leaving place to a pure final
accelerating trend in the final approach to the critical time $t_c$.

Figure \ref{fg:basin_def_n3m15g10} shows that the oscillations
with varying frequency and amplitude seen in figure
\ref{Figm1.5n3} are nothing but the projection on one axis of a spiraling 
structure in the plane. Actually, figure \ref{fg:basin_def_n3m15g10} shows more
than that: in the plane of the reduced price $y_1$ and its ``velocity'' $y_2$,
it shows two special trajectories that connect exactly the origin 
$y_1=0, y_2=0$ to infinity. From general mathematical theorems of dynamical systems,
one can then show that any trajectory starting close to the origin will never
be able to cross any of these two orbits. As a consequence, any real trajectory
will be guided within the spiraling channel, winding around the
central point $0$ many times
before exiting towards the finite-time singularies. The approximately log-periodic
oscillations result from the oscillatory structure of the fundamental reversal term
associated with the acceleration driven by the trend-following term. The conjunction
of the two leads to the beautiful spiral, governing 
a hierarchical organization of the 
spiralling trajectories around the origin in the price-velocity space.
See [Ide and Sornette, 2002; Sornette and Ide, 2001] for a detailed mathematical
study of this system. 

In sum, the simple two-dimensional dynamical system 
(\ref{dyn1bis},\ref{dyn2bis}) embodies two 
nonlinear terms, exerting respectively positive 
feedback and reversal, which compete to create a singularity in finite time decorated by
accelerating oscillations.
The power law singularity results from the increasing growth rate.
The oscillations result from
the restoring mechanism. As a function of the order of 
the nonlinearity of the
growth rate and of the restoring term, a rich variety of behavior is
observed.
The dynamical behavior is traced back fundamentally to
the self-similar spiral structure of trajectories in phase space
unfolding around an unstable spiral point at the origin. The interplay
between the restoring mechanism and the nonlinear growth rate leads
to approximately log-periodic oscillations with
remarkable scaling properties.

\section{Autopsy of major crashes: universal exponents and log-periodicity}

\vskip-0.3cm
\subsection{The crash of October 1987}

As discussed in section 2, the crash of Oct. 1987 and its black monday on Oct.~19 
remains one of the most striking drops ever seen on stock markets,
both by its overwhelming amplitude and its encompassing sweep over most markets
worldwide. It was preceded by a remarkably strong ``bull'' regime epitomized
by the following quote from Wall Street Journal,
on Aug. 26, 1987, the day after the 1987 market peak:
``In a market like this, every story is a positive one. Any
news is good news. It's pretty much taken for granted
now that the market is going to go up.'' 
Investors were thus largely unaware
of the forthcoming risk happenings [Grant, 1990].

\vskip-0.3cm
\subsubsection{Precursory pattern}

\begin{technical}
Time is often converted into decimal year units\,: for non-leap years,
$365$ days $=1.00$ year which leads to $1$ day = $0.00274$ years. 
Thus $0.01$ year $= 3.65$ days and $0.1$ year $= 36.5$ days
or $5$ weeks. For example, Oct. 19, 1987 corresponds to $87.800$.
\end{technical}

\begin{figure}
\begin{center}
\epsfig{file=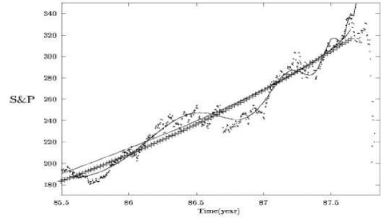,width=8cm,height=6cm}
\caption{\protect\label{fig1first} Evolution as a function of time 
of the New York stock exchange
index S\&P500 from July $1985$ to the end of Oct. $1987$ ($557$ trading days).
The $+$ represent a constant return increase of $\approx 30 \%$/year and
gives $var\lp F_{exp} \rp \approx 113$ (see text for definition).
The best fit to the power-law (\ref{faka}) gives $A_1 \approx 327$, $B_1  \approx -79$,
$t_c  \approx 87.65$, $m_1  \approx 0.7$ and $\mbox{var}_{pow} \approx 107$.
The best fit to expression (\ref{kflala}) gives $A_2 \approx 412$, $B_2 \approx -165$,
$t_c \approx 87.74$, $C \approx 12$, $\omega \approx 7.4$, $T =2.0$,
$m_2 \approx 0.33$ and $\mbox{var}_{lp} \approx 36$. One can observe
four well-defined oscillations fitted by the expression (\ref{kflala}), 
before finite size effects
limit the theoretical divergence of the acceleration, at which point the
bubble ends in the crash. All the fits are carried over the whole time interval
shown, up to $87.6$. The fit with eq.(\ref{kflala}) turns out to be very robust with
respect to this upper bound which can be varied significantly.
Reproduced from [Sornette et al., 1996].
}
\end{center}
\end{figure}

Fig.~\ref{fig1first} shows the evolution of the New York stock exchange
index S\&P500 from July $1985$ to the end of Oct. $1987$ after the crash.
The plusses ($+$) represent the best fit to an exponential growth obtained
by assuming that the market is given an average return of about $30 \%$ per year.
This first representation does not describe the apparent overall
acceleration before the crash, occurring already more than a year in advance.
This acceleration ({\it cusp}-like shape) is better represented by using
power law functions that sections 5 and 6 showed to be signatures of a critical
behavior of the market. The monotonic line corresponds to the following
power law parameterization:
\be
\label{lp1} F_{pow}\lp t \rp  = A_1+B_1\lp t_c-t \rp ^{m_1} ~,  \label{faka}
\ee
where $t_c$ denotes the time at which the powerlaw fit of the S\&P500
presents a (theoretically) diverging slope, announcing an imminent crash. 
In order to qualify and compare the fits, the variances, denoted 
$\mbox{var}$ equal to the mean of the squares of the errors
between theory and data, or its square-root called the root-mean-square
(r.m.s.) are calculated. The ratio
of two variances corresponding to two different hypotheses is taken as a
qualifying statistic. The ratio of the variance
of the constant rate hypothesis to that of the
power-law is equal to $\mbox{var}_{exp}/\mbox{var}_{pow} \approx 1.1$
indicating only a slightly better performance of the power law in capturing
the acceleration, the number of free variables being the same and 
equal to $2$.

However, already to the naked eye, the most striking feature in
this acceleration is the presence of systematic oscillatory-like deviations. 
Inspired by the insight given in section 5 and especially section 6, the oscillatory
continuous line is obtained by fitting the data by the
following mathematical expression
\be
\label{lp2} F_{lp} \lp t \rp  = A_2+B_2\lp t_c-t \rp ^{m_2} \left[
1+C\cos \lp \omega \log \lp (t_c-t)/T \rp \rp \right]~. \label{kflala}
\ee 
This equation is the simplest example of a log-periodic correction to a pure power law
for an observable exhibiting a singularity at the time $t_c$ at which the crash
has the highest probability to occur.
The log-periodicity here stems from the cosine function of the logarithm
of the distance $t_c-t$ to the critical time $t_c$.
Due to log-periodicity, the evolution of the financial index
becomes (discretely) scale-invariant close to the critical point.

\begin{figure}
\begin{center}
\epsfig{file=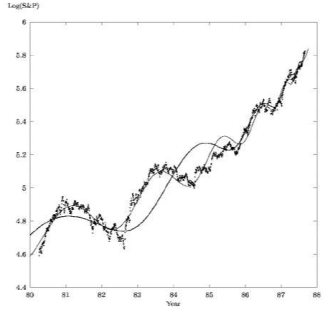,width=8cm,height=6cm}
\caption{\protect\label{figlong87} Time dependence of the logarithm of the New
York stock exchange index S\&P500 from january 1980 to september 1987
and best fit by the improved nonlinear log-periodic formula 
developed in [Sornette and Johansen, 1997] (thin line). The exponent and log-periodic
angular frequency are $m_2=0.33$ and $\omega^{1987}=7.4$.
The crash of October 19, 1987
corresponds  to $1987.78$ decimal years. The thick line is the fit by
(\ref{kflala}) on the subinterval from July $1985$ to the end of $1987$ and
is represented on the full time interval starting in $1980$. The comparison
with the thin line allows one to visualize the frequency shift described by
the nonlinear theory. Reproduced from [Sornette and Johansen, 1997].
}
\end{center}
\end{figure}

The log-periodic correction to scaling implies the existence of a hierarchy of
characteristic time intervals $t_c - t_n$, 
given by the expression 
\be
T_n = T_c - (T_c-T_0) \lambda^{-n}~, 
\label{rerfdffghw}
\ee
with a prefered scaling ratio
denoted $\lambda$.
For the October 1987 crash, we find
$\lambda \simeq 1.5-1.7$ (this value is remarkably universal and is
found approximately the same for other crashes as we shall see). 
We expect a cut-off at short time scales (i.e. above $n
\sim $ a few units) and also
at large time scales due to the existence of finite size effects. These
time scales $t_c - t_n$ are not universal but depend upon the specific market.
What is expected to be universal are the ratios $\frac{t_c - t_{n+1}}{t_c - t_n}
= \lambda$. For details on the fitting procedure, we refer to [Sornette et al., 1996].

It is possible to generalize the simple log-periodic power law formula used in
figure \ref{fig1first} by using a mathematical tool, called bifurcation
theory, to obtain its generic nonlinear correction, that allows one to 
account quantitatively for the behavior
of the Dow Jones and S\&P500 indices up to $8$ years prior
to the Oct. 1987. 
The result of this theory presented in [Sornette and Johansen, 1997]
is used to generate the fit shown in figure \ref{figlong87}. One sees
clearly that the new formula accounts remarkably well for almost eight years of market
price behavior compared to only a little more than two years 
for the simple log-periodic formula shown in figure \ref{fig1first}.
The nonlinear theory developed in [Sornette and Johansen, 1997] 
leads to ``log-frequency modulation'',
an effect first noticed empirically in [Feigenbaum and Freund, 1996].
The remarkable quality of the fits shown in figures \ref{fig1first} and \ref{figlong87}
have been assessed in [Johansen and Sornette, 1999b].

In a recent reanalysis, J.A. Feigenbaum [2001] examined the data
in a new way by taking the 
first differences for the logarithm of the S\&P 500 from
1980 to 1987. The rational for taking the price variation rather than the price
itself is that the fluctuations, noises or deviations are expected to be more
random and thus  more innocuous than for the price which is a cumulative quantity.
By rigorous hypothesis testing, Feigenbaum
found that the log-periodic component cannot be rejected at the $95\%$-confidence level:
in plain words, this means that the probability that the log-periodic
component results from chance is about or less than $0.05$. 

\vskip-0.3cm
\subsubsection{Aftershock patterns}

\begin{figure}
\parbox[l]{7.5cm}{
\epsfig{file=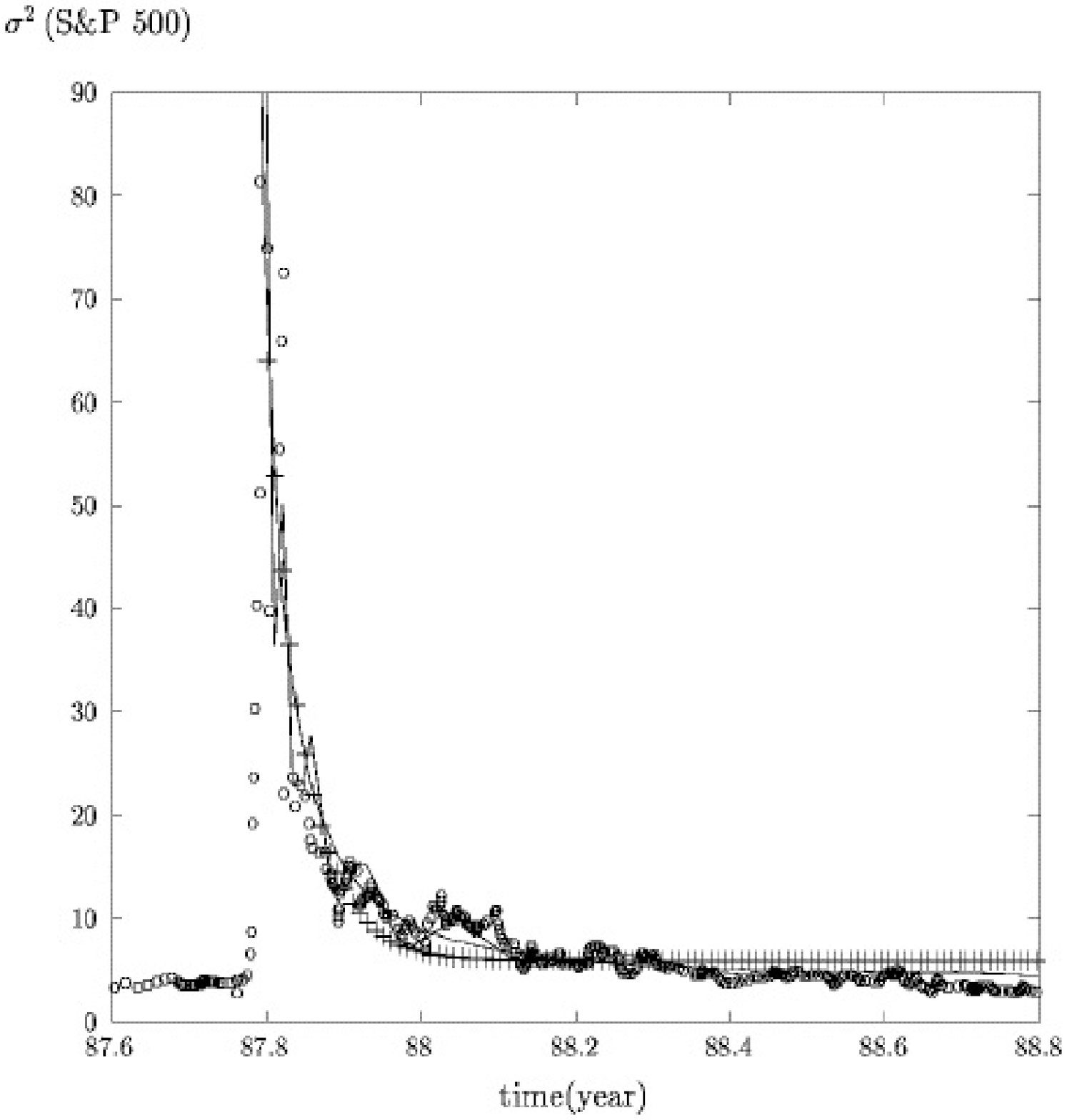,height=5cm,width=7cm}
\caption{\protect\label{fig2first} Time evolution of the implied volatility of the
S\&P500 index (in logarithmic scale) after the Oct.1987 crash, 
taken from [Chen et al., 1995]. 
The $+$ represent an
exponential decrease with $var\lp F_{exp} \rp \approx 15$.
The best fit to a power-law, represented by the monotonic line, gives $A_1
\approx 3.9$, $B_1 \approx 0.6$,  $t_c = 87.75$, $m_1 \approx -1.5 $ and
$\mbox{var}_{pow} \approx 12$. The best fit to expression (\ref{kflala})
with $t_c - t$ replaced by
$t-t_c$ gives  $A_2 \approx 3.4$, $B_2 \approx 0.9$, $t_c \approx 87.77$, $C
\approx 0.3$,  $\omega \approx 11$, $m_2 \approx -1.2$ and $\mbox{var}_{lp}
\approx 7$. One can observe six well-defined oscillations fitted by (\ref{kflala}).
Reproduced from [Sornette et al., 1996].}}
\hspace{5mm}
\parbox[r]{7.5cm}{
\epsfig{file=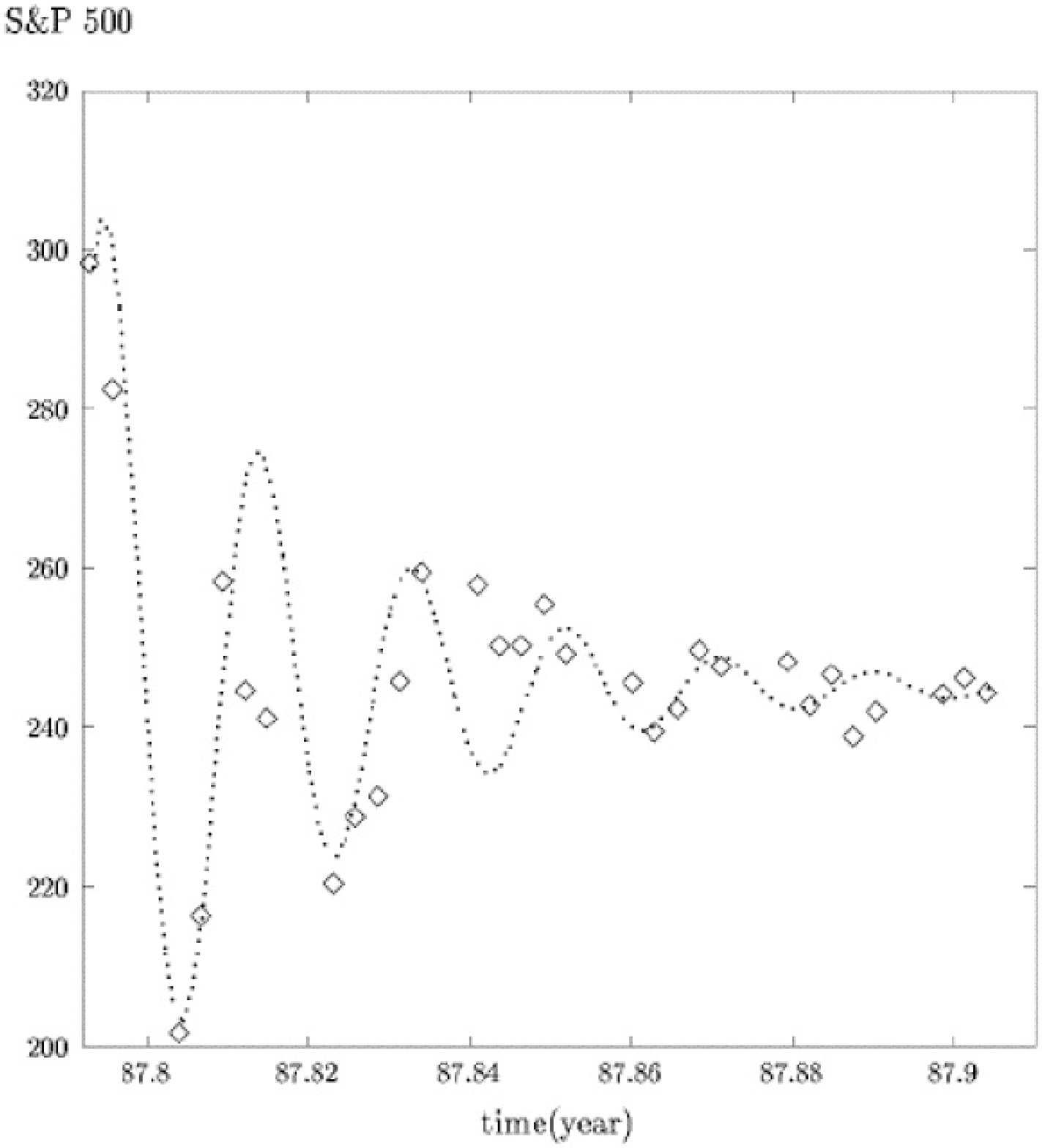,height=5cm,width=7cm}
\caption{\protect\label{fig3first} Time evolution of the S\&P500 index over
a time window of a few weeks after the Oct. 19, 1987 crash. The fit 
with an exponentially decaying sinusoidal function shown
in dashed line suggests that a good
model for the short-time response of the US market is a {\it single}
dissipative harmonic oscillator or damped pendulum.
Reproduced from [Sornette et al., 1996].}}
\end{figure}

If the concept of a crash as a kind of critical point has any value, we
should be able to identify post-crash signatures of the underlying
cooperativity. In fact, we should expect an at least qualitative symmetry
between patterns before and after the crash. In other words, we should be
able to document the existence of a critical exponent as well as
log-periodic oscillations on relevant quantities after the crash. 
Such a signature in the volatility of the S\&P500 index (a measure
of the market risk perceived by investors), 
implied from the price of S\&P500 options, can indeed be seen in 
figure \ref{fig2first}.

Figure \ref{fig2first} presents the time evolution of the implied volatility 
of the S\&P500, taken from [Chen et al., 1995]. The perceived market risk is small
prior to the crash, jumps up abruptly at the time of the crash and then
decays slowly over several months. This decay to ``normal times'' of perceived risks
is compatible with a slow power law decay decorated by 
log-periodic oscillations, which can be fitted by expression (\ref{kflala})
with $t_c - t$ (before the crash) replaced by $t-t_c$ (after the crash). 
Our analysis with (\ref{kflala}) with
$t_c-t$ replaced by $t-t_c$ gives again an estimation
of the position of the critical time $t_c$, which is found correctly within a few days.
Note the long time scale covering a period of the order of a year
involved in the relaxation of the volatility after the
crash to a level comparable to the one before the crash. This  implies
the existence of a ``memory effect'':  market participants remain nervous for quite
a long time after the crash, after being burned out by the dramatic event. 

It is also noteworthy that the S\&P500 index as well as other markets worldwide have remained
close to the after-crash level for a long time. For instance,
by February 29, 1988, the world index stood at $72.7$ (reference $100$ on
September 30, 1987). Thus, the price level established in the October crash
seems to have been a virtually unbiased estimate of the average price level over
the subsequent months (see also figure \ref{fig3first}). 
This is in support of the idea of a critical point,
according to which the event is an intrinsic signature of a
self-organization of the markets worldwide. 

There is another striking signature of the cooperative behavior of the
US market, found by analyzing the time evolution of the S\&P500 index over
a time window of a few weeks after the Oct. 19, 1987 crash. A fit shown in
figure \ref{fig3first} with an exponentially decaying sinusoidal function suggests that the US
market behaved, for a few weeks after the crash, as a {\it single}
dissipative harmonic oscillator, 
with a characteristic decay time of about one week
equal to the period of the oscillations.
 In other words, the price followed
the trajectory of a pendulum moving back and forth
with damped oscillations around an equilibrium position. 

This signature strengthens the
view of a market as a cooperative self-organizing system. The basic story 
suggested by these figures is the following. Before the crash,
imitation and speculation
were rampant and led to a progressive ``aggregation'' of the multitude
of agents into a large effective ``super-agent'', as illustrated in 
figures \ref{fig1first} and \ref{figlong87}; right after the crash, the
market behaved as a single ``super-agent'' finding rapidly the equilibrium price
through a return to ``equilibrium'', as shown in figure \ref{fig3first}. 
On longer time scales, the ``super-agent'
progressively was fragmented and the diversity of behaviors was rejuvenated as
seen from figure \ref{fig2first}.

\vskip-0.3cm
\subsection{The crash of October 1929}

\begin{figure}
\parbox[l]{7.5cm}{
\epsfig{file=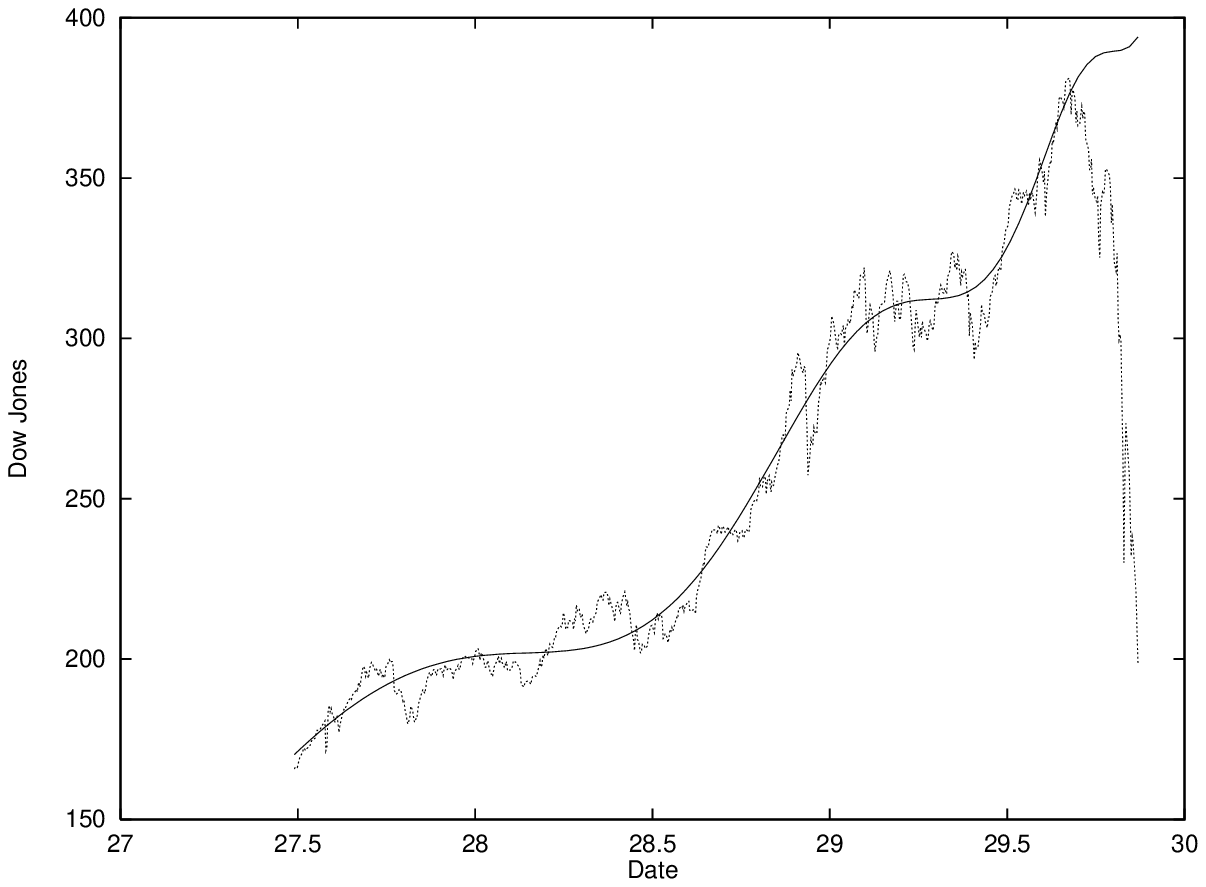,height=5cm,width=7cm}
\caption{\protect\label{figcourt29} The Dow Jones index prior to the October 1929
crash on Wall Street. The fit shown as a continuous line is the equation (\ref{kflala}) with
$A_2 \approx  571, B_2 \approx -267, B_2 C \approx 14.3, 
m_2 \approx  0.45, t_c \approx  1930.22, \omega\approx 7.9$ and $\phi\approx  1.0$.
reproduced from [Johansen and Sornette, 1999a].}}
\hspace{5mm}
\parbox[r]{7.5cm}{
\epsfig{file=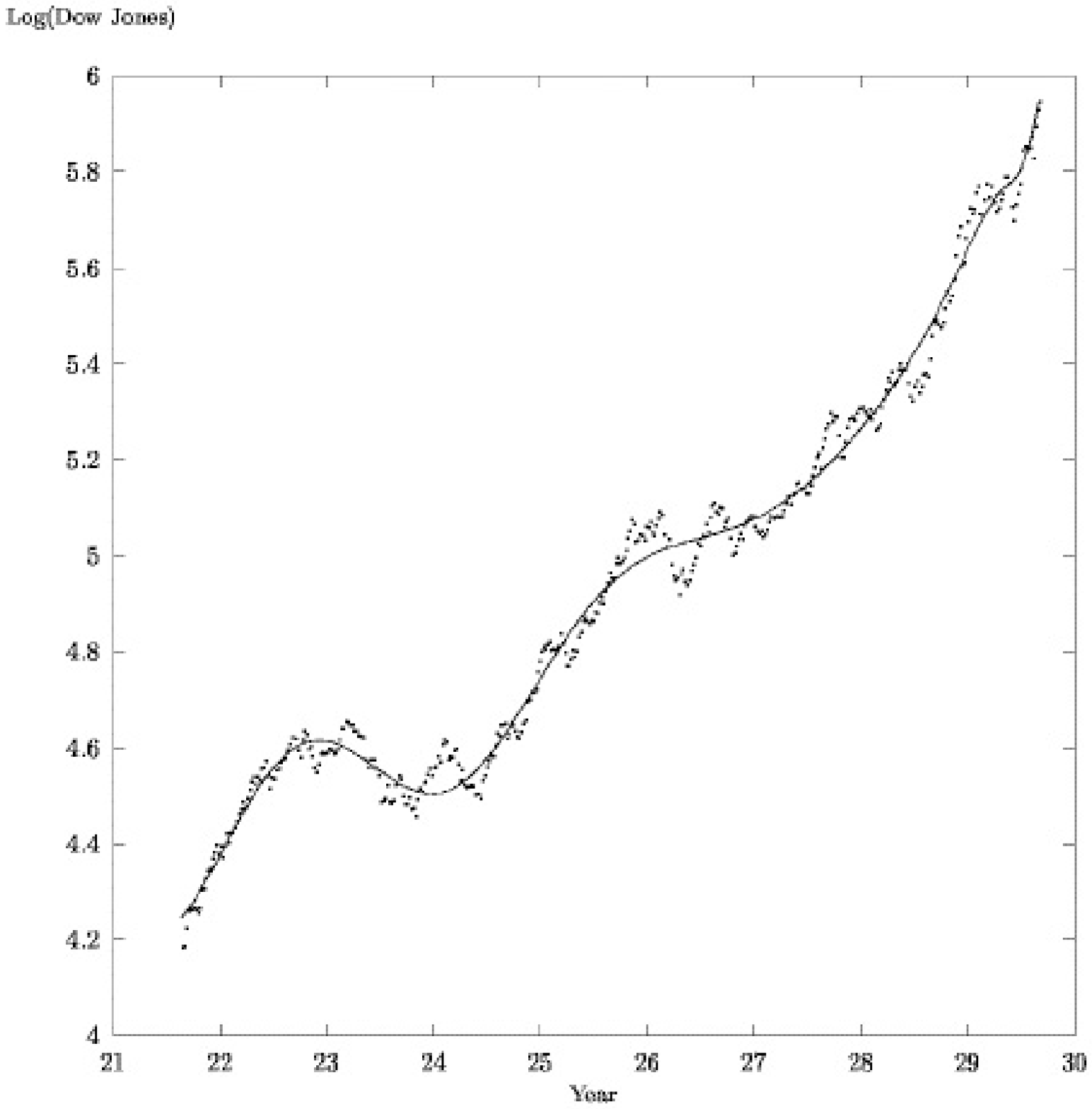,height=5cm,width=7cm}
\caption{\protect\label{figlong29} Time dependence of the logarithm of the Dow
Jones stock exchange index from June 1921 to September 1929 and best fit by 
the  improved nonlinear log-periodic formula  developed in [Sornette and Johansen, 1997]. 
The crash of October 23, 1929 corresponds to $1929.81$ 
decimal years. The parameters of the fit are:  
r.m.s.$=0.041$, $t_c = 1929.84$ year, $m_2 = 0.63$, $\omega = 5.0$,
$\Delta \omega = -70$, $\Delta t = 14$ years, $A_2 = 61$, $B_2 = -0.56$,
$C = 0.08$. $\Delta \omega$ and $\Delta t$ are two new parameters introduced in
[Sornette and Johansen, 1997]. Reproduced from [Sornette and Johansen, 1997].}}
\end{figure}

The crash of Oct. 1929 is the other major historical market event of the twentieth
century. Notwithstanding the differences in technologies and the absence of 
computers and other modern means of information transfer, the Oct. 1929 crash
exhibits many similarities with the Oct. 1987 crash, so much so as shown
in figures \ref{figcourt29} and \ref{figlong29}, 
that one can wonder about the similitudes: what has
not changed over the history of mankind is the interplay between
human's crave for exchanges and profits, 
and their fear of uncertainty and losses. The similarity between the
two situations in 1929 and 1987 was in fact noticed at a qualitative level in
an article in the {\it Wall Street Journal} on october 19, 1987, the very morning of
the day of
the stock market crash (with a plot of stock prices in the 1920s and the
1980s). See the discussion in [Shiller, 1989].

The similarity between the two crashes can be made quantitative
by comparing the fit of the Dow Jones index with formula (\ref{kflala}) from June 1927
till the maximum before the crash in October 1929, as shown in figure \ref{figcourt29}, to 
the corresponding fit for the October 1987 crash shown in figure \ref{fig1first}.
Notice the similar widths of the two time windows, the similar acceleration and oscillatory
structures, quantified by similar exponents $m_2$ and log-periodic
angular frequency $\omega$: $m_2^{1987}=0.33$ compared to 
$m_2^{1929}=0.45$; $\omega^{1987}=7.4$ compared to $\omega^{1987}=7.9$. These
numerical values are remarkably close and can be considered 
equal to within their uncertainties.

Figure \ref{figlong29} for the October 1929 crash is the analog of figure
\ref{figlong87} for the October 1987 crash. It uses the 
improved nonlinear log-periodic formula 
developed in [Sornette and Johansen, 1997] over 
a much larger time window starting in June 1921.
Also according to this improved theoretical formulation,
the values of the exponent
$m_2$ and of the log-periodic angular frequency $\omega$ for the two great crashes
are quite close to each other: $m_2^{1929} = 0.63$ and $m_2^{1987} = 0.68$. 
This is in agreement with
the universality of the exponent $m_2$ predicted from the
renormalization group
theory for log-periodicity [Saleur and Sornette, 1996; Sornette, 1998].
A similar universality is also expected for the log-frequency,
albeit with a
weaker strength as it has been shown [Saleur and Sornette, 1996] that fluctuations
and noise will modify $\omega$ differently depending on their nature. 
The fits indicate that $\omega_{1929} = 5.0$ and $\omega_{1987} = 8.9$. These values
are not unexpected and fall within the range found for other crashes (see below).
They correspond to a prefered scaling ratio equal respectively to 
$\lambda_{1929}=3.5$ compared to $\lambda_{1987}=2.0$.

The Oct. 1929 and Oct. 1987 thus exhibit two similar precursory patterns
on the Dow Jones index, starting respectively 2.5 and 8 years before them.
It is thus a striking observation that essentially similar
crashes have punctuated this century, notwithstanding
tremendous changes in all imaginable ways of life and work. The only thing
that has probably changed little are the way humans think and behave.
The concept that emerges here is that the organization of traders in
financial markets leads intrinsically to 
``systemic instabilities'', that probably result in
a very robust way
from the fundamental nature of human beings, including our gregarious
behavior, our
greediness, our instinctive psychology during panics and crowd behavior and
our risk aversion.
The global behavior of the market, with its log-periodic structures that emerge
as a result of the cooperative behavior of traders, is reminiscent of the
process of the emergence of intelligent behavior at a macroscopic scale
that individuals at the  microscopic scale cannot perceive. This process has
been discussed in biology for instance in animal populations such as ant
colonies or in connection with the emergence of consciousness
[Anderson et al., 1988].

There are however some differences between the two crashes.
An important quantitative difference between the great crash
of 1929 and the collapse of stock prices in October 1987 was that stock
price variability in the year following the crash was much higher in 1929 than in 1987
[Romer, 1990].
This has led economists to argue that the collapse of stock prices in October 1929
generated significant temporary increased uncertainty 
about future income that led consumers to
forgo purchases of durable goods.  Forecasters were then much more
uncertain about the course of future income following the stock market
crash than was typical even for unsettled times. Contemporary
observers believed that consumer uncertainty was an important force
depressing consumption, that may have been an important factor in the 
strengthening of the great depression. The increase of uncertainty after the 
Oct. 1987 crash has led to a smaller effect, as no depression ensued.
However, figure \ref{fig2first} clearly quantifies an increased uncertainty 
and risk, lasting months after the crash. 

\vskip-0.3cm
\subsection{The three Hong Kong crashes of 1987, 1994 and 1997}

Hong Kong has a strong free-market attitude, characterized by very few
restrictions on both residents and non-residents, private persons or companies,
to operate, borrow, repatriate profit and capital. This goes on even after 
Hong Kong reverted to Chinese
sovereignty on July 1st, 1997 as a Special Administrative Region (SAR) of the
People's Republic of China, as it was promised a ``high degree of autonomy'' for at
least 50 years from that date according to the terms of the 
Sino-British Joint Declaration. The SAR is ruled according to a mini-constitution,
the Basic Law of the Hong Kong SAR.
Hong Kong has no exchange controls and crossborder remittances are readily
permitted. These rules have not changed since July 1st, 1997 when China took over
sovereignty from the UK. Capital can thus flow in and out of the Hong Kong stock
market in a very fluid manner. There are no 
restrictions on the conversion and remittance of
dividends and interest. Investors bring their capital into Hong Kong through the
open exchange market and remit it the same way.

\begin{figure}
\parbox[l]{7.5cm}{
\epsfig{file=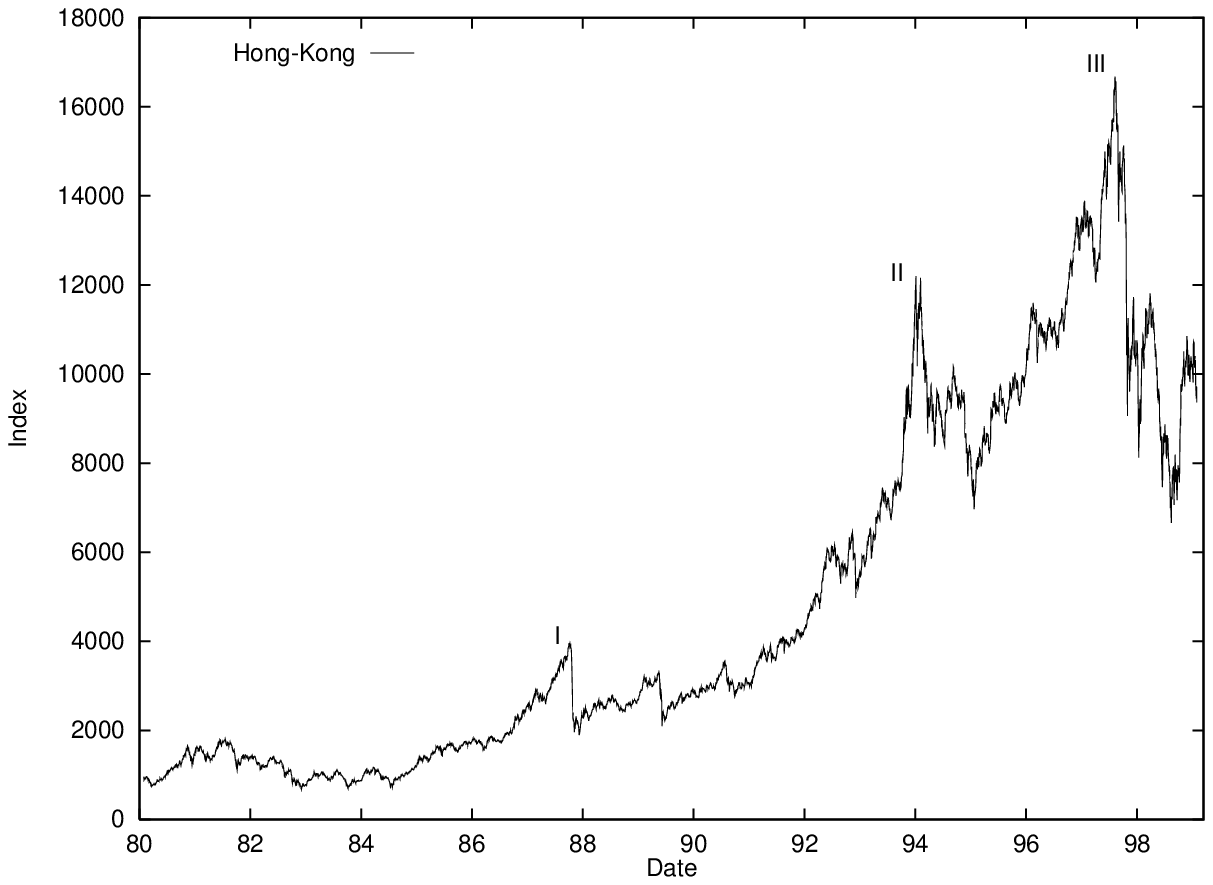,height=5cm,width=7cm}
\caption{\protect\label{hk} The Hong Kong stock market index as a function
of time. Three extended bubbles followed by large crashes can be identified.
The approximate dates of the crashes are  Oct. 87 (I),  Jan 94 (II)
and Oct 97 (III). Reproduced from [Johansen and Sornette, 2001b].}}
\hspace{5mm}
\parbox[r]{7.5cm}{
\epsfig{file=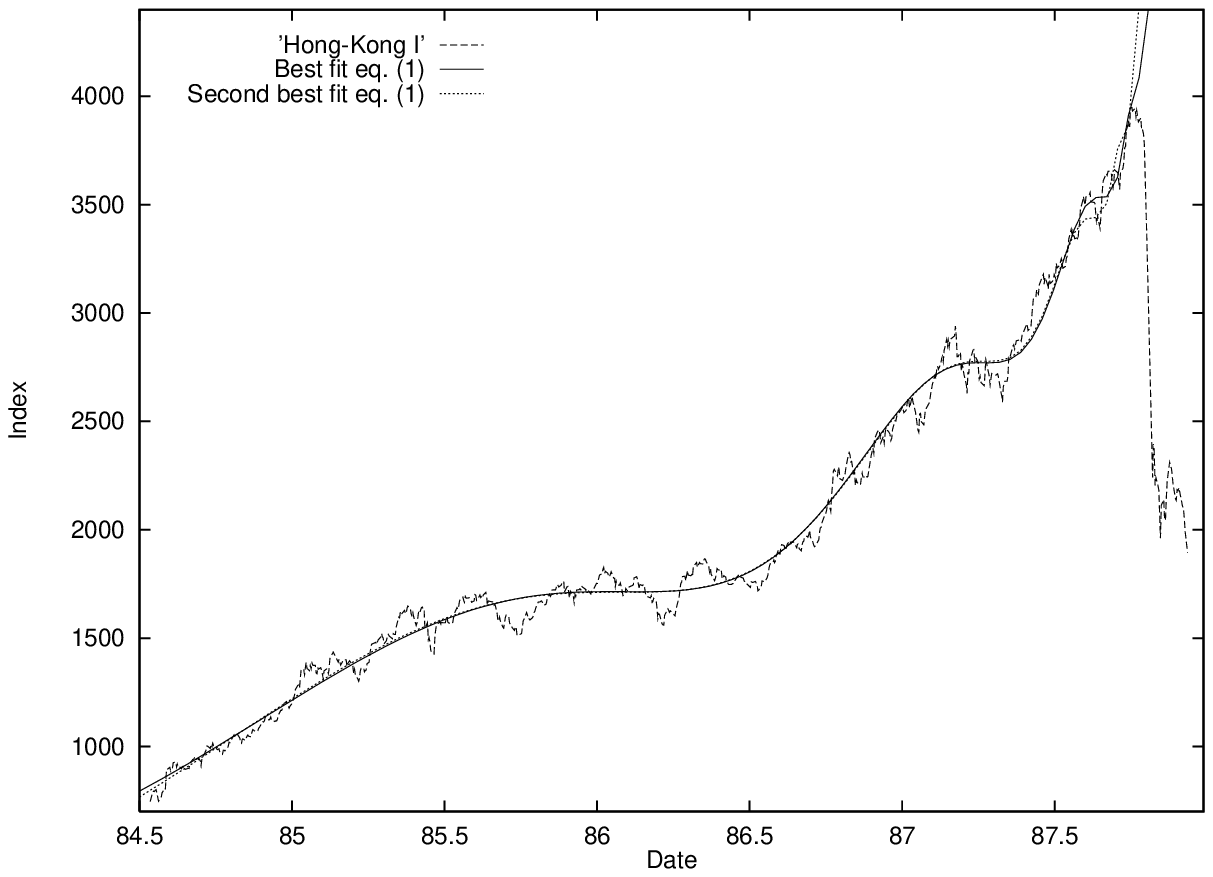,height=5cm,width=7cm}
\caption{\protect\label{hkbub1} Hong Kong stock market bubble ending with the
crash of Oct. 87. On Oct. 19, 1987, the Hang Seng index
closed at $3362.4$. On oct. 26, it closed at $2241.7$, corresponding
to a loss of $33.3\%$.
See table \protect\ref{asitab2} for the parameter
values of the fit with equation (\ref{kflala}). Reproduced from 
[Johansen and Sornette, 2001b].}}
\end{figure}

Accordingly, we may expect speculative behavior and crowd effects to be
free to express themselves in their full force. Indeed, the Hong Kong stock
market provides maybe the best textbook-like examples of speculative bubbles decorated by
log-periodic power law accelerations followed by crashes. Over the last
15 years only, one can identify three major bubbles and crashes. They are 
indicated as I, II and III in figure \ref{hk}.

\begin{figure}
\parbox[l]{7.5cm}{
\epsfig{file=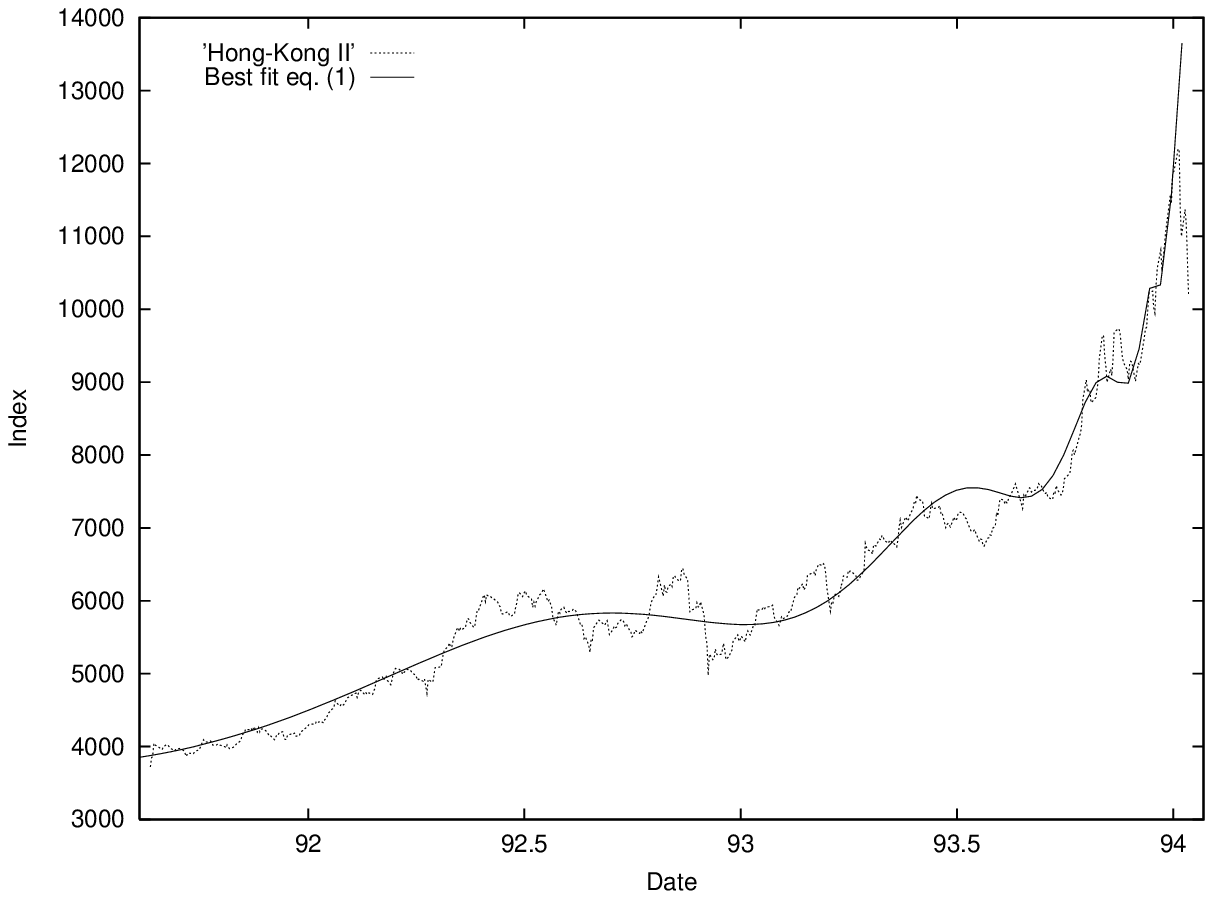,height=5cm,width=7cm}
\caption{\protect\label{hkbub2} Hong Kong stock market bubble ending with the
crash of early 94. On Feb. 4, 1994, the Hang Seng index
topped at $12157.6$. A month later, on March 3rd, 1994, it closed at
$9802$, corresponding to a cumulative loss of $19.4\%$. It went even
further down two months later, with a close at $8421.7$ on May, 9, 1994,
corresponding to a cumulative loss since the high on  Feb. 4 of $30.7\%$.	
See table \protect\ref{asitab2} for the parameter
values of the fit with equation (\ref{kflala}). Reproduced from [Johansen and Sornette, 2001b].}}
\hspace{5mm}
\parbox[r]{7.5cm}{
\epsfig{file=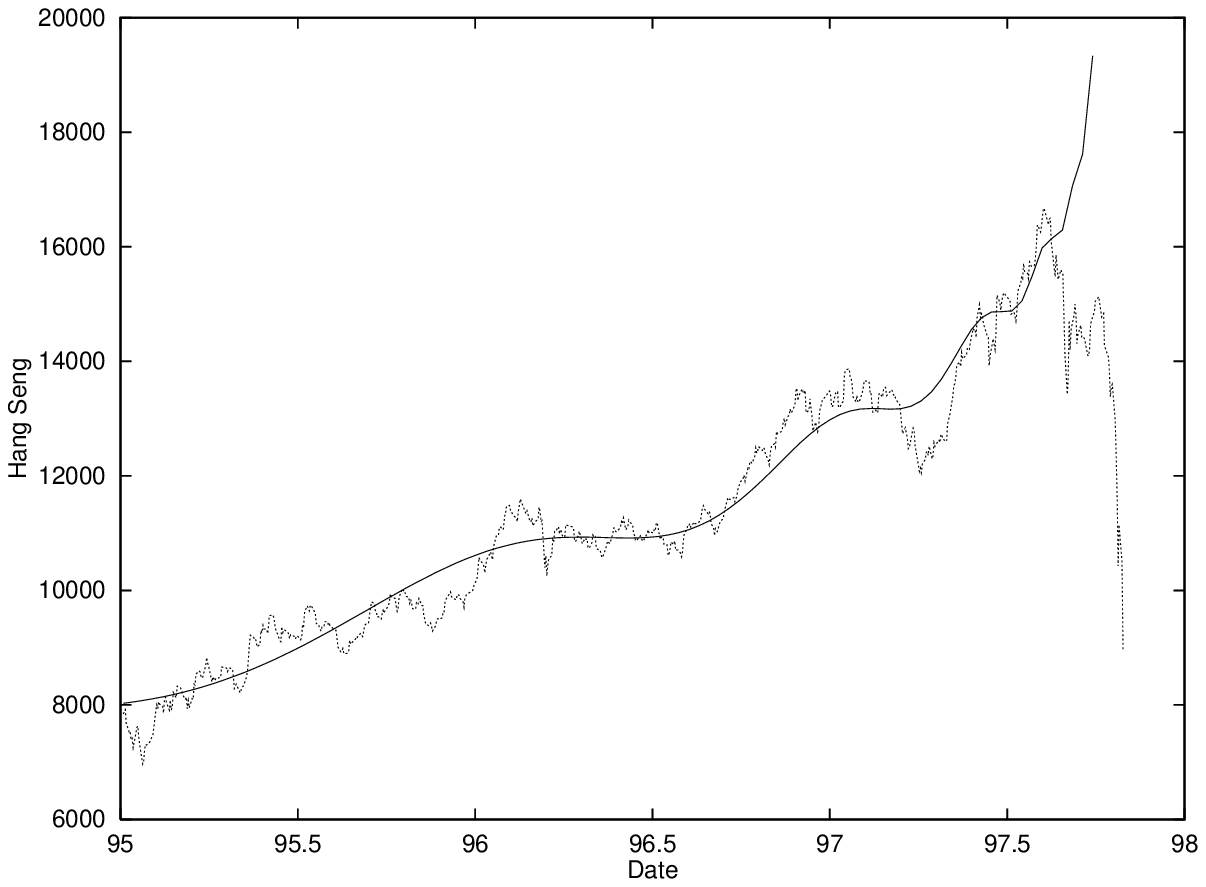,height=5cm,width=7cm}
\caption{\protect\label{hkbub3} The Hang Seng index prior to the October 1997
crash on the Hong Kong Stock Exchange. The index topped at 
$16460.5$ on Aug. 11, 1997. It then regularly decayed
to $13601$ reached on Oct. 17, 1997. It then crashed
abruptly reaching a close of $9059.9$ on Oct. 28, 1997, with an
intra-day low of $8775.9$. The amplitude of the total cumulative
loss since the high on Aug. 11 is $45\%$. The amplitude of the crash
from Oct. 17 to Oct. 28 is $33.4\%$.
The fit is equation (\ref{kflala})
with $A_2 \approx  20077 $, $B_2 \approx  -8241 $, $C \approx -397$, 
$m_2 \approx  0.34 $, $t_c\approx 1997.74 $, $\omega\approx 7.5$ and
$\phi\approx  0.78 $. 
Reproduced from [Johansen and Sornette, 1999a, 2001b].}}
\end{figure}

\begin{enumerate}
\item The first bubble and crash are shown in figure \ref{hkbub1} and 
are synchronous to the worldwide Oct. 1987 crash already discussed.
On Oct. 19, 1987, the Hang Seng index
closed at $3362.4$. On oct. 26, it closed at $2241.7$, corresponding
to a cumulative loss of $33.3\%$.

\item The second bubble ends in early 1994 and is shown in figure \ref{hkbub2}.
The bubble ends by what we could call a ``slow crash'': 
on Feb. 4, 1994, the Hang Seng index
topped at $12157.6$ and, a month later on March 3rd, 1994, it closed at
$9802$, corresponding to a cumulative loss of $19.4\%$. It went even
further down over the next two months, with a close at $8421.7$ on May, 9, 1994,
corresponding to a cumulative loss since the high on  Feb. 4 of $30.7\%$.

\item The third bubble, shown in figure \ref{hkbub3} 
ended in mid-august 1997 by a slow and regular
decay until Oct. 17, 1997, followed by an abrupt crash: the drop from
$13601$ on Oct. 17 to $9059.9$ on Oct. 28 corresponds to a $33.4\%$ loss.
The worst daily plunge of $10\%$ was the third biggest percentage fall following the
$33.3\%$ crash in Oct. 1987 and $21.75\%$ fall after the
Tiananmen Square crackdown in June 1989.

\end{enumerate}

\begin{figure}
\begin{center}
\epsfig{file=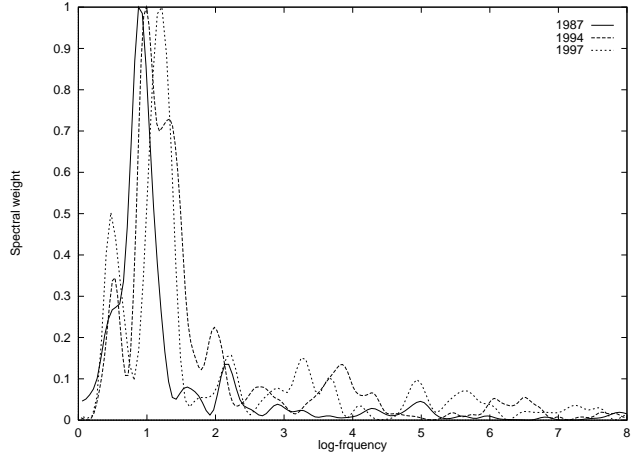,height=6cm}
\caption{\protect\label{Figlomballhk} Lomb spectral analysis of the three bubbles
preceding the three crashes on the
Hong Kong market shown 
in figures \protect\ref{hkbub1}-\protect\ref{hkbub3}. See [Press et al., 1992] 
for explainations 
on the Lomb spectral analysis. All three bubbles are characterized
by almost the same ``universal'' log-frequency $f \approx 1$ corresponding
to a prefered scaling ratio of the discrete scale invariance equal to
$\lambda = \exp \left( 1/f \right) \approx 2.7$. Courtesy of A. Johansen
}
\end{center}
\end{figure}

The table \ref{asitab2} gives the parameters of the fits 
with equation (\ref{kflala}) of the bubble
phases of the three events I, II and III shown in figures \ref{hkbub1}-\ref{hkbub3}.
It is quite remarkable that
the three bubbles on the Hong Kong stock market have essentially
the same log-periodic angular frequency $\omega$ within
$\pm 15$\%. 
These values are also quite similar to what has been
found for bubbles on the USA market and for the FOREX (see below).
In particular, for the Oct. 1997 crash on the Hong Kong market, we have
$m_2^{1987}=0.33 < m_2^{HK1997}=0.34 < m_2^{1929}=0.45$ and
$\omega^{1987}=7.4 <  \omega^{HK1997}=7.5 < \omega^{1929}=7.9$; 
the exponent $m_2$ and the log-periodic angular frequency $\omega$
for the October 1997 crash on the Hong Kong Stock Exchange are perfectly
bracketed by the two main crashes on Wall Street!
Figure \ref{Figlomballhk} demonstrates the ``universality''
of the log-periodic component of the signals in the three bubbles
preceding the three crashes on the Hong Kong market.

\begin{table}[b]
\begin{center}
\begin{tabular}{|c|c|c|c|c|c|c|c|c|c|} \hline
Stock market & $A_2$ & $B_2$ & $B_2 C$ & $m_2$  & $t_c$ & $\omega$ & $\phi$ \\ \hline
Hong Kong I &  $5523;4533$ & $-3247;-2304$ & $171;-174$ & $0.29;0.39$ &
$87.84;87.78$ & $5.6;5.2$ & $-1.6;1.1$
\\ \hline
Hong Kong II&  $21121$ & $-15113$ & $-429$ & $0.12$ & $94.02$ & $6.3$ & $-0.6$
\\ \hline
Hong Kong III&  $20077$ & $-8241$ & $-397$ & $0.34$ & $97.74$ & $7.5$ & $0.8$
\\ \hline
\end{tabular}
\vspace{5mm}
\caption[]{Fit parameters of the three
speculative bubbles on the Hong Kong stock market shown 
in figures \ref{hkbub1}-\ref{hkbub3}
leading to a large crash. Multiple entries correspond to the two best fits.
Reproduced from [Johansen and Sornette, 2001b].}
\label{asitab2}
\end{center}
\end{table}

\vskip-0.3cm
\subsection{The crash of Oct. 1997 and its resonance on the US market}

The Hong Kong market crash of Oct. 1997 has been presented as a textbook
example where contagion and speculation took a course of their own.
When Malaysian Prime Minister Dr Mahathir Mohamad 
made his now famous address to
the World Bank-International Monetary Fund seminar in Hong Kong 
in September 1997, many critics
pooh-poohed his proposal to ban currency speculation as an attempt to hide the
fact that Malaysia's economic fundamentals were weak.
They pointed to the fact that the currency turmoil had not affected Hong Kong,
whose economy was basically sound. Thus, if Malaysia and other countries
were affected, that's because their economies were weak.
At that time, it was easy to point out the deficits in the then current account
of Thailand, Malaysia and Indonesia. In contrast, Hong Kong had a good current
account situation and moreover had solid foreign reserves worth US\$88 billion.
This theory of the strong-won't-be-affected already suffered a setback when the
Taiwan currency's peg to the US dollar had to be removed after the Taiwan
authorities spent US\$5 billion to defend their currency from speculative attacks,
and then gave up. The ``coup de grace'' came with the meltdown in Hong Kong in Oct. 1997
which shocked the analysts and the media as this
high-flying market was considered the safest haven in Asia. In contrast to the
meltdown in Asia's lesser markets as country after country, led
by Thailand in July 1997, succumbed to economic and currency problems,
Hong Kong was supposed to be different. With its Western-style markets, the
second largest in Asia after Japan, it was thought to be immune to the financial
flu that had swept through the rest of the continent. It is clear from our
analysis of section 5 and from
the lessons of the two previous bubbles ending in Oct. 1987 and in early 1994
that those assumptions naively overlooked the contagion leading to over-investments
in the build-up period preceding the crash
and the resulting instability, which left the Hong Kong
market vulnerable to speculative attacks. Actually, hedge funds 
in particular are known to have taken positions consistent with a possible crisis
on the currency and on the stock market, by ``shorting'' (selling) the currency
to drive it down, forcing the Hong Kong government to raise interest rates to defend it
by increasing the currency liquidity but as a consequence having equities suffer,
making the stock market more unstable. 

As we have already stressed, one should not mix the 
``local'' cause from the fundamental cause of the instability. 
As the late George Stigler once put it, 
to blame 'the markets' for an outcome we don't like is
like blaming the waiters in restaurants for obesity. Within the framework
defended here (see also [Sornette, 2003]), crashes occur as possible (but not necessary) outcomes
of a long preparation, that we refer for short as ``herding'', which makes
the market enter into a more and more unstable regime. 
When in this state, there are many possible
``local'' causes that may cause it to stumble. Pushing the argument to the extreme
to make it crystal clear, it is as if the responsability for the collapse of the
infamous Tacoma Narrows Bridge that once connected mainland Washington with the Olympic
peninsula was attributed to strong wind.
It is true that, on November 7, 1940, at approximately 11:00 AM,
it suddenly collapsed after developing a
remarkably ``ordered'' sway in response to a strong wind after it
had been open to traffic for only a few months
(see Tacoma Narrows Bridge
historical film footage showing in 250 frames (10 seconds) the
maximum torsional motion shortly before failure of this immense
structure:  {\tt http://cee.carleton.ca/Exhibits/Tacoma\_Narrows/}). However, the strong
wind of that day is only the ``local'' cause while there is a more
fundamental cause: the bridge, like
most objects, has a small number of characteristic vibration
frequencies, and one day the wind was exactly the strength needed to
excite one of them. The bridge responded by vibrating at this
characteristic frequency so strongly, i.e., by ``resonating'', 
that it fractured the supports holding it together. The fundamental cause
of the collapse of the Tacoma Narrows Bridge thus lies in an error of
conception that enhanced the role of one specific mode of resonance.
We propose that, analogously to the collapse of the Tacoma Narrows Bridge,
many stock markets crash as the results of built-in or acquired instabilities. These
instabilities may in turn be revealed by ``small'' perturbations that lead to
the collapse. 

The speculative attacks in periods of market instabilities are sometimes 
pointed at as possible causes of serious
potential hazards for developing countries when allowing the global financial
markets to have free play, especially when these countries
come under pressure to open up their financial sectors to large
foreign banks, insurance companies, stock broking firms and other institutions,
under the World Trade Organisation's financial services negotiations.
We argue that the problem comes in fact fundamentally from the over-enthusiastic
initial in-flux of capital as a result of herding,
 that initially profits the country, but at the risk
of future instabilities: developing countries as well as investors ``can not have
the cake and eat it too!'' From an efficient market view point, the speculative
attacks are nothing but the revelation of the instability and the means by which
the markets are forced back to a more stable dynamical state.

Interestingly, the Oct. 1997 crash on the Hong Kong market had important
echos in other markets worldwide and in particular in the US markets.
The story is often told as if a ``wave of selling'', starting in 
Hong Kong, has spread first to other southeast
Asian markets based on negative sentiment - which served to reaffirm the deep
financial problems of the Asian tiger nations - then to the European
markets, and finally to the US market.
The shares that were hardest hit in Western markets were the
multinational companies, which receive part of their earnings from the southeast
Asian region. The reason for their devaluation is that the region's economic
slowdown would lower corporate profits.
It is estimated that the 25 companies which make up one third of Wall Street's S\&P500 
index market capitalisation earn roughly half of their income from non-US
sources. Lower growth in southeast Asia heightened one of the biggest concerns of Wall
Street investors. To carry on the then present ``bull'' run, the market needed sustained
corporate earnings - if they were not forthcoming, the cycle of rising share prices
would whither into one of falling share prices.
Concern over earnings might have proved to be the straw that broke Wall Street's
six-year bull run. 

\begin{figure}
\begin{center}
\epsfig{file=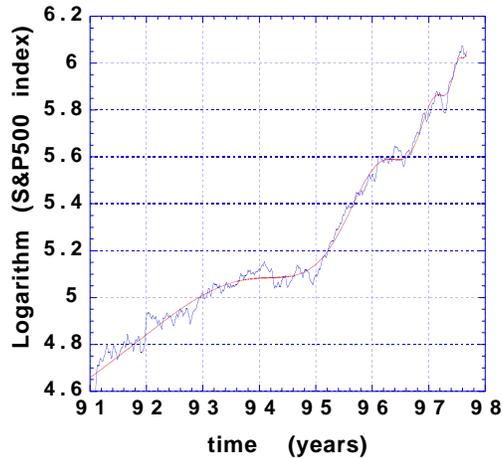,height=6cm}
\caption{\protect\label{SPlong1997} The best fit shown
as the smooth continuous line of the 
logarithm of the S\&P500 index from Jan. 1991 till Sept. 4, 1997 (1997.678)
by the improved nonlinear log-periodic formula 
developed in [Sornette and Johansen, 1997], already used in figures \ref{figlong87}
and \ref{figlong29}. The exponent $m_2$ and log-periodic angular frequency
$\omega$ are respectively $m_2=0.73$ (compared to $0.63$ for Oct. 1929 and
$0.33$ for Oct. 1987) and $\omega=8.93$ (compared to $5.0$ for Oct. 1929 and
$7.4$ for Oct. 1987). The critical time 
predicted by this fit is $t_c=1997.948$, i.e., mid-december 1997.
Courtesy of A. Johansen
}
\end{center}
\end{figure}

Fingerprints of
herding and of an incoming instability were detected by several groups
independently and announced publicly. According to our
theory, the turmoil on the financial US market in Oct. 1997 should not
be seen only as a passive reaction to the Hong Kong crash. The log-periodic
power law signature observed on the US market over several years before Oct. 1997
(see figure \ref{SPlong1997})
indicates that a similar ``herding'' instability was also developing simultaneously.
In fact, the detection of log-periodic structures and a
prediction of a stock market correction or a crash at the end of october 1997 was 
formally issued jointly ex-ante on Sept. 17, 1997 by A. Johansen and the author,
to the French office for the  protection of
proprietary softwares and inventions with registration number 94781.
In addition, a trading strategy
was been devised using put options in order to provide an experimental test of
the theory. A $400\%$ profit has been obtained in a two week period covering
the mini-crash of October 28, 1997. The proof of this profit is available from
a Merrill Lynch client cash management account released in November 1997. 
Using a variation of our theory which turns
out to be slightly less reliable (see the
comparative tests in [Johansen and Sornette, 1999b]),
a group of physicists and economists [Vandewalle et al., 1998a] also
made a public announcement published on Sept. 18, 1997
in a Belgium journal [Dupuis, 1997]
and communicated afterwards their methodology in a scientific publication
[Vandewalle et al., 1998b]. Two other groups have also analyzed, after the fact, the
possibility to predict this event. Feigenbaum and Freund [1998] analyzed
the log-periodic oscillations in the S\&P500
and the NYSE in relation to the October 27'th ``correction''
seen on Wall Street. Gluzman and Yukalov [1998]
proposed a new approach based on the algebraic self-similar renormalization
group to analyze the time series corresponding to
the Oct. 1929 and 1987 crashes and the Oct. 1997 
correction of the New York Stock Exchange (NYSE) [Gluzman and Yukalov, 1998].

The best fit of the 
logarithm of the S\&P500 index from Jan. 1991 till Sept. 4, 1997
by the improved nonlinear log-periodic formula 
developed in [Sornette and Johansen, 1997], already used in figures \ref{figlong87}
and \ref{figlong29} is shown in figure \ref{SPlong1997}.
This result and many other
analyses led to the prediction alluded to above.
It turned out that the crash did not really occur\,: what happened was
that the Dow plunged 554.26 points, finishing the day
down 7.2\%, and NASDAQ posted its biggest-ever (up to that time)
one-day point loss. In
accordance with a new rule passed after Oct. 1987 ``Black Monday'', trading was halted on all
major U.S. exchanges. Private communications
from professional traders to the author indicate 
that many believed that a crash was coming but
this turns out to be incorrect. This sentiment has also to be put
in the perspective of the earlier sell-off at the beginning
of the month triggered by Greenspan's statement that the boom in
the U.S. economy was unsustainable and that the current rate of gains in the
stock market was unrealistic. 

It is actually interesting that the critical time $t_c$ identified around this data
indicated a change of regime rather than a real crash: after this
turbulence, the US market remained more or less flat, thus breaking the previous
``bullish'' regime, with large volatility until
the end of January 1998, and then started again a new ``bull'' phase stopped
in its course in August 1998, that we shall analyze below. The observation of a change of
regime after $t_c$ is in full agreement with the rational expectation model of a
bubble and crash described in section 5\,:
the bubble expands, the market believes that a crash may be more and more
probable, the prices develop characteristic structures of speculation and herding
but the critical time passes
without the crash happening.
This can be interpreted as the non-zero probability scenario also predicted by the
rational expectation model of a bubble and crash described in section 5,
that it is possible that no crash occurs over the whole lifetime of the bubble including $t_c$.

The simultaneity of the critical times $t_c$ of the Hong Kong crash and of the end of the
US and European speculative bubble phases at the end of Oct. 1997 
may be neither a lucky occurrence
nor a signature of a causal impact of one market (Hong Kong) onto others, as has
been often discussed too naively. This simultaneity can actually be predicted 
in a model of rational expectation bubbles allowing the coupling and interactions
between stock markets. For general interactions, if a critical time appears in one
market, it should also be present in other markets as a result of the nonlinear
interactions existing between the markets [Johansen and Sornette, 2001a].

\vskip-0.3cm
\subsection{Currency crashes}

\begin{figure}
\begin{center}
\epsfig{file=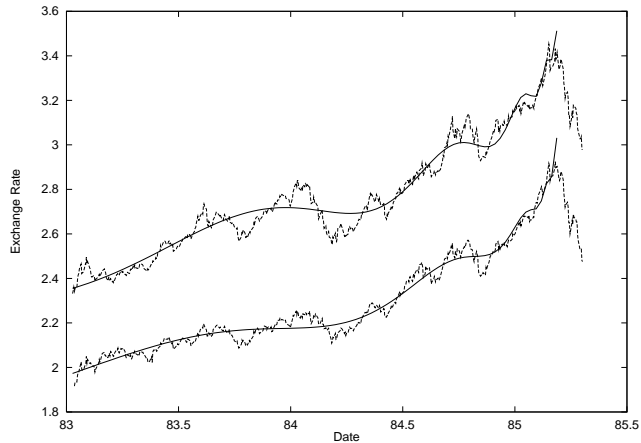,height=6cm}
\caption{\protect\label{DEMCHF1985} The US \$ expressed in German Mark DEM (top curve) and in
Swiss franc CHF (bottom curve) prior to its
collapse on mid-1985. 
The fit to the DEM currency against the US dollar 
with equation (\ref{kflala}) is shown as the continuous and
smooth line and give
$A_2\approx  3.88  $, $B_2 \approx  -1.2 $, $B_2 C \approx   0.08 $, 
$m_2 \approx   0.28$, $t_c \approx 1985.20$,  $\omega \approx 6.0$ and
$\phi \approx -1.2 $. The fit to the Swiss franc against the US dollar 
with equation (\ref{kflala}) gives
$A_2 \approx  3.1$, $B_2 \approx  -0.86 $, $B_2 C \approx   0.05 $, 
$m_2 \approx 0.36 $, $t_c \approx 1985.19$, $\omega \approx 5.2$ 
and $\phi \approx -0.59 $. Note the
small fluctuations in the value of the scaling ratio $2.2 \leq \lambda \leq
2.7$, which constitutes one of the key test of our ``critical herding'' theory.
Reproduced from [Johansen and Sornette, 1999a].
}
\end{center}
\end{figure}

Currencies can also develop bubbles and crashes. The bubble on the dollar
starting in the early 1980s and ending in 1985
 is a remarkable example shown in figure \ref{DEMCHF1985}.

The US dollar experienced an unprecedented cumulative appreciation 
against the currencies of the major industrial countries starting around 1980,
with several consequences: loss of competitiveness with important
implications for domestic industries, increase of the US merchandise trade
deficit by as much as \$45 billion by the end of 1983, with export sales about
\$35 billion lower and the import bill \$10 billion higher. For instance, in 1982,
it was already expected that, through its effects on
export and import volume, the appreciation would reduce real gross national
product by the end of 1983 to a level 1\%-1.5\% lower than the 3rd quarter 1980
pre-appreciation level [Feldman, 1982]. 
The appreciation of the US dollar from 1980-1984 was accompanied
by substantial decline in prices  
for the majority of manufactured imports from Canada, Germany, and Japan. However, for 
a substantial minority of prices, the imported items' dollar
prices rose absolutely and in relation to the general US price level.  The
median change was a price decline of 8\% for imports from Canada and Japan,
and a decrease of 28\% for goods from Germany [Fieleke, 1985].  
As a positive effect, 
the impact on the US inflation outlook 
was to improve it very significantly.
There is also evidence that the strong dollar in the first half of the 1980s
forced increased competition in U.S. product markets,
especially vis-a-vis continental Europe [Knetter, 1994].

As we explained in section 5, according to the rational expectation theory of speculative
bubbles, prices can be driven up by an underlying looming risk of 
a strong correction or crash. Such a possibility has been advocated as an explanation
for the strong appreciation of the US dollar from 1980 to early 1985 [Kaminsky and Peruga, 1991].
If the market believes that a discrete event may
occur when the event does not materialize for some time, 
this may have two consequences: drive price up and lead to an apparent 
inefficient predictive performance of forward exchange rates (forward 
and future contracts
are financial instruments that track closely ``spot'' prices as they embody the
best information on the expectation of market participants on near-term spot price
in the future). Indeed, from October 1979 to February 1985, forward
rates systematically underpredicted the strength of the US dollar. 
Two discrete events could be identified as governing market expectations 
[Kaminsky and Peruga, 1991]:
1) change in monetary regime in October 1979 and the resulting
private sector doubts about the Federal Reserve's commitment to 
lower money growth and inflation; 2) private sector anticipation of the
dollar's depreciation beginning in March 1985. i.e., anticipation of a 
strong correction, exactly as in the bubble-crash model of section 5.  
The corresponding
characteristic power law acceleration of bubbles decorated by log-periodic oscillations
is shown in figure \ref{DEMCHF1985}.

Expectations of future exchange rate have been shown to be excessive in the
posterior period from 1985.2 to 1986.4, indicating 
bandwagon effects at work and the
possibility of a rational speculative bubble [MacDonald and Torrance, 1988].
As usual before a strong correction or a crash, analysts were showing 
over-confidence and there were many reassuring talks of the absence of a significant danger of
collapse of the dollar, which has risen to unprecedented heights against
foreign currencies [Holmes, 1985]. On the long term however, it was clear
that such a strong dollar was unsustainable and there were indications 
that the dollar was overvalued, in particular because
foreign exchange markets generally hold that a
nation's currency can remain strong over the longer term, only if the
nation's current account is healthy: in constrast, for the first half of 1984, the US
current account suffered a seasonally adjusted deficit of around \$44.1 billion.
    
\begin{figure}
\begin{center}
\epsfig{file=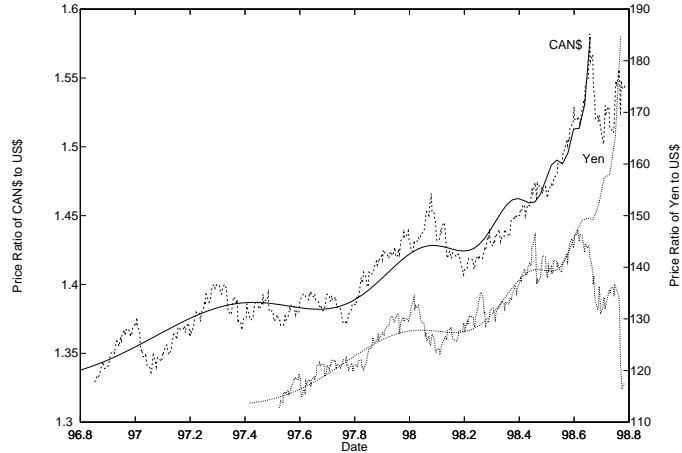,height=6cm}
\caption{\protect\label{forex98} The US dollar expressed in CAN\$ and YEN currencies 
prior to its drop starting in Aug. 1998. The fit with equation
(\ref{kflala}) to the two exchange rates gives $A_2\approx 1.62$,
$ B_2\approx  -0.22 $, $B_2 C \approx   -0.011 $, $m_2 \approx   0.26 $, $
t_c \approx   98.66  $, $ \phi \approx   -0.79 $, $ \omega \approx
8.2$ and $A_2\approx  207  $, $ B_2\approx  -85 $, $B_2 C \approx   2.8 $,
$m_2 \approx 0.19 $, $ t_c \approx   98.78  $, $ \phi \approx   -1.4 $,
$ \omega \approx  7.2$, respectively. Reproduced from [Johansen et al., 1999].
}
\end{center}
\end{figure}

A similar but somewhat attenuated bubble 
of the US dollar expressed respectively in Canadian dollar
and Japanese Yen, extending over slightly less than a year and bursting
in the summer of 1998, is shown in figure \ref{forex98}.
Paul Krugman has suggested that   
this run-up on the Yen and Canadian dollar, 
as well as the near collapse of U.S. financial markets at the end of the summer
of 1998, which is discussed
in the next section, are the un-wanted ``byproduct
of a vast get-richer-quick scheme by a handful of shadowy financial operators'' which
backfired [Krugman, 1998]. The remarkable quality of the fits of the
data with our theory does indeed give credence to the role of speculation, imitation and
herding, be them spontaneous, self-organized or manipulated in part. Actually,
Frankel and Froot [1988; 1990] have found that, 
over the period 1981-1985, the market shifted away from the
fundamentalists and toward the chartists or trend-followers.

\vskip-0.3cm
\subsection{The crash of August 1998}

From its top on mid-June 1998 ($1998.55$) to its bottom
on the first days of Sept. 1998 ($1998.67$), the US S\&P500 stock market
lost $19\%$. This ``slow'' crash and in particular the turbulent behavior
of the stock markets worldwide starting mid-august are widely associated with
and even attributed to the plunge of the Russian
financial markets, the devaluation of its currency and the default of
the government on its debts obligations. 

\begin{figure}
\begin{center}
\epsfig{file=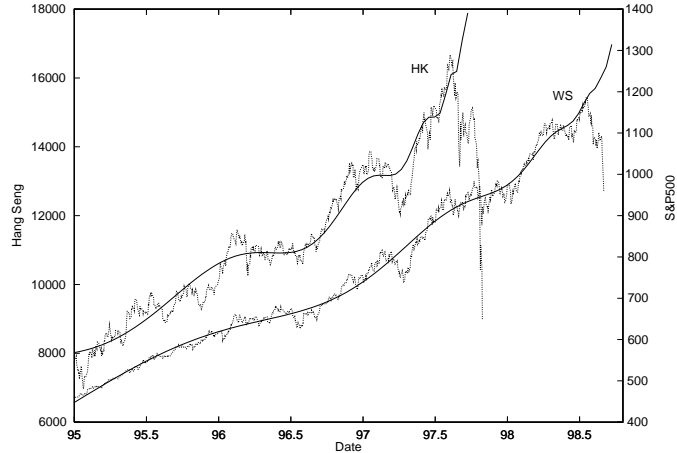,height=6cm}
\caption{\protect\label{hk97ws98} The Hang Seng index prior to the October 1997
crash on the Hong-Kong Stock Exchange already shown in figure \ref{hkbub3}
and the S\&P500 stock market index prior to the
crash on Wall Street in August 1998. 
The fit to the S\&P500 index is equation (\ref{kflala})
with $A_2 \approx  1321$, $ B_2\approx -402 $, $
B_2 C \approx  19.7 $, $m_2 \approx  0.60$, $
t_c\approx  98.72 $, $ \phi \approx  0.75$ and $ \omega \approx  6.4$. 
Reproduced from [Johansen et al., 1999].
}
\end{center}
\end{figure}

The analysis presented in figure
\ref{hk97ws98} suggests a different story: the Russian event may have been the
triggering factor but not the fundamental cause! One can observe clear
fingerprints of a kind of speculative herding, starting more than three years
before, with its characteristic power law acceleration decorated by log-periodic
oscillations. The table \ref{tablesummaryrus} gives a summary of the parameters
of the log-periodic power law fit to the main bubbles and crashed
discussed until now. The crash of Aug. 1998 is seen to fit nicely in the 
family of crashes with ``herding'' signatures.

This indicates that the stock market was again developing an unstable
bubble which would have culminated at some critical time $t_c \approx 1998.72$, close
to the end of Sept. 1998. According to the rational expectation bubble models
of section 5, the probability for a strong correction or a crash was increasing
as $t_c$ was approached, with a raising susceptibility to ``external'' perturbations, such as
news or financial difficulties occurring somewhere in the ``global village''. 
The Russian meltdown was just such a perturbation. What is remarkable is that 
the US market contained somehow the information of an upcoming instability through its
unsustainable accelerated growth and structures!
The financial world being an extremely complex system of interacting components, it
is not farfetched to imagine that Russia was led to take actions 
against its unsustainable debt policy at the time of a strongly
increasing concern by many about risks on investments made in developing countries.

The strong correction starting mid-august was not specific
to the US markets. Actually, it was much stronger in some other markets,
such as the German market. Indeed, within the period of only 9 months 
preceding July 1998, the German DAX index went up from about $3700$
to almost $6200$ and then quickly declined over less than one month to
below $4000$. Precursory log-periodic structures have been documented 
for this event over
the nine months preceding July 1998 [Drozdz et al., 1999], with the addition that
analogous log-periodic oscillations occurred also on smaller time scales
as precursors of smaller intermediate decreases, with 
similar prefered scaling ratio $\lambda$ at the various levels of resolution.
However, the reliability of these observations at smaller time scales established
by visual inspection in [Drozdz et al., 1999] remain to be established with
rigorous statistical tests.

\begin{table}[h]
\begin{center}
\begin{tabular}{|c|c|c|c|c|c|c|c|c|c|c|c|} \hline
crash & $t_c$ & $t_{max}$ & $t_{min}$ & drop & $m_2$ & $\omega$ &
$\lambda$ & $A_2$ & $B_2$ & $B_2 C$ & $Var$ \\ \hline
1929 (WS)&  $30.22$ & $29.65$ & $29.87$ & $47\%$ & $0.45$ & $7.9$ &
$2.2$ & $571$ & $-267$ & $14.3$ & $56$  \\ \hline
1985 (DEM) &  $85.20$ & $85.15$ & $85.30$ & $14\%$ & $0.28$ & $6.0$ &
$2.8$ & $3.88$ & $1.16$ & $-0.77$ & $0.0028$ \\ \hline
1985 (CHF) &  $85.19$ & $85.18$ & $85.30$ & $15\%$ & $0.36$ & $5.2$ &
$3.4$ & $3.10$ & $-0.86$ & $-0.055$ & $0.0012$ \\ \hline
1987 (WS)&  $87.74$ & $87.65$ & $87.80$ & $30\%$ & $0.33$ & $7.4$ &
$2.3$ & $411$ & $-165$ & $12.2$ & $36$ \\ \hline
1997 (H-K) &  $97.74$ & $97.60$ & $97.82$ & $46\%$ & $0.34$ & $7.5$ &
$2.3$ & $20077$ & $-8241$ & $-397$ & $190360$ \\ \hline
1998 (WS)&  $98.72$ & $98.55$ & $98.67$ & $19\%$ & $0.60$ & $6.4$ &
$2.7$ & $1321$ & $-402$ & $19.7$ & $375$ \\ \hline
1998 (YEN)& $98.78$ & $98.61$ & $98.77$ & $21\%$ & $0.19$ & $7.2$ &
$2.4$ & $207$ & $-84.5$ & $2.78$ & $17$\\ \hline
1998 (CAN\$)&$98.66$& $98.66$ & $98.71$ & $5.1\%$& $0.26$ & $8.2$ &
$2.2$ & $1.62$ & $-0.23$ & $-0.011$ & $0.00024$\\ \hline
1999 (IBM) & $99.56$ &$99.53$ & $99.81$  & $34\%$ & $0.24$ & $5.2$ & $3.4$ & & & &\\
\hline
2000 (P\&G) & $00.04$ & $00.04$ & $00.19$ & $54\%$ & $0.35$ & $6.6$ & $2.6$  & & & &\\
\hline
2000 (Nasdaq) & $00.34$ & $00.22$ & $00.29$ & $37\%$ & $0.27$ & $7.0$ & $2.4$  & & & &\\
\hline
\end{tabular}
\end{center}
\caption[]{Summary of the parameters
of the log-periodic power law fit to the main bubbles and crashes
discussed in this section (see figures \ref{Nasdaqcrash}, \ref{IBMcrashfit}
and \ref{PGcrashfit} below for the April 2000 crash on the Nasdaq and the 
two crashes on IBM and on Procter \& Gamble). $t_c$ is the critical time
predicted from the fit of each financial time series
to the equation (\ref{kflala}). The other
parameters of the fit are also shown. $\lambda=\exp \left[{2 \pi \over \omega}\right]$
is the prefered scaling ratio of the log-periodic oscillations.
The error $Var$ is the variance between
the data and the fit and has units of $price \times price$. 
Each fit is performed up to the time $t_{max}$ at which the market index 
achieved its highest maximum before the crash. $t_{min}$ is the time of the 
lowest point of the market disregarding smaller ``plateaus''. The percentage 
drop is calculated as the total loss from $t_{max}$ to $t_{min}$.
Reproduced from [Johansen et al., 1999].}
\label{tablesummaryrus}
\end{table}

\vskip-0.3cm
\subsection{The Nasdaq crash of April 2000}

In the last few years of the second Millenium,
there was a growing divergence in the stock market between``New Economy'' and
``Old Economy'' stocks, between technology and almost everything else. 
Over 1998 and 1999, stocks in the Standard \& Poor's technology sector have risen nearly
fourfold, while the S\&P500 index has gained just 50\%. And without technology, the
benchmark would be flat. In January 2000 alone, 30\% of net inflows into mutual funds
went to science and technology funds, versus just 8.7\% into S\&P500 index funds.
As a consequence, 
the average price-over-earning ratio P/E for Nasdaq companies was above 200
(corresponding to a ridiculous earning yield of $0.5\%$), a stellar
value above anything that serious economic valuation theory would consider reasonable.
It is worth recalling that the very same concept and wording of a ``New Economy''
was hot in the minds and mouths of investors in the 1920s and in the early 1960s
as already mentioned. In the 1920s, the new technologies
of the time were General Electric, ATT and other electric and communication companies,
and they also exhibited impressive price appreciations of the 
order of hundreds of percent in an 18 month
time intervals before the 1929 crash. In the early 1960s, the growth stocks were
in the new
electronic industry like Texas Instruments and Varian Associates, which expected to
exhibit a very fast rate of earning growth, were highly prized and far outdistanced
the standard blue-chip stocks. Many companies associated with the esoteric
high-tech of space travel and electronics sold in 1961 for over $200$ times
their previous year's earning. The ``tronics boom'',
as it was called, has actually remarkably 
similar features to the ``new economy'' boom preceeding the Oct. 1929 crash
or the ``new economy'' boom of the late 1990s, ending in the April 2000 crash on 
the Nasdaq index.

The Nasdaq Composite index
dropped precipiteously with a low of 3227 on April 17, 2000, corresponding to a 
cumulative loss of $37 \%$ counted from its all-time high of 5133 
reached on March 10, 2000. 
The Nasdaq Composite consists mainly of stock related to the so-called
``New Economy'', {\it i.e.}, the Internet, software, computer hardware,
telecommunication and so on. A main characteristic of these companies is that their
price-earning-ratios (P/E's), and even more so their price-dividend-ratios,
often came in three digits prior to the crash. Some companies, such as VA LINUX,  actually
had a {\it negative} Earning/Share of -1.68. Yet they were traded
around \$40 per share which is close to the price of Ford in early March
2000. Opposed to this, ``Old
Economy'' companies, such as Ford, General Motors and DaimlerChrysler, had
P/E $\approx 10$. The
difference between ``Old Economy'' and ``New Economy'' stocks is
thus the expectation of {\it future earnings} [Sornette, 2000b]:
investors, who expect an enormous increase in for example the sale of Internet and
computer related products rather than in car sales, are hence more
willing to invest in Cisco rather than in Ford notwithstanding the fact
that the earning-per-share of the former is much smaller than for the
later. For a similar price per share (approximately \$60 for Cisco and \$55
for Ford), the earning per share was \$0.37 for Cisco compared to
\$6.0 for Ford (Cisco has a total market capitalisation of \$395 billions
(close of April, 14, 2000) compared to \$63 billions for Ford).
In the standard fundamental valuation formula, in which the expected return of
a company is the sum of the dividend return and of the growth rate, ``New
Economy'' companies are supposed to compensate for their lack of present
earnings by a fantastic potential growth. In essence, this means that the bull market
observed in the Nasdaq in 1997-2000 has been fueled by
expectations of increasing future earnings rather than economic
fundamentals (and by the expectation that others will expect the same thing
and will help increase the capital gains): the
price-to-dividend ratio for a company such as Lucent Technologies (LU) with a
capitalization of over $\$300$ billions prior to its crash on the 5 Jan.
2000 is
over $900$ which means that you get a higher return on your checking
account(!)
unless the price of the stock increases. Opposed to this, an ``Old
Economy'' company
such as DaimlerChrysler gives a return which is more than thirty times higher.
Nevertheless, the shares of Lucent Technologies rose by more than $40\%$
during 1999 whereas the share of DaimlerChrysler declined by more than $40$\%
in the same period. The recent crashes of IBM, LU and
Procter \& Gamble (P\&G) correspond to a loss equivalent to many countries
state budget. And this is usually attributed to a ``business-as-usual''
corporate statement of a slightly revised smaller-than-expected earnings!

These considerations make it credible that it is the {\it expectation} of
future earnings and future capital gains rather than present
economic reality that motivates the average investor, 
thus creating a speculative bubble. It has also been proposed 
[Mauboussin and Hiler, 1999] that better
business models, the network effect, 
first-to-scale advantages and real options effect could account for the
apparent over-valuation, providing a sound justification for the high
prices of dot.com and other new-economy companies. 
These interesting views expounded in early 1999 were in synchrony with the 
bull market in 1999 and preceding years. They participated in the general optimistic view
and added to the herding strength. They seem less attractive in the context of the 
bearish phase of the Nasdaq market that has followed its crash in April 2000 
and which is still running more than two years later: Koller and Zane [2001] argue 
that the traditional triumvirate, 
earnings growth, inflation, and interest rates,
explains most of the growth and decay of US indices (while not excluding
the existence of a bubble of hugely capitalized new-technology companies).

Indeed, as already
stressed, history provides many examples
of bubbles, driven by unrealistic expectations of future earnings, followed by
crashes [White, 1996; Kindleberger, 2000]. The same basic ingredients are found repeatedly:
fueled by initially well-founded economic fundamentals, investors
develop a self-fulfilling enthusiasm by an imitative process or crowd
behavior that leads to an unsustainable accelerating overvaluation. 
We propose that the fundamental
origin of the crashes on
the U.S. markets in 1929, 1962, 1987, 1998 and 2000 belongs to the same category,
the difference being mainly in which sector the bubble was created: in 1929, it
was utilities; in 1962, it was the electronic sector;
in 1987, the bubble was supported by a general deregulation with
new private investors with high
expectations; in 1998, it was strong expectation
on investment opportunities in Russia that collapsed; in 2000,
it was the expectations on the Internet, telecommunication and so on.
that have fueled the bubble. 
However, sooner or later, investment values always revert to a
fundamental level based on real cash flows. 

\begin{figure}
\begin{center}
\epsfig{file=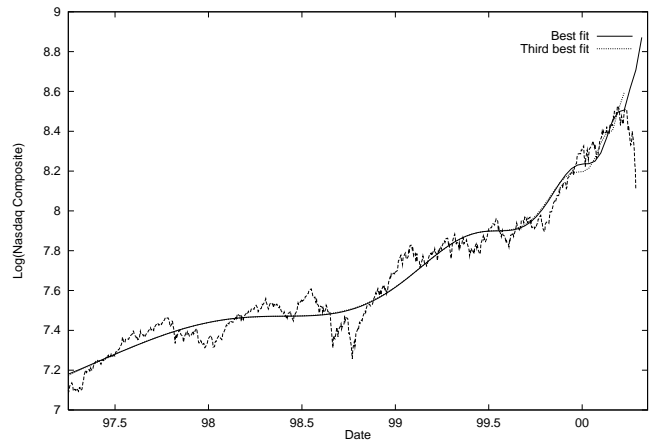,height=6cm}
\caption{\protect\label{Nasdaqcrash} Best (r.m.s. $\approx 0.061$) and third best
(r.m.s. $\approx 0.063$) fits with equation (\ref{kflala}) to the natural
logarithm of the Nasdaq Composite. The parameter
values of the fits are $A_2\approx 9.5$, $B_2\approx -1.7$, $B_2 C\approx 0.06$,
$m_2 \approx 0.27$, $t_c\approx 2000.33$, $\omega \approx 7.0$, $\phi \approx
-0.1$ and $A_2\approx 8.8$, $B_2\approx -1.1$, $B_2 C\approx 0.06$ ,$m_2\approx 0.39$,
$t_c\approx 2000.25$, $\omega \approx 6.5$, $\phi \approx -0.8$, respectively.
Reproduced from [Johansen and Sornette, 2000a].
}
\end{center}
\end{figure}

Figure \ref{Nasdaqcrash} shows the logarithm of the Nasdaq Composite fitted
with the log-periodic power law equation (\ref{kflala}). 
The data interval to fit was identified using the same
procedure as for the other crashes: the first point is the lowest value of
the index prior to the onset of the bubble and the last point is that of the
all-time high of the index. There exists some subtelty with respect to
identifying the onset of the bubble, the end of the bubble being objectively
defined as the date where the market reached its maximum. A bubble signifies
an acceleration of the price. In the case of Nasdaq, it tripled from 1990 to
1997. However, the increase was a factor 4 in the 3 years preceding
the current crash thus defining an ``inflection point'' in the index. In
general,
the identification of such an ``inflection point'' is quite straightforward on
the most liquid markets, whereas this is not always the case for the emergent markets
[Johansen and Sornette, 2001b].
With respect to details of the methodology of the fitting procedure, we refer
the reader to [Johansen et al., 1999].

Undoubtedly, observers and analysts have forged post-mortem stories
linking the April 2000 crash in part with the  effect of the crash of Microsoft Inc. resulting
from the breaking of negotiations during the weekend of April 1st
with the US federal government on the antitrust issue, as well as from many other
factors. Here, we interpret
the Nasdaq crash as the natural death of a speculative bubble, anti-trust
or not, the results presented here strongly suggesting that the bubble would
have collapsed anyway.
However, according to our analysis based on the probabilistic model of bubbles
described in sections 5 and 6,
the exact timing of the death of the bubble is not fully deterministic and allows
for stochastic influences, but within the remarkably tight bound of about
one month (except for the slow 1962 crash).

\begin{figure}
\parbox[l]{7.5cm}{
\epsfig{file=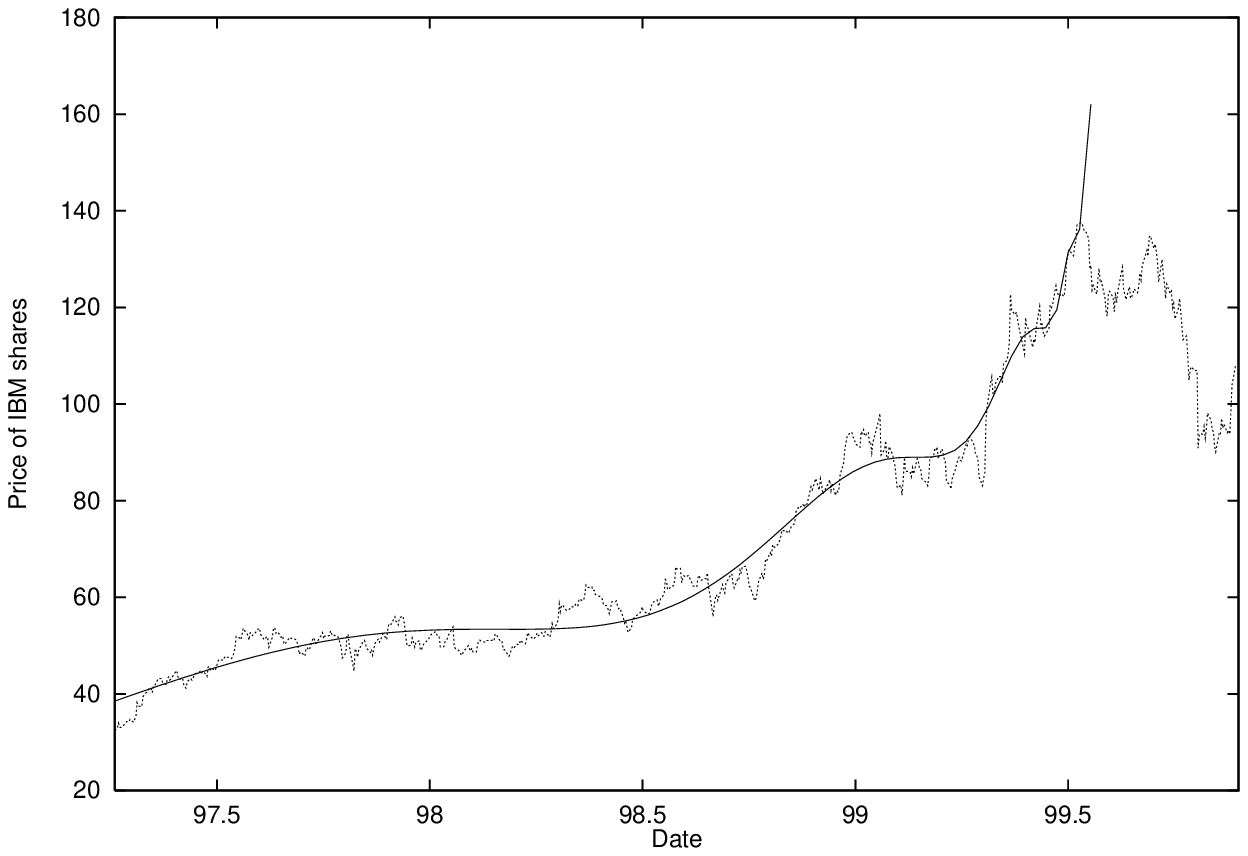,height=5cm,width=7cm}
\caption{\protect\label{IBMcrashfit} Best (r.m.s. $\approx 3.7$) fit with equation (\ref{kflala})
to the price of IBM shares. The parameter
values of the fits are $A_2 \approx 196, B_2\approx -132, B_2 C\approx -6.1,
m_2\approx 0.24, t_c\approx 99.56, \omega \approx 5.2$ and
$\phi \approx 0.1$.
Reproduced from [Johansen and Sornette, 2000a].}}
\hspace{5mm}
\parbox[r]{7.5cm}{
\epsfig{file=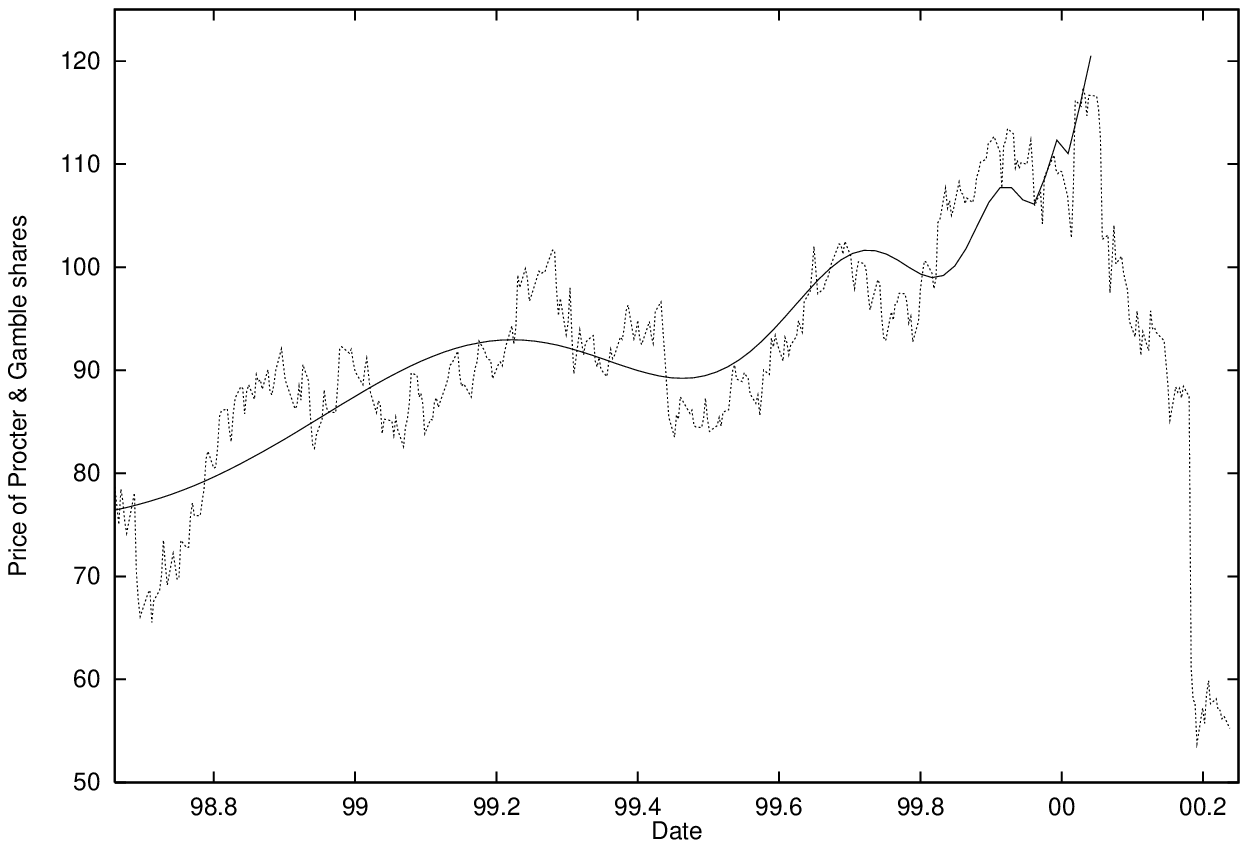,height=5cm,width=7cm}
\caption{\protect\label{PGcrashfit} Best (r.m.s. $\approx 4.3$) fit with equation (\ref{kflala})
to the price of Procter \& Gamble shares. The parameter values of the fit are
$A_2 \approx 124$, $B_2\approx -38$, $B_2 C\approx 4.8$ ,$m_2 \approx 0.35$, $t_c\approx
2000.04$, $\omega \approx 6.6$ and $\phi \approx -0.9$.
Reproduced from [Johansen and Sornette, 2000a].}}
\end{figure}

Log-periodic critical signatures can also be detected on individual
stocks as shown in figures \ref{IBMcrashfit} for IBM and \ref{PGcrashfit} for
Procter \& Gamble. These two figures offer a quantification of the precursory signals.
The signals are more noisy than for large indices but nevertheless clearly present.
There is a weaker degree of generality for individual stocks as
the valuation of a company is also a function of many other idiosynchratic factors
associated with the specific course of the company. Dealing with broad market 
indices averages out all these specificities to mainly keep track of the overall
market ``sentiment'' and direction. This is the main reason why the log-periodic
power law precursors are stronger and more significant for aggregated financial series
in comparison with individual assets. If speculation, imitation and herding become
at some time the strongest force driving the price of an asset, we should then 
expect the log-periodic power law signatures to emerge again strongly above
all the other idiosynchratic effects.

Generalization of this analysis to emergent markets, including
six Latin-American
stock market indices (Argentina, Brazil, Chile, Mexico, Peru and Venezuela) and
six Asian stock market
indices (Hong-Kong, Indonesia, Korea, Malaysia, Philippines and Thailand) has
been performed in [Johansen and Sornette, 2001b]. This work 
also discusses the existence of intermittent and strong correlation between
markets following major international events. 

\vskip-0.3cm
\subsection{``Anti-bubbles''}

We now summarize the
evidence that imitation between traders and their herding behavior not
only lead to speculative bubbles with accelerating over-valuations of
financial markets possibly followed by crashes,
but also to ``anti-bubbles'' with decelerating market devaluations
following all-time highs [Johansen and Sornette, 1999c]. 
There is thus a certain degree of symmetry of 
the speculative behavior
between the ``bull'' and ``bear'' market regimes. This behavior is documented
on the Japanese Nikkei stock index from  1. Jan 1990 until 31 Dec. 1998,
on the Gold future prices after 1980, and on the recent 
behavior of the US S\&P500 index from mid-2000 to Aug. 2002, all
of them after their all-time highs.

The question we ask is whether the cooperative herding behavior of
traders might also produce market evolutions that are symmetric
to the accelerating speculative
bubbles often ending in crashes. This symmetry is performed with respect to
a time inversion
around a critical time $t_c$ such that $t_c-t$ for $t<t_c$ is changed into
$t-t_c$ for $t>t_c$.
This symmetry suggests looking at {\it decelerating} devaluations instead of
accelerating valuations. A related observation has been reported 
in figure \ref{fig2first} in relation to the Oct. 1987 crash 
showing that the implied volatility of traded
options has relaxed {\it after} the Oct. 1987 crash to its
long-term value, from a maximum at the time of the crash, according 
to a decaying power law with
decelerating log-periodic oscillations. It is this type of
behavior that we document now but for real prices.

The critical time $t_c$ then corresponds to the culmination of
the market, with either a power law increase with accelerating
log-periodic oscillations preceding it or a power law decrease with decelerating
log-periodic oscillations after it. In
the Russian market, both structures appear
simultaneously for the same $t_c$ [Johansen and Sornette, 1999c]. 
This is however a rather rare occurrence, probably
because accelerating
markets with log-periodicity almost inevitably end-up in a crash, a market
rupture that
thus breaks down the symmetry ($t_c-t$ for $t<t_c$ into  $t-t_c$ for
$t>t_c$). Herding behavior can occur and progressively
weaken from a maximum in ``bearish'' (decreasing) market phases, even if the
preceding ``bullish'' phase ending at $t_c$ was not characterised by a
strengthening imitation. The symmetry
is thus statistical or global in general and holds in the ensemble
rather than for each single case individually. 

\vskip-0.3cm
\subsubsection{The ``bearish'' regime on the  Nikkei starting from 1st Jan. 1990}

\begin{figure}
\begin{center}
\epsfig{file=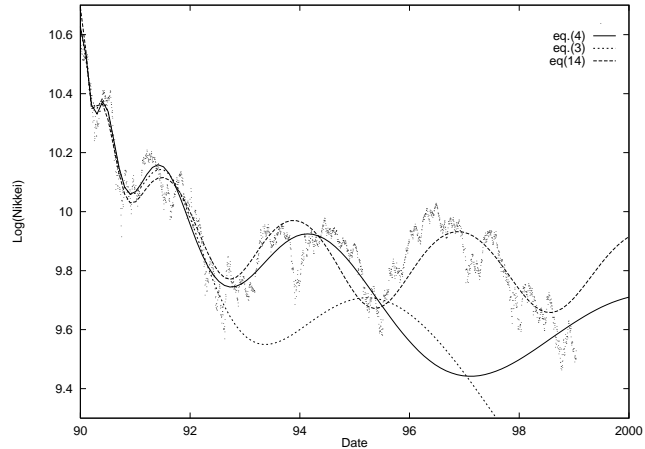,height=6cm}
\caption{\protect\label{decnikkei} Natural logarithm of the Nikkei stock market
index after the start of the decline from 1st Jan 1990 until 31 Dec. 1998. 
The dotted line is the simple log-periodic
formula (\ref{kflala}) used to fit adequately the interval of
$\approx 2.6$ years starting from 1st Jan 1990.
The continuous line is the improved nonlinear log-periodic formula 
developed in [Sornette and Johansen, 1997] and 
already used for the 1929 and 1987 crashes over 8 years
of data. It is used to fit adequately the interval of $\approx 5.5$ years
starting from 1st Jan 1990.
The dashed line is an extension of the previous nonlinear log-periodic formula
to the next-order of description which was developed in [Johansen and Sornette, 1999c]
and is used to fit adequately the interval of $\approx 9$ years
starting from 1st Jan 1990 to Dec. 1998.  Reproduced from [Johansen and Sornette, 1999c].
}
\end{center}
\end{figure}

The most recent example of a genuine long-term depression comes from Japan,
where the Nikkei has decreased by more than $60$ \% in the 12 years following
the all-time high of 31 Dec. 1989. In figure \ref{decnikkei}, we see (the
logarithm of)  the Nikkei from 1 Jan. 1990
until 31 Dec. 1998. The three fits, shown as the undulating lines,
use three mathematical expressions of
increasing sophistication: the dotted line is the simple log-periodic
formula (\ref{kflala}); the continuous line is the improved nonlinear log-periodic formula 
developed in [Sornette and Johansen, 1997] and already used for the 
1929 and 1987 crashes over 8 years
of data; the dashed line is an extension of the previous nonlinear log-periodic formula
to the next-order of description which was developed in [Johansen and Sornette, 1999c].
This last most sophisticated mathematical formula 
predicts the transition from the log-frequency 
$\omega_1$ close to $t_c$ to $\omega_1 + \omega_2$ for $T_1 < \tau < T_2$ and 
to the log-frequency $\omega_1 + \omega_2 + \omega_3$ for $T_2
< \tau$.  Using indices $1$, $2$ and $3$ respectively for 
the simplest to the most sophisticated formulas, 
the parameter values of the first fit of the Nikkei are $A_1 \approx
10.7 , B_1\approx -0.54 , B_1 C_1\approx  -0.11 , m_1 \approx  0.47 , t_c
\approx  89.99 , \phi_1 \approx -0.86 , \omega_1 \approx  4.9$ for equation
(\ref{kflala}).
The  parameter values of the second fit of the Nikkei are $A_2 \approx
10.8 ,  B_2 \approx  -0.70 , B_2 C_2 \approx  -0.11 , m_2 \approx  0.41 , t_c
\approx  89.97 ,   \phi_2 \approx  0.14 , \omega_1 \approx  4.8 , T_1
\approx 9.5$ years, $\omega_2 \approx 4.9$.
The third fit uses the entire time interval and is performed by adjusting only
$T_1$, $T_2$,
$\omega_2$ and $\omega_3$, while $m_3=m_2$, $t_c$ and $\omega_1$ are
fixed at the values obtained from the previous fit. The values obtained for
these four parameters are $T_1 \approx 4.3$ years, $T_2 \approx 7.8$ years,
$\omega_2 \approx -3.1 $ and $T_2 \approx 23$ years. Note that
the values obtained for the two time scales $T_1$ and $T_2$
confirms their ranking. This last fit predicts a change of regime and that
the Nikkei should increase in 1999. 

Not only do the first two equations agree remarkably well with respect to the parameter
values produced by the fits, but they are also in good agreement with previous
results obtained from stock market and Forex bubbles with respect to the
values of exponent $m_2$. What lends credibility to
the fit with the most sophisticated formula is that, despite its complex form,
we get values for the two cross-over time scales $T_1$, $T_2$
which correspond to what is expected from the ranking and from
the 9 year interval of the data. We refer to [Johansen and Sornette, 19999c]
for a detailed and rather technical discussion.

The prediction summarized by figure \ref{decnikkei} was made public
on Jan. 25, 1999 by posting a preprint on the Los Alamos www
internet server, see 
http://xxx.lanl.gov/abs/cond-mat/9901268. The preprint was later published
as [Johansen and Sornette, 1999c]. The prediction stated that
the Nikkei index should
recover from its 14 year low ($13232.74$ on Jan. 5, 1999) and reach 
$\approx 20500$ a year later corresponding to an increase in the index of 
$\approx 50\%$. This prediction was mentioned in a 
wide-circulation journal in physical sciences which appeared 
in May 1999 [Stauffer, 1999].

\begin{figure}
\begin{center}
\epsfig{file=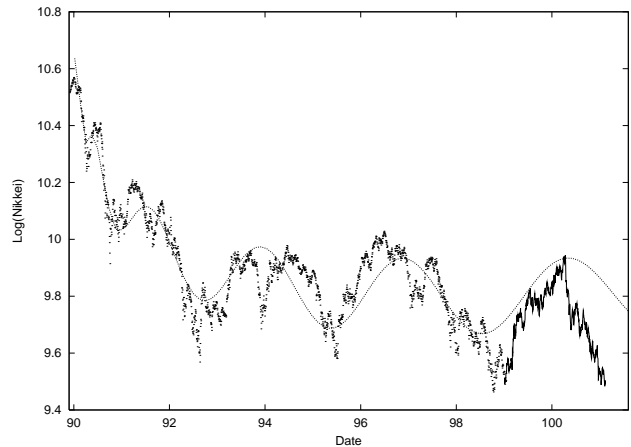,height=6cm}
\caption{\protect\label{decnikkeinew} Natural logarithm of the Nikkei stock market
index after the start of the decline from 1st Jan 1990 until Feb. 2001. 
The continuous smooth line is the extended nonlinear log-periodic formula
which was developed in [Johansen and Sornette, 1999c]
and is used to fit adequately the interval of $\approx 9$ years
starting from 1st Jan 1990. The Nikkei data is separated in two parts.
The dotted line shows the data used to perform the fit with formula 
developed in [Johansen and Sornette, 1999c]
and to issue the prediction in Jan. 1999 (see figure \protect\ref{decnikkei}).
Its continuation as a continuous line gives the behavior of the Nikkei index
after the prediction has been made. Reproduced from [Johansen and Sornette, 2000b].
}
\end{center}
\end{figure}

In figure \ref{decnikkeinew}, the actual and
predicted evolution of the Nikkei over 1999 and later are compared [Johansen and Sornette, 2000b]. 
Not only did the 
Nikkei experience a trend reversal as predicted, but it has also followed 
the quantitative prediction with rather impressive precision. In particular,
the prediction of the $50\%$ increase at the end of 1999 is validated accurately.
The prediction of another trend reversal is also accurately predicted, with
the correct time for the reversal occuring beginning of 2000: the predicted
maximum and observed one match closely.
It is important to note that the error between the curve 
and the data has not grown after the last point used in the fit over 1999. This tells 
us that the prediction has performed well for more than one year. 
Furthermore, since the
relative error between the fit and the data is within $\pm 2\%$ over a
time period of 10 years, not only has the prediction performed well, but also
the underlying model. 

The fulfilling of this prediction is even more
remarkable than the comparison between the curve and the data indicates, because
it included a change of trend: at the time when the
prediction was issued, the market was declining and showed no tendency to
increase. Many economists were at that time very pessimistic and could not 
envision when Japan and its market would rebounce. For instance, 
P. Krugman wrote July 14, 1998 in the Shizuoka Shimbun
at the time of the banking scandal ``the central problem with Japan right now 
is that there just is not enough demand to go around - that consumers and 
corporations are saving too much and borrowing too little... So seizing these 
banks and putting them under more responsible management is, if anything, 
going to further reduce spending; it certainly will not in and of itself 
stimulate the economy... But at best this will get the economy back to where 
it was a year or two ago - that is, depressed, but not actually plunging.''
Then, in the Financial Times, January, 20th, 1999, P. Krugman wrote in an
article entitled ``Japan heads for the edge'' the following: ``...the story 
is starting to look like a tragedy. A great economy, which does not deserve 
or need to be in a slump at all, is heading for the edge of the cliff -- and 
its drivers refuse to turn the wheel.'' In a poll of thirty economists performed 
by Reuters (one of the major news and finance data provider in the world)
in October 1998 reported in Indian Express on the 15 Oct. (see
http://www.indian-express.com/fe/daily/19981016/28955054.html),
only two economists predicted growth for
the fiscal year of 1998-99. For the year 1999-2000 the prediction was a 
meager 0.1\% growth. This majority of economists said that
``a vicious cycle in the economy was unlikely to disappear any
time soon as they expected little help from the government's economic stimulus
measures... Economists blamed moribund domestic demand, falling prices, weak capital spending
and problems in the bad-loan laden banking sector for dragging down the economy.''

It is in this context that we predicted
an approximately $50\%$ increase of the market in the 
12 months following Jan. 1999,
assuming that the Nikkei would stay within the error-bars of
the fit. Predictions of trend reversals is 
noteworthy difficult and unreliable, especially in the linear framework of 
auto-regressive models used in standard economic analyses. The present 
nonlinear framework is well-adapted to the forecasting of change of trends, 
which constitutes by far the most difficult challenge posed to forecasters.
Here, we refer to our prediction of a trend reversal 
within the strict confine of our extended formula: trends are limited
periods of times when the oscillatory behavior shown in figure \ref{decnikkeinew}
is monotonous. A change of trend thus corresponds to crossing a local maximum
or minimum of the oscillations. Our formula seems to have predicted
two changes of trends, bearish to bullish at the beginning of 1999
and bullish to bearish at the beginning of 2000.

\vskip-0.3cm
\subsubsection{The gold deflation price starting mid-1980}

\begin{figure}
\begin{center}
\epsfig{file=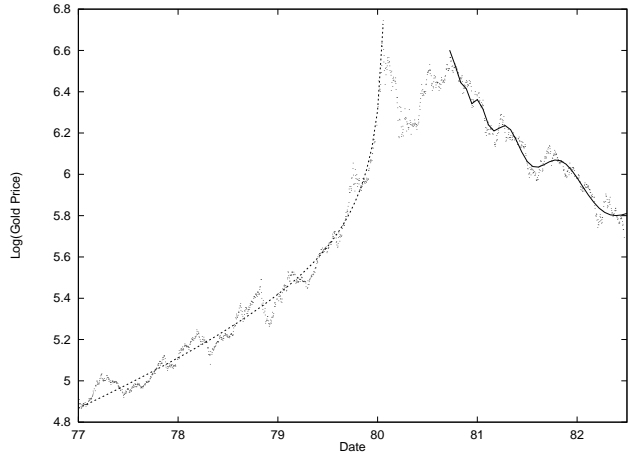,height=6cm}
\caption{\protect\label{gold} Natural logarithm of the gold 100 Oz Future
price in US\$ showing a power law acceleration followed by a 
decline of the price in the early eighties. The line
after the peak is expression (\ref{kflala}) fitted over an interval of
$\approx 2$ years. Reproduced from [Johansen and Sornette, 1999c].
}
\end{center}
\end{figure}

Another example of log-periodic decay is that of the Gold price after the
burst of the  bubble in 1980 as shown in figure \ref{gold}. The bubble has
an {\it average} power law acceleration as shown in the figure but
{\it without} any visible log-periodic structure. A pure power law
fit will however not ``lock in'' on the true date of the crash, but insists on an earlier
date than the last data point. This suggests that the  behavior of the price
might be different in some sense in the last few weeks prior to the burst of
the bubble. The continuous line before the peak is
expression (\ref{kflala}) fitted over an interval of $\approx 3$ years.
The parameter values of this fit are $A_2\approx 8.5 ,
B_2 \approx  -111 , B_2 C\approx  -110 , m_2 \approx  0.41 , t_c \approx  80.08 ,
\phi \approx  -3.0 , \omega \approx  0.05$.
The price of gold after its peak is fitted by expression (\ref{kflala})
and the result is shown as the undulating continuous line.
Again, we obtain a reasonable agreement with previous results for the exponent
$m_2$ with a good prefered scaling ratio $\lambda \approx 1.9$.
The strength of the log-periodic oscillations compared to the leading
behavior is $\approx 10$\%. The parameter values of the fit in this
anti-bubble regime are $A_2\approx 6.7 ,
B_2 \approx  -0.69 , B_2 C\approx  0.06 , m_2 \approx  0.45 , t_c \approx  80.69 ,
\phi \approx  1.4 , \omega \approx  9.8$. 

\vskip-0.3cm
\subsubsection{ The US 2000-2002 Market Descent: How Much Longer and Deeper?}

\begin{figure}
\begin{center}
\epsfig{file=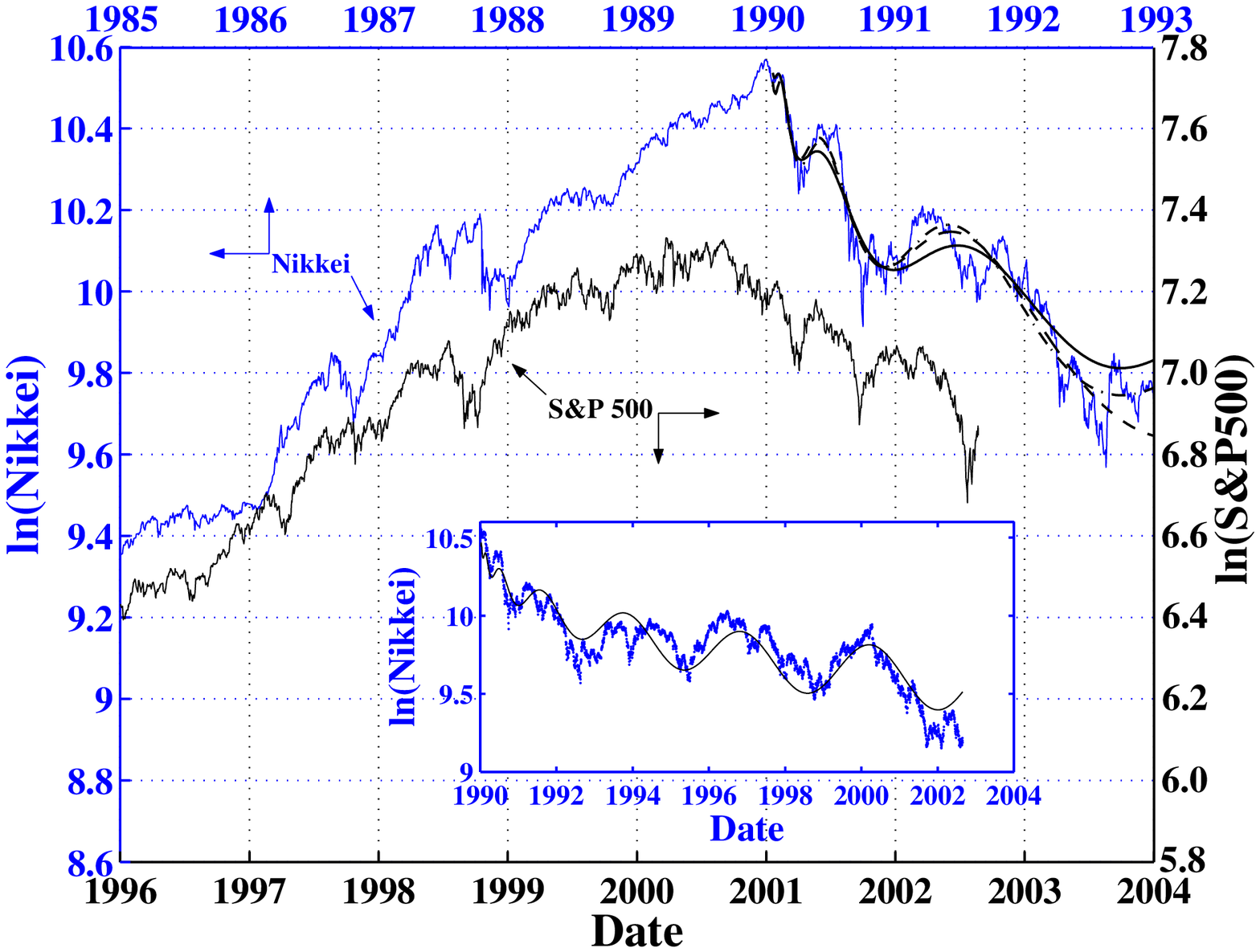,width=13cm, height=10cm}
\end{center}
\caption{Comparison between the evolutions of the US S\&P500 index
from 1996 till August, 24, 2002 (bottom and right axes) and the
Japanese Nikkei index from 1985 to 1993 (top and left axes). 
The years are written on the 
horizontal axis (and marked by a tick on the axis) where January 1
of that year occurs. The
dashed line is the simple log-periodic formula (\ref{lp2})
fitted to the Nikkei index (with $t_c-t$ replaced
by $t-t_c$). The data used in this fit goes from
01-Jan-1990 to 01-Jul-1992 [Johansen and Sornette, 1999c]. The parameter values
are $t_c=$ 28-Dec-1989, $\alpha=0.38$, $\omega=5.0$, $\phi=2.59$,
$A=10.76$, $B=-0.067$ and $C=-0.011$. The root-mean-square residue is $\chi =
0.0535$. The dash-dotted line is the improved nonlinear
log-periodic formula developed in [Sornette and Johansen, 1997]
fitted to the Nikkei index. The Nikkei index data used in this fit
goes from 01-Jan-1990 to 01-Jul-1995 [Johansen and Sornette, 1999c]. The
parameter values are $t_c=$ 27-Dec-1989, $\alpha=0.38$,
$\omega=4.8$, $\phi=6.27$, $\Delta_t=6954$, $\Delta_\omega=6.5$,
$A=10.77$, $B=-0.070$, $C=0.012$. The root-mean-square residue is $\chi =
0.0603$. The continuous line is the fit of the Nikkei index with
the third-order formula developed in [Johansen and Sornette, 1999c]. The
Nikkei index data used in the fit goes from 01-Jan-1990 to
31-Dec-2000. The fit is performed by fixing $t_c$, $\alpha$ and
$\omega$ at the values obtained from the second-order fit and
adjusting only $\Delta_t$, $\Delta'_t$, $\Delta_\omega$,
$\Delta'_\omega$ and $\phi$. The parameter values are
$\Delta_t=1696$, $\Delta'_t=5146$, $\Delta_\omega=-1.7$,
$\Delta'_\omega=40$, $\phi=6.27$, $A=10.86$, $B=-0.090$,
$C=-0.0095$. The root-mean-square residue of the fit is $\chi = 0.0867$. In the three fits,
$A$, $B$ and $C$ are slaved to the other variables by multiplier
approach in each iteration of optimization search. The inset shows
the 13-year Nikkei anti-bubble with the fit with the third-order
formula over these 13 years shown as the continuous line. The parameter values 
slighly different:
$\Delta_t=52414$, $\Delta'_t= 17425$, $\Delta_\omega=23.7$,
$\Delta'_\omega=127.5$, $\phi=5.57$, $A=10.57$, $B=-0.045$,
$C=0.0087$. The root-mean-square residue of the fit is $\chi = 0.1101$. In all the fits,
times are expressed in units of days, in contrast with the yearly
unit used in [Johansen and Sornette, 1999c]). Thus, the parameters $B$ and $C$
are different since they are unit-dependent, while all the other
parameters are independent of the units. Reproduced from
[Sornette and Zhou, 2002].} \label{Fig:NikkeiSP}
\end{figure}

Sornette and Zhou [2002] have recently analyzed the
remarkable similarity in the behavior of the US S\&P500 index from 1996 to August 2002 and 
of the Japanese Nikkei index from 1985 to 1992, corresponding
to an 11 years shift. In particular, the structure of the price trajectories
in the bearish or anti-bubble phases are strikingly similar, as seen in figure
\ref{Fig:NikkeiSP}.

\begin{figure}
\begin{center}
\epsfig{file=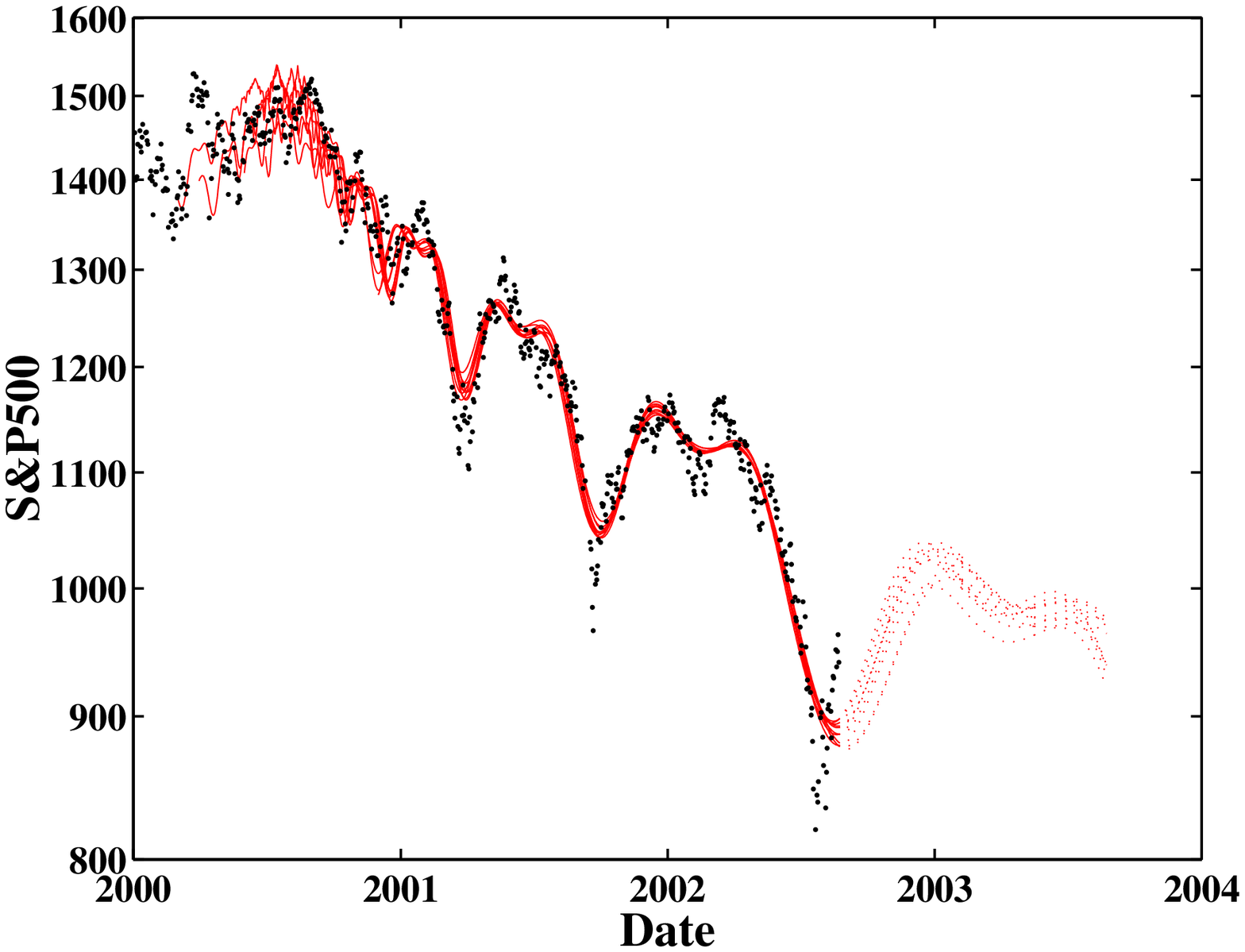,width=10cm, height=8cm}
\end{center}
\caption{Fitted trajectories using Eq.~(\ref{lp2})
(with $t_c-t$ replaced by $|t-t_c|$), each curve corresponding to a
different starting time from Mar-01-2000 to Dec-01-2000 with
one month interval. The different fits are obtained
as a sensitivity test with respect to the starting time of the anti-bubble
which is consistently found to start at $t_c \approx$ July 15-August 15, 2000.
The dotted lines show the predicted future trajectories. One sees that
the fits are quite robust with respect to different starting date
$t_{\mathtt{start}}$ from Mar-01-2000 to Dec-01-2000. Reproduced from
[Sornette and Zhou, 2002].}
\label{Fig:AllFit4}
\end{figure}

Sornette and Zhou [2002] have performed a battery of 
tests, starting with parametric fits of the index with two
log-periodic power law formulas, followed by the so-called Shank's transformation
applied to characteristic times. They also carried out two spectral analysis,
the Lomb periodogram applied to the parametrically detrended index
and the non-parametric $(H,q)$-analysis of fractal signals
[Zhou and Sornette, 2002b,c]. These 
different approaches complement each other and confirm the presence
of a very strong log-periodic structures. A rather novel feature is
the detection of a
significant second-order harmonic which provides a statistically significant
improvement of the description of the data by the theory, as tested
using the statistical theory of nested hypotheses. The 
description of the S\&P500 index since mid-2000 to end of Aug. 2002
based on the combination of the first and second log-periodic harmonics
is shown in figure \ref{Fig:AllFit4}. 

In the next two years, Sornette and Zhou [2002]
predict an overall continuation of
the bearish phase, punctuated by local rallies; specifically, they
predict an overall increasing market until the end of the year 2002 or 
until the first quarter of 2003;
they predict a severe following descent (with maybe one or two 
severe ups and downs in the middle) which stops during the first 
semester of 2004. Beyond this, they can not be very certain due to the possible
effect of additional nonlinear collective effects and of a real departure from the anti-bubble
regime. The similarities between the two stock market
indices may reflect deeper similarities between the fundamentals of
two economies which both went through over-valuation with strong speculative phases
preceding the transition to bearish phases 
characterized by a surprising number of bad surprises (bad loans for Japan
and accounting frauds for the US) sapping investors' confidence.

\section{Synthesis}

\vskip-0.3cm
\subsection{``Emergent'' behavior of the stock market}
    
In this paper, we have synthesized a large body of evidence in 
favor of the hypothesis
that large stock market crashes are analogous to
critical points studied in the statistical physics community
in relation to magnetism, melting, and so on. Our main assumption is the
existence of a cooperative behavior of traders imitating each other
described in sections 5 and 6. A general result of the theory is the existence of
log-periodic structures decorating the time evolution of the system.
The main point is that the market anticipates the crash in a
subtle self-organized and cooperative fashion, hence releasing precursory
``fingerprints'' observable in the stock market prices. In other words, this
implies that market prices contain information on impending crashes. If
the traders were to learn how to decipher and use this information, they
would act on it and
on the knowledge that others act on it, nevertheless the crashes would still probably
happen. Our results suggest a weaker form of the ``weak efficient market
hypothesis'' [Fama, 1991], according to which the market prices contain, in
addition to the information generally available to all, subtle information
formed by the global market that most or all individual traders have not yet
learned to decipher and use. Instead of the usual interpretation of the
efficient market hypothesis in which traders extract and incorporate
consciously (by their action) all information contained in the market prices,
we propose that the market as a whole can exhibit an ``emergent'' behavior not
shared by any of its constituants. In other words, we have in mind the process of
the emergence of intelligent behaviors at a macroscopic scale that
individuals at
the  microscopic scale cannot perceive. This process has been discussed in
biology for instance in animal populations such as ant colonies 
[Wilson, 1971;  Holldobler and Wilson, 1994] or in
connection with the emergence of conciousness [Anderson et al., 1988; Holland, 1992].

Let us mention another realization of this concept, which is found in the
information contained in option prices on the
fluctuations of their underlying asset. Despite the fact
that the prices do not follow geometrical brownian motion, whose existence is a
prerequisite for most option pricing models, traders have apparently
adapted to empirically incorporate subtle information in the correlation of
price distributions with fat tails [Potters et al., 1998].
In this case and in contrast to the crashes, the traders have had time to adapt.
The reason is probably that traders have been exposed for decades to option
trading in which the characteristic time scale for option lifetime is in
the range
of month to years at most. This is sufficient for an extensive learning
process to
occur. In contrast, only a few great crashes occur typically during a lifetime and this
is certainly not enough to teach traders how to adapt to them. The situation
may be compared to the ecology of biological species which constantly 
strive to adapt. By the forces of evolution, they generally succeed to survive
by adaptation under
slowly varying constraints. In constrast, life
may exhibit successions of massive extinctions
and booms probably associated with dramatically fast-occuring
events, such as meteorite impacts and
massive volcanic eruptions. The response of a
complex system to such extreme events is a problem of
outstanding importance that is just beginning to be studied 
[Commission on Physical Sciences, Mathematics, and Applications, 1990].

Most previous models proposed for crashes have pondered the possible mechanisms
to explain the collapse of the price at very short time scales. Here in
contrast, we propose that the underlying cause of the crash must be
searched years before it in the progressive accelerating ascent of the market
price, reflecting an increasing build-up of the market cooperativity. From
that point of view, the specific manner by which prices collapsed is not
of real importance since, according to the concept of the critical point, any
small disturbance or process may have triggered the instability, once ripe. The
intrinsic divergence of the sensitivity and the growing instability of the
market
close to a critical point might explain why attempts to unravel the local origin
of the crash have been so diverse. Essentially all would work once the system is
ripe. Our view is that the crash has an endogeneous origin and that exogeneous
shocks only serve as triggering factors. We propose that the origin of the
crash is
much more subtle and is constructed progressively by the market as a whole. In
this sense, this could be termed a systemic instability.

\vskip-0.3cm
\subsection{Implications for mitigations of crises}

Economists,
J.E. Stiglitz and recently P. Krugman in particular 
as well as financier Soros, have argued that markets should
not be left completely alone. The mantra of the free-market purists requiring that
markets should be totally free may not always be the best solution,
because it overlooks two key problems: (1) the tendency of 
investors to develop strategies that may destabilize markets in
a fundamental way and (2) the non-instantenous adjustment of 
possible imbalance between countries. Financier George
Soros has argued that real world international
financial markets are inherently volatile and unstable since 
``market participants
are trying to discount a future that is itself shaped by market expectations''. 
This question is of course at the center of the debate on whether
local and global markets are able to stabilize on their own after a crisis
such as the Asian crisis which started in 1997. In this example, to justify the intervention
of the IMF (international monetary fund), Treasury Secretary Rubin warned 
in Jan. 1998 that global markets would not be able to stabilize in
Asia on their own, and that a strong role on the part of the IMF and other
international institutions, and governments, was necessary, lest the crisis spread
to other emerging markets in Latin America and Eastern Europe.

The following analogy with forest fires is useful to illustrate
the nature of the problem. In many areas
around the world, the dry season sees 
numerous large wildfires, sometimes with deaths of 
firefighters and other people, the destruction of many structures
and of large forests.
It is widely accepted that livestock grazing, timber harvesting, and fire
suppression over the past century have led to unnatural conditions -- excessive
biomass (too many trees without sufficient biodiversity
and dead woody material) and altered species mix -- in
the pine forests of the West of the U.S.A., 
in the Mediterraneen countries and elsewhere. 
These conditions make the forests more susceptible
to drought, insect and disease epidemics, and other forest-wide catastrophes
and in particular large wildfires [Gorte, 1995]. 
Interest in fuel management, to reduce fire control costs and damages, has thus been
renewed with the numerous, destructive wildfires spread across the West of the
U.S.A. The most-often used technique of fuel management is fire
suppression.
Recent reviews comparing Southern California on the one hand, where management is active
since 1900,
and Baja California (north of Mexico) on the other hand where
management is essentially absent (a ``let-burn'' strategy)
highlight a remarkable fact [Minnich and Chou, 1997; Moreno, 1998]: 
only small and relatively moderate patches of fires
occur in Baja California, compared to a wide distribution of fire sizes in
Southern California including huge destructive fires. The selective 
elimination of small fires (those that can be controlled) in normal weather in
Southern California  restricts large fires to extreme weather episodes, 
a process that encourages broad-scale high spread rates and intensities.
It is found that the danger of fire suppression is the inevitable development
of coarse-scale bush fuel patchiness and large instance fires in contradistinction
with the natural self-organization of small patchiness in left-burn areas.
Taken at face value,
the ``let-burn'' theory seems paradoxically the correct strategy which maximizes
the protection of property and of resources, at minimal cost.

This conclusion seems to be correct when the fuel is left on 
its own to self-organize
in a way consistent with the dynamics of fires. In other words, the fuel-fire
constitutes a complex non-linear system with negative and positive feedbacks that
may be close to optimal: 
more fuel favors fire; fires decreases the instantaneous
level of fuel but may accelerate its future production; many small fires
create natural barriers for the development and extension of large fires;
fires produce rich nutrients in the soil; fires
have other benefits, for instance,
a few species, notably lodgepole pine and jack pine,
are serotinous -- their cones will only open and spread their seeds when they
have been exposed to the heat of a wildfire. 
The possibility for complex nonlinear systems to find the ``optimal'' or
to be close to the optimal solution have been stressed before in several contexts
[Crutchfield and Mitchell, 1995; Miltenberger et al., 1993; Sornette et al., 1994]. 
Let us mention for instance
a model of fault networks interacting through the elastic deformation
of the crust and rupturing during earthquakes which finds
that faults are the optimal
geometrical structures accomodating the tectonic deformation: they result
from a global mathematical optimization problem that the dynamics of the system solves
in an analog computation, i.e., by following its self-organizing
dynamics (as opposed to digital computation performed by digital
computers). One of the notable levels of organization is called self-organized
criticality [Bak, 1996; Sornette, 2000a] and has been applied in 
particular to explain forest fire distributions [Malamud et al., 1998].

Baja California could be a representative of this 
self-organized regime of the fuel-fire complex left to itself, leading
to many small fires and few big ones. Southern California could
illustrate the situation where interference both in the production of fuel
and also in its combustion by fires (by trying to stop fires)
leads to a very broad distribution with many
small and moderate controlled fires and too many uncontrollable very large ones.

Where do stock markets stand in this picture? The proponents of 
the ``left-alone'' approach could get ammunition from the 
Baja-Southern California comparison, but they would forget an essential element:
stock markets and economies are more like Southern California than Baja California.
They are not isolated. Even if no government or regulation interfere, 
they are ``forced'' by 
many external economic, political, climatic
influences that impact them and on which they may also have some impact. If the example
of the wildland fires has something to teach us, it is that we must
incorporate in our understanding
both the self-organizing dynamics of the fuel-fire complex
as well as the different exogeneous sources of randomness (weather and wind
regimes, natural lightning strike distribution, and so on.).

The question of whether some regulation could be useful is translated into
whether Southern California fires would be better left alone. Since the
management approach fails to function fully satisfactorily, one may wonder
whether the let-burn scenario would not be better. This has in fact been
implemented in Yellowstone park as the ``let-burn'' policy but
was abandoned following the huge Yellowstone fires of 1988. Even the
``leave-burn'' strategy may turn out to be unrealistic from a societal point-of-view
because allowing a specific fire to burn down may lead to socially unbearable risks
or emotional sensitivity, often
discounted over a very short time horizon (as opposed to the long-term view
of land management implicit in the left-burn strategy).

We suggest that the most momentous events in stock markets, the
large financial crashes, can indeed be seen as the response of a self-organized
system forced by a multitude of external factors in the presence of
regulations. The external forcing is
an essential element to consider and it modifies the perspective on the ``left-alone''
scenario. For instance, during the recent Asian crises,
the International Monetary Fund and the 
U.S. government considered that
controls on the international flow of capital were counterproductive or
impractical. J.E. Stiglitz, 
the chief economist of the IMF until 2000, has argued that 
in some cases it was justified to restrict short-term
flows of money in and out of a developing economy and that 
industrialized
countries sometimes pushed developing nations too fast to deregulate their
financial systems.
The challenge remains, as always, to
encourage and work with countries that are ready and able
to implement strong corrective actions and to cooperate
toward finding the financial solutions best suited to the
needs of the individual case and the broader functioning
of the global financial system when difficulties arise [Checki and Stern, 2000].

Another important issue concerns the endogeneous versus exogeneous nature
of shocks. Sornette et al. (2002) have shown that it is possible in some
cases to distinguish the effects on the financial volatility of the
Sept. 11, 2001 attack or of the coup against Gorbachev on
Aug., 19, 1991 (exogeneous shocks) from financial crashes such as Oct. 1987 as well as
smaller volatility bursts (endogeneous shocks). 
Using a parsimonious autoregressive process (the ``multifractal random walk'')
with long-range memory defined on the logarithm of the volatility,
they predict strikingly different response functions of the price volatility
to great external shocks compared to endogeneous shocks, i.e.,
which result from the cooperative accumulation of many small shocks.
This approach views the origin of endogeneous shocks as the coherent
accumulations of tiny bad news, and thus provides a natural unification of previous
explanations of large crashes including Oct. 1987. 
Sornette and Helmstetter (2002) have suggested that these results
are generally valid for
systems with long-range persistence and memory, which can
exhibit different precursory as well as recovery patterns 
in response to shocks of exogeneous versus endogeneous origins.
By endogeneous, one can consider either fluctuations resulting
from an underlying chaotic dynamics or from a stochastic forcing origin
which may be external or be an effective coarse-grained description
of the microscopic fluctuations.
In this scenario, endogeneous shocks result from a kind of 
constructive interference of
accumulated fluctuations whose impacts survive longer than the
large shocks themselves. As a consequence, the recovery after
an endogeneous shock is in general slower at early times and can be
at long times
either slower or faster than after an exogeneous perturbation.
This offers the tantalizing possibility of distinguishing between
an endogeneous versus exogeneous cause of a given shock, even
when there is no ``smoking gun.'' This could help in 
investigating the exogeneous versus self-organized origins in problems
such as the causes of major biological extinctions, of 
change of weather regimes and of the climate,
in tracing the source of social upheaval and wars, and so on.

\vskip-0.3cm
\subsection{Predictions}

Ultimately, only forward predictions can demonstrate the usefulness of
a theory, thus only time will tell. 
However, as we have suggested by the
many examples reported in section 7, 
the analysis points to an interesting
predictive potential. However, a
fundamental question concerns the use of a reliable crash
prediction scheme, if any. Assume that a crash prediction is issued stating that a
crash of an amplitude between $20\%$ and $30\%$ will occur 
between one and two months from now. At least three different scenarios
are possible [Johansen and Sornette, 2000a]:
\begin{itemize}

\item Nobody believes the prediction which was then futile and, assuming
that the prediction was correct, the market crashes. One may consider
this as a victory for the ``predictors'' but as we have experienced in relation
to our quantitative prediction of the change in regime of the Nikkei index
[Johansen and Sornette, 1999c; 2000b], this would only be considered by some
critics just another
``lucky one'' without any statistical significance.

\item Everybody believes the warning, which causes panic and the market
crashes as consequence. The prediction hence seems self-fulfilling and the
success is attributed more to the panic effect than to a real predictive power.

\item Sufficiently many investors believe that the prediction {\it may} be correct,
investors make reasonable adjustments and the steam
goes off the bubble. The prediction hence disproves itself.

\end{itemize}

None of these scenarios are attractive. In the first two, the crash is not
avoided and in the last scenario the prediction disproves itself and as
a consequence the theory looks unreliable. This seems to be the inescapable
lot of scientific investigations of systems with learning and reflective
abilities, in contrast with the usual inanimate and unchanging physical laws
of nature. Furthermore, this touches the key-problem of scientific
responsibility. Naturally, scientists have a responsibility to publish their
findings. However, when it comes to the practical implementation of those
findings in society, the question becomes considerably more complex, as
history has taught us. We believe however that increased awareness of the
potential for market instabilities, offered in particular by our 
approach, will help in constructing a more stable and efficient stock market.

Specific guidelines for prediction and careful tests are presented in 
[Sornette and Johansen, 2001a] and especially in [Sornette, 2003]. In particular,
Sornette [2003]
explains how and to what degree crashes as well as other large market events,
may be predicted. This work examines in details
what are the forecasting skills of the proposed methodology 
and their limitations, in particular in terms
of the horizon of visibility and expected precision. Several cases studies
are presented in details, with a careful count of successes and failures.
See also [Johansen and Sornette, 2001b]
for applications to emergent markets and [Sornette and Zhou, 2002] for
a live prediction on the future evolution of the US stock market in the
next two years, from Aug. 2002 to the first semester of 2004.

\newpage

{\bf References}

\vskip 0.5cm

Adam, M.C. and A. Szafarz, Oxford Economic Papers 44, 626-640 (1992).

Andersen, J.V. and D. Sornette, 
Have your cake and eat it too: increasing returns while lowering large risks!
Journal of Risk Finance 2 (3), 70-82 (2001).

Andersen, J.V., S. Gluzman and D. Sornette,
Fundamental Framework for Technical Analysis, European Physical 
Journal B {\bf 14}, 579-601 (2000).

Anderson, P. W., K. J. Arrow and D. Pines, Editors, The economy
as an evolving complex system (Addison-Wesley, New York, 1988).

Arad, I., Biferale, L., Celani, A., Procaccia, I. and M. Vergassola,
Statistical conservation laws in turbulent transport - art. no. 164502,
Phys. Rev. Lett. 8716 N16:4502,U62-U64 (2001).

Assoe, K.G.,
Regime-switching in emerging stock market returns,
Multinational Finance Journal 2, 101-132  (1998).

Bak, P., How nature works : the science of self-organized criticality, 
New York, NY, USA : Copernicus (1996).

Barber, B.M. and Lyon, J.D.,
Detecting long-run abnormal stock returns: The empirical power and
specification of test statistics, Journal of Financial Economics 43, N3,
341-372 (1997).

Barra, F., Davidovitch, B. and Procaccia, I.,
Iterated conformal dynamics and Laplacian growth - art. no. 046144,
Phys. Rev. E 6504 N4 PT2A:U486-U497 (2002).

Barro, R.J., E.F. Fama, D.R. Fischel, A.H. Meltzer, R. Roll
and L.G. Telser, Black monday and the future of financial markets, edited by
R.W. Kamphuis, Jr., R.C. Kormendi and J.W.H. Watson (Mid American Institute for
Public Policy Research, Inc. and Dow Jones-Irwin, Inc., 1989).

Basle Committee on Banking Supervision,
Core Principles for Effective Banking Supervision, Basle
September 1997. 

Bassi, F., P.Embrechts, and M.Kafetzaki,
Risk Management and Quantile Estimation, in: Adler, R.J., R.E.Feldman, 
M.Taqqu, eds., A Practical Guide to Heavy Tails, Birkhauser, Boston, 111-30 (1998).

Bikhchandani, S., Hirshleifer, D. and Welch, I.,
A theory of fads, fashion, custom, and cultural change as informational cascades, 
Journal of Political Economy 100, 992Ð1026 (1992).

Blanchard, O.J., 1979, Economics Letters 3, 387 - 389.

Blanchard, O.J. and M.W. Watson, Bubbles, Rational Expectations and
Speculative Markets, in: Wachtel, P. ,eds., Crisis in Economic and
Financial Structure: Bubbles, Bursts, and Shocks. Lexington Books: Lexington (1982)

Boissevain, J. and Mitchell, J.,
Network analysis: Studies in human interaction (Mouton, 1973).

Bouchaud, J.-P. and R. Cont,
A Langevin approach to stock market fluctuations and crashes,
Eur. Phys. J. B 6, 543-550 (1998).

Cai, J.,
A Markov model of switching-regime ARCH,
Journal of Business \& Economic Statistics 12, 309-316  (1994).

Callen,~E. and Shapero,~D., A theory of social
imitation, Physics Today, July,~23-28 (1974).

Camerer, C., Bubbles and Fads in Asset Prices,
Journal of Economic Surveys {3, 3-41 (1989).

Campbell, J.Y., A.W. Lo, A.C. MacKinlay,
The econometrics of financial markets, Princeton, N.J. : Princeton
University Press (1997).

Chaitin, G.J., {\it Algorithmic Information Theory}
(Cambridge University Press, Cambridge and New York, 1987).

Chauvet, M.,
An econometric characterization of business cycle dynamics with factor
structure and regime switching,
International Economic Review 39, 969-996 (1998).

Checki, T.J. and Stern, E.,
Financial Crises in the Emerging Markets: The Roles of the Public and Private
Sectors, Current Issues in Economics and Finance 
(Federal Reserve Bank of New York)  6, (13), 1-6 (2000).

Chen, N.-F., C.J. Cuny and R.A. Haugen,
Stock volatility and the levels of the basis and open interest in futures
contracts, Journal of Finance 50, 281-300 (1995). 

Chowdhury, D. and D. Stauffer,
A generalized spin model of financial markets, Eur. Phys. J. B 8, 477-482 (1999).

Coe, P.J.,
Financial crisis and the great depression:  A regime switching approach,
Journal of Money, Credit, \& Banking 34 (1), 76-93 (2002).

Commission on Physical Sciences, Mathematics, and Applications,
Computing and Communications in the Extreme  Research for Crisis Management and
Other Applications, Steering Committee, Workshop Series on High Performance
Computing and Communications, Computer Science and Telecommunications Board
National Academy Press, Washington, D.C, 1990.

Cont., R. and Bouchaud, J.-P.,
Herd behavior and aggregate fluctuations in financial markets, Macroeconomic
Dynamics 4, 170-196 (2000).

Cootner, P.H., ed.,
The random character of stock market prices 
(Cambridge, Mass., M.I.T. Press, 1967).

Corcos, A., J.-P. Eckmann, A. Malaspinas, Y. Malevergne and D. Sornette,
Imitation and contrarian behavior: hyperbolic bubbles, crashes and chaos,
Quantitative Finance 2, 264Ð281 (2002)

Crutchfield,~J.P. and Mitchell,~M.,
The evolution of emergent computation,
Proc. Nat. Acad. Sci. U.S.A.  92,~10742-10746 (1995) .

De Bandt, O. and P. Hartmann, Systemic risk: a survey, Financial economics
and internation macroeconomics, Discussion paper series No. 2634 (2000).

Devenow, A. and Welch, I., Rational herding 
in financial markets, European Eco-nomic Review 40, 603Ð616 (1996).

Diebold, F.X., Schuermann, T. and Stroughair, J.D.,
Pitfalls and opportunities in the use of extreme value theory in risk
management, preprint (2001).

Driffill, J. and Sola, M.,
Intrinsic bubbles and regime-switching,
Journal of Monetary Economics 42, 357-373 (1998).

Drozdz, S., Ruf, F., Speth, J. and Wojcik, M.,
Imprints of log-periodic self-similarity in the stock market,
European Physical Journal 10, 589-593 (1999).

Dubrulle, B., F. Graner and D. Sornette, eds.,
Scale invariance and beyond (EDP Sciences and Springer, Berlin, 1997).

Dunning, T. J., `Trades' Unions and Strikes, London (1860).

Dupuis H.  {\em Un krach avant Novembre} 
in Tendances the 18. September 1997 page 26.

Embrechts, P., C.P.Kluppelberg, and T.Mikosh, Modelling Extremal Events, 
Springer-Verlag, Berlin, 645 pp (1997).

Falkovich, G., Gawedzki, K. and Vergassola, M.,
Particles and fields in fluid turbulence, Rev. Mod. Phys. 73 N4:913-975 (2001).

Fama, E.F., Efficient capital markets. 2., Journal of Finance 46, 1575-1617 (1991).

Farmer, J.D., Market force, ecology and  evolution, preprint
at adap-org/9812005 (1998).

Feigenbaum, J.A.,
A statistical analysis of log-periodic precursors to 
financial crashes, Quantitative Finance 1, 346Ð360 (2001).

Feigenbaum J.A. and Freund P.G.O.,
Discrete scale invariance in stock markets before crashes, 
Int. J. Mod. Phys. B 10, 3737-3745 (1996).

Feigenbaum J.A. and Freund P.G.O.,
Discrete scale invariance and the ''second black Monday'',
Modern Physics Letters B 12, 57-60 (1998).

Feldman, R.A.,
Dollar Appreciation, Foreign Trade, and the U.S. Economy,
Federal Reserve Bank of New York Quarterly Review 7, 1-9 (1982).

Fieleke, N.S.,
Dollar Appreciation and U.S. Import Prices,
New England Economic Review (Nov/Dec 1985), 49-54 (1985).

Frankel, J.A. and Froot, K.A.,
Chartists, Fundamentalists and the Demand for Dollars,
Greek Economic Review 10, 49-102 (1988).

Frankel, J.A. and Froot, K.A.,
Chartists, Fundamentalists, and Trading in the Foreign Exchange Market,
American Economic Review 80, 181-185  (1990).

Galbraith, J.K.,
 The great crash, 1929 (Boston : Houghton Mifflin Co., 1997).
 
Gaunersdorfer, A.,
Endogenous fluctuations in a simple asset pricing model with heterogeneous agents,
Journal of Economic Dynamics \& Control 24, 799-831 (2000).

Geller, R.J., Jackson, D.D., Kagan, Y.Y. and Mulargia, F.,
Geoscience - Earthquakes cannot be predicted, Science 275 N5306,1616-1617 (1997).

Geller, R.J., Jackson, D.D., Kagan, Y.Y. and Mulargia, F.,
Cannot earthquakes be predicted? - Responses, Science 278 N5337, 488-490 (1997)

Gluzman, S. and Yukalov, V.I.,
Booms and crashes in self-similar markets, Modern Physics Letters B 12, 575-587 (1998).

Goldenfeld, N.,
Lectures on phase transitions and the renormalization group
(Addison-Wesley Publishing Company, Reading, Massachussets, 1992).

Gorte, R.W., Forest Fires and Forest Health,
Congressional Research Service Report, The Committee for the National
Institute for the Environment, 1725 K Street, NW, Suite
212, Washington, D.C. 20006 (1995).

Gould, S.J. and Eldredge, N., Punctuated equilibrium
comes of age, Nature 366, 223-227 (1993). 

Graham, J.R. (1999)  Herding among Investment Newsletters: 
Theory and Evidence, Journal of Finance 54, 237-268.

Grant, J.L.,
Stock Return Volatility During the Crash of 1987,
Journal of Portfolio Management 16, 69-71 (1990).

Grassia, P.S., Delay, feedback and quenching in financial 
markets, Eur. Phys. J. B 17, 347-362 (2000).

Gray, S.F.,
Regime-switching in Australian short-term interest rates,
Accounting \& Finance 36, 65-88 (1996).

Grinblatt, M., Titman, S. and Wermers, R.,
Momentum investment strategies, portfolio performance, and herding: A
study of mutual fund behavior,
American Economic Review 85, 1088-1105 (1995).

Hamilton, J.B.,
A New Approach to the Economic Analysis of Nonstationary Time
Series and the Business Cycle, Econometrica 57, 357-384 (1989).

Harris, L.,
Circuit Breakers and Program Trading Limits: What Have We Learned?
in ``The 1987 Crash, Ten Years Later: Evaluating the Health of the Financial Markets'',
October 1997 conference, published in volume II
of the annual Brookings-Wharton Papers on Financial Services, The Brookings
Institution Press, Washington, D.C., 1997)

Helbing, D., Farkas, I. and Vicsek, T. (2000) 
Simulating Dynamical Features of Escape Panic, Nature 407, 487-490.

Holland, J.H., Complex adaptive systems, Daedalus 121, 17-30 (1992).

Holldobler, B. and E.O. Wilson, Journey to the ants : a story of scientific exploration
(Cambridge, Mass.  Belknap Press of Harvard University Press, 1994).

Holmes, P.A. (1985)
How Fast Will the Dollar Drop? Nation's Business {\bf 73}, 16.

Hsieh, D.A.,
Nonlinear dynamics in financial markets: evidence and implications,
Financial Analysts Journal (july-August), 55-62 (1995).

Huberman, G. and Regev, T., Contagious speculation
and a cure for cancer: a nonevent that made stock prices soar, J. Finance 56, 387-396 (2001) .

Ide, K. and D. Sornette,
Oscillatory Finite-Time Singularities in Finance, Population and Rupture, 
Physica A  307 (1-2), 63-106 (2002).

Johansen A, Ledoit O, Sornette D., Crashes as critical points,
International Journal of Theoretical and Applied Finance 3, 219-255 (2000).

Johansen, A. and D. Sornette, Stock market crashes are outliers, 
European Physical Journal B 1, 141-143 (1998).

Johansen, A. and D. Sornette, Critical Crashes, Risk 12 (1), 91-94 (1999a),

Johansen, A. and D. Sornette,
Modeling the stock market prior to large crashes,
Eur. Phys. J. B 9 (1), 167-174 (1999b).

Johansen, A. and D. Sornette,
Financial ``anti-bubbles'': log-periodicity in Gold and Nikkei collapses,
Int. J. Mod. Phys. C 10, 563-575 (1999c).
 
Johansen, A. and D. Sornette,
The Nasdaq crash of April 2000: Yet another example of
log-periodicity in a speculative bubble ending in a crash,
European Physical Journal B 17, 319-328 (2000a). 

Johansen, A. and Sornette, D.,
Evaluation of the quantitative prediction of a trend reversal on
the Japanese stock market in 1999, Int. J. Mod. Phys. C 11, 359-364 (2000b).

Johansen, A. and D. Sornette (2001a)
Finite-time singularity in the dynamics of the world population and economic indices,
Physica A 294, 465-502.

Johansen, A. and D. Sornette, 
Bubbles and anti-bubbles in
Latin-American, Asian and Western stock markets: An empirical study, 
International Journal of Theoretical and
Applied Finance 4 (6), 853-920 (2001b).

Johansen, A. and D. Sornette,
Large Stock Market Price Drawdowns Are Outliers, 
Journal of Risk 4(2), 69-110 (2002).

Johansen A., Sornette D. and  Ledoit O.,
Predicting Financial Crashes Using Discrete Scale Invariance, Journal of Risk 1, 5-32 (1999).

Kadanoff, L.P., Wolfram on Cellular Automata; A Clear and
Very Personal Exposition, Physics Today, July (2002)

Kaminsky, G. and Peruga, R.,
Credibility Crises: The Dollar in the Early 1980s,
Journal of International Money \& Finance 10, 170-192 (1991).

Karplus, W.J., The Heavens are Falling: The Scientific Prediction of
Catastrophes in Our Time (New York: Plenum Press, 1992).

Keynes, J.M., The general theory of employment, interest
and money (Harcourt, Brace, New York, Chap. 12, 1936).

Kindleberger, C.P.,
Manias, panics, and crashes: a history of financial crises (4th ed.  New York: Wiley, 2000).

Kirman, A.,
Epidemics of opinion and speculative bubbles in financial markets,
In M. Taylor (ed.), Money and financial markets, Macmillan (1991).

Knetter, M.M.,
Did the strong dollar increase competition in U.S. product markets?
Review of Economics \& Statistics 76, 192-195 (1994).

Knuth, D.E., The art of computer programming, vol.2, 1-160,
Addison-Wesley Publ. (1969).

Koller, T. and D.W. Zane,
What happened to the bull market? The McKinsey Quarterly Newsletter 4,
(August 2001), http://www.mckinseyquarterly.com

Krawiecki, A., J.A. Holyst and D. Helbing, 
olatility Clustering and Scaling for Financial Time Series due to Attractor Bubbling,
Phys. Rev. Lett. 89 (15), art. 158701 (2002).

Krugman, P.,
I Know What the Hedgies Did Last Summer, Fortune, December issue (1998).
                            
Laherr\`ere, J. and Sornette, D.,
Stretched exponential distributions in Nature and Economy: ``Fat tails''
with characteristic scales, European Physical Journal B 2, 525-539 (1998).

Lamont, O., Earnings and expected returns,
The Journal of Finance LIII, 1563-1587 (1988).

Levy, M., H. Levy and S. Solomon,
Microscopic simulation of the stock market -- the effect of microscopic
diversity, J. Physique I 5, 1087-1107 (1995).

Levy, M., H. Levy and S. Solomon, 
The microscopic simulation of financial markets: from investor behavior
to market phenomena (Academic Press, San Diego, 2000).

Liggett, T.M., Interacting particle systems
(New York: Springer-Verlag, 1985).

Liggett, T.M.,
Stochastic models of interacting systems, The Annals of Probability 25, 1-29 (1997).

Lux, T.,
Herd behaviour, bubbles and crashes,
Economic Journal: The Journal of the Royal Economic Society 105, 881-896 (1995).

Lux, T.,
The socio-economic dynamics of speculative markets: interacting agents,
chaos, and the fat tails of return distributions,
Journal of Economic Behavior \& Organization 33, 143-165 (1998).

Lux, T. and Marchesi, M.,
Scaling and criticality in a stochastic multi-agent model of a financial
market, Nature 397, 498-500 (1999).

Lux, T. and Marchesi, M.,
 Volatility clustering in financial markets: a micro-simulation of interacting agents,
International Journal of Theoretical and Applied Finance 3, 675-702 (2000). 

L'vov, V.S., Pomyalov, A. and Procaccia, I.,
Outliers, Extreme Events and Multiscaling, Phys. Rev. E 6305, 
PT2:6118, U158-U166 (2001).

MacDonald, R. and Torrance, T.S.,
On Risk, Rationality and Excessive Speculation in the Deutschmark-US
Dollar Exchange Market: Some Evidence Using Survey Data,
Oxford Bulletin of Economics \& Statistics 50, 107-123 (1988).

Malamud, B.D., Morein, G. and Turcotte, D.L.,
Forest fires: An example of self-organized critical behavior,
Science 281, 1840-1842  (1998).

Malkiel B.G., A random walk down
Wall Street (WW Norton \& Company: New York, 1999).

Mauboussin, M.J. and Hiler, R.,
Rational Exuberance? Equity research report of Credit Suisse First Boston,
January 26 (1999).

Maug, E. and Naik, N., Herding and delegated portfolio management: The
impact of relative performance evaluation on asset allocation, Working paper, Duke University
(1995).

McNeil, A.J., Extreme value theory for risk managers, preprint
ETH Zentrum Zurich (1999).

Megginson, W.L., The impact of privatization on 
capital market development and individual share
 ownership, presentation at the 3rd FIBV global emerging markets conference and exhibition,
 Istanbul, April 5-7 (2000)\\
 $http://www.oecd.org/daf/corporate-affairs/privatisation/capital-markets/megginson/sld001.htm$

Miltenberger,~P.,~Sornette,~D. and Vanneste,
Fault self-organization as optimal random paths selected by 
critical spatio-temporal dynamics of
earthquakes,  Phys. Rev. Lett. 71,~3604-3607 (1993).

Minnich, R.A. and Chou, Y.H.,
Wildland fire patch dynamics in the chaparral of southern California and
northern Baja California, International Journal of Wildland Fire 7, 221-248 (1997).

Montroll, E.W. and Badger, W.W.,
{\it Introduction to quantitative aspects of social phenomena} (New York: Gordon and Breach, 1974).

Mood, A., The distribution theory of runs, Annals of Mathematical Statistics 11,
367-392 (1940).

Moreno, J.M., ed.,
Large forest fires (Leiden: Backhuys Publishers, 1998).

Moss de Oliveira, S., de Oliveira, P.M.C and Stauffer, D.,
Evolution, Money, War and Computers (Teubner, Stuttgart-Leipzig, 1999).

Mulligan, C.B. and Sala-i-Martin, X.,
Extensive margins and the demand for money at low interest rates, Journal
of Political Economy (2000).

Nature debates, Is the reliable prediction of individual earthquakes a
realistic scientific goal?   $http://helix.nature.com/debates/earthquake/$ (1999)

Onsager, L., Crystal statistics. I. 
A two-dimensional model with an order-disorder transition,
Physics Review 65, 117-149 (1944).

Orl\'ean, A., Mim\'etisme et anticipations rationnelles:
une perspective keynesienne, Recherches Economiques de Louvain 52, 45-66  (1984).

Orl\'ean, A., L'auto-r\'ef\'erence dans la th\'eorie 
keynesienne de la sp\'eculation, Cahiers d'Economie Politique 14-15 (1986).

Orl\'ean, A., Comportements mim\'etiques et diversit\'e
des opinions sur les march\'es financiers, Chap. III, in Th\'eorie \'economique
et crises des march\'es financiers, Bourguinat, H. and Artus, P., editors
(Economica, Paris) pp. 45-65 (1989).

Orl\'ean, A., Mimetic contagion and speculative bubbles,
Theory and Decision 27, 63-92 (1989).

Orl\'ean, A., Disorder in the stock market (in french)
La Recherche 22, 668-672 (1991).

Orl\'ean, A., Bayesian interactions and collective
dynamics of opinion - Herd behavior and mimetic contagion,
Journal of Economic Behavior \& Organization 28, 257-274 (1995).

Pandey, R. B., and Stauffer, D.,
Search for log-periodicity oscillations in stock market simulations,
International Journal of Theoretical and Applied Finance 3, 479-482 (2000).

Phoa, W., Estimating credit spread risk using extreme value theory -- Application of
actuarial disciplines to finance, Journal of Portfolio Management 25, 69-73 (1999).

Potters, M., R. Cont and J.-P. Bouchaud,
Financial markets as adaptative ecosystems, Europhysics Letters 41, 239-244 (1998).

Press, W.H. {\it et al.~}, Numerical Recipes Cambridge University Press (1992).

Roehner, B.M. and D. Sornette,
The sharp peak-flat trough pattern and critical speculation, 
European Physical Journal B 4, 387-399 (1998).

Roehner, B.M.  and D. Sornette,
``Thermometers'' of Speculative Frenzy, European Physical Journal B 16, 729-739 (2000).

Roll, R., The International Crash of October 1987,
Financial Analysts Journal 4 (5), 19-35 (1988).

Romer, C.D.,
The Great Crash and the Onset of the Great Depression,
Quarterly Journal of Economics 105, 597-624 (1990).

Saleur, H. and D. Sornette,
Complex exponents and log-periodic corrections in frustrated systems, 
J.Phys.I France 6, 327-355 (1996).

Sato, A.H. and Takayasu, H.,
Dynamic numerical models of stock market price: from microscopic
determinism to macroscopic randomness, Physica A 250, 231-252 (1998).
      
Schaller, H. and van Norden, S., Regime switching in stock market returns,
Applied Financial Economics 7, 177-191 (1997).

Scharfstein, D. and Stein, J.,
Herd behavior and investment, American Economic Review 80, 465Ð479 (1990).

Shefrin, H.,
Beyond greed and fear : understanding behavioral finance and the
psychology of investing (Boston, Mass.: Harvard Business School Press, 2000).

Shiller, R.J.,
{\it Market volatility} (Cambridge, Mass.: MIT Press, 1989).

Shiller, R.J., {\it Irrational exuberance} (Princeton University Press, Princeton, N.J., 2000).

Shleifer, A.,
Inefficient markets : an introduction to behavioral finance
(Oxford; New York : Oxford University Press, 2000).

Sircar, R. and G. Papanicolaou,
General Black-Scholes models accounting for increased market volatility 
from hedging strategies, Applied Mathematical Finance 5, 45-82 (1998). 

Sornette D., Discrete scale invariance and complex dimensions,
Physics Reports 297, 239-270 (1998).

Sornette, D.,
Complexity, catastrophe and physics, Physics World 12, (N12), 57-57 (1999).

Sornette, D., Critical Phenomena in Natural Sciences,
Chaos, Fractals, Self-organization and Disorder: Concepts and Tools,
(Springer Series in Synergetics, Heidelberg, 2000a).

Sornette, D., Stock Market Speculation: Spontaneous Symmetry 
Breaking of Economic Valuation, Physica A 284, 355-375 (2000b).

Sornette, D., 
Predictability of catastrophic events: material rupture, earthquakes, turbulence, 
financial crashes and human birth, Proceedings of the National Academy of 
Sciences USA, V99 SUPP1:2522-2529 (2002).

Sornette, D., Why Stock Markets Crash: Critical Events in Complex Financial Systems
(Princeton University Press, Princeton, N.J., 2003) 456 pages, 165 figures, 21 tables. 

Sornette, D. and J.V. Andersen,
A Nonlinear Super-Exponential Rational Model of Speculative Financial Bubbles,
Int. J. Mod. Phys. C 13 (2), 171-188 (2002).

Sornette, D., J.V. Andersen and P. Simonetti,
Portfolio Theory for ``Fat Tails,''
International Journal of Theoretical and Applied Finance 3 (3), 523-535 (2000).

Sornette, D. and A. Helmstetter,
Endogeneous Versus Exogeneous Shocks in Systems with Memory, in press
in Physica A
(http://arXiv.org/abs/cond-mat/0206047)

Sornette, D. and K. Ide,
Theory of self-similar oscillatory finite-time singularities
in Finance, Population and Rupture, Int. J. Mod. Phys. C 14 (3) (2002)
(e-print at http://arXiv.org/abs/cond-mat/0106054).

Sornette D. and Johansen A. 
Large financial crashes, Physica A 245, 411-422 (1997).

Sornette D. and A. Johansen, A Hierarchical Model of Financial Crashes,
Physica A 261, 581-598 (1998).

Sornette, D. and A. Johansen,
Significance of log-periodic precursors to financial crashes,
Quantitative Finance 1 (4), 452-471 (2001).

Sornette, D., A. Johansen and J.-P. Bouchaud,
Stock market crashes, Precursors and Replicas, J.Phys.I France 6, 167-175 (1996).

Sornette, D., Y. Malevergne and J.F. Muzy,
Volatility fingerprints of large shocks: Endogeneous versus exogeneous,
preprint at http://arXiv.org/abs/cond-mat/0204626 (2002).

Sornette,~D.,~Miltenberger,~P. and Vanneste,~C.,
Statistical physics of fault patterns self-organized by repeated earthquakes,~
Pure and Applied Geophysics 142,~491-527 (1994).

Sornette, D., P. Simonetti and J. V. Andersen,
 $\phi^q$-field theory for Portfolio optimization: ``fat tails'' and
non-linear correlations, Physics Reports 335, 19-92 (2000).

Sornette, D. and W.-X. Zhou,
The US 2000-2002 Market Descent: How Much Longer and Deeper?
in press in Quantitative Finance (2002)
(http://arXiv.org/abs/cond-mat/0209065)

Stauffer, D., Monte-Carlo-Simulation mikroskopischer
B\"{o}rsenmodelle, Physikalische Bl\"{a}tter 55, 49 (1999).

Stauffer,~D. and Aharony,~A., Introduction to Percolation
Theory, 2nd ed. (Taylor \& Francis,~London; Bristol,~P.A., 1994).

Stauffer, D. and D. Sornette,
Self-Organized Percolation Model for Stock Market Fluctuations,
Physica A 271, 496-506 (1999).

Takayasu, H., Miura, H., Hirabayshi, T. and Hamada K.,
Statistical properties of deterministic threshold elements -- The
case of the market price, Physica A 184, 127-134 (1992).

Thaler, R.H., editor,
Advances in behavioral finance (New York : Russell Sage Foundation, 1993).

Trueman, B., Analyst forecasts and herding behavior, 
The Review of Financial Studies 7, 97Ð124 (1994).

Van Norden, S. and Schaller, H.,
The predictability of stock market regime: Evidence from the Toronto Stock
Exchange, Review of Economics \& Statistics 75, 505-510 (1993).

Van Norden, S., Regime switching as a test for exchange rate bubbles,
Journal of Applied Econometrics 11, 219-251 (1996).

Vandewalle, N., Boveroux, P., Minguet, A. and Ausloos, M.,
The crash of October 1987 seen as a phase transition: amplitude and
universality, Physica A 255, 201-210 (1998a).  

Vandewalle, N., Ausloos, M., Boveroux, P. and Minguet, A.,
How the financial crash of October 1997 could have been predicted,
European Physical Journal B 4, 139-141 (1998b).

Welch, I.,
 Sequential sales, learning, and cascades, Journal of Finance 47, 695Ð732 (1992).
 see also http://welch.som.yale.edu/cascades for
an annotated bibliography and resource reference on ``information cascades''.

Welch, I.,
Herding among security analysts, Journal of Financial Economics 58 (3), 369-396 (2000).

White E.N., Stock market crashes and
speculative manias. In {\em The international library of macroeconomic and
financial history} 13. An Elgar Reference Collection, Cheltenham, UK;
Brookfield, US (1996).

White, E.N. and Rappoport, P.,
The New York Stock Market in the 1920s and 1930s:
Did Stock Prices Move Together Too Much?
Anglo-American Financial Systems: Institutions and Markets in the
TwentiethCentury, ed. by M. Bordo and R. Sylla, Burr Ridge Irwin,
pp. 299-316 (1995).
   
Wilson, E.O., The insect societies (Cambridge, Mass., Belknap
Press of Harvard University Press, 1971).

Wilson, K.G.,
Problems in Physics with many scales of length,
Scientific American 241 (2), 158-179 (1979).

Wolfram, S., A new kind of science (Wolfram Media, Inc.; ISBN: 1579550088; 2002) 

Youssefmir, M. and B.A. Huberman and T. Hogg,
Bubbles and Market Crashes, Computational Economics 12, 97-114 (1998).

Zwiebel, J., Corporate conservatism and 
relative compensation, Journal of Political Economy 103, 1Ð25 (1995).

Zhou, W.-X. and D. Sornette,
Statistical Significance of Periodicity and Log-Periodicity 
with Heavy-Tailed Correlated Noise, Int. J. Mod. Phys. C 13 (2), 137-170 (2002a).

Zhou, W.-X. and D. Sornette, Generalized q-Analysis of
Log-Periodicity: Applications to Critical Ruptures, in press in Phys. Rev. E (2002b)
in press, http://arXiv.org/abs/cond-mat/0201458.

Zhou, W.-X. and D. Sornette,
Non-Parametric Analyses of Log-Periodic Precursors to Financial Crashes
(preprint at http://arXiv.org/abs/cond-mat/0205531) (2002c).

\end{document}